\documentclass[conference,compsoc]{IEEEtran}

\usepackage{ifthen} %% provides `\newboolean`, `\setboolean`, and `\ifthenelse`

\newboolean{our_draft}
\setboolean{our_draft}{false}

\newboolean{colored}
\setboolean{colored}{false}

\def\conferenceversion{0}
\def\authorversion{1}
\def\extendedversion{2}
\def\paperversion{\extendedversion}
\newcommand{\ifisconferenceversion}[1]{\ifthenelse{\equal{\paperversion}{\conferenceversion}}{#1}{}}
\newcommand{\ifisauthorversion}[1]{\ifthenelse{\equal{\paperversion}{\authorversion}}{#1}{}}
\newcommand{\ifisextendedversion}[1]{\ifthenelse{\equal{\paperversion}{\extendedversion}}{#1}{}}
\usepackage{microtype}
\usepackage[nocompress]{cite}
\usepackage{tikz}
\usepackage{amsmath}
\usepackage{amssymb} %% for \triangleq
\usepackage{balance} %% provides command `\balance` to balance the last page of the appendix
\usepackage{glossaries}
\usepackage{glossaries-extra}
\usepackage{xspace}
\usepackage{xfrac} %% for nice fractions
\usepackage{orcidlink} %% for ORCIDs in author list
\usepackage{comment} %% for comment environment
\usepackage{newtxtext} % Provides Times font with small caps & italics support
\usepackage{anyfontsize} % fixes warning "Size substitutions with differences"
\usepackage{enumitem} % allows to configure enumerate environments
\usepackage{pifont} % provides \ding{...} for checkmarks and crosses (used to itemize our contributions)
\usepackage{siunitx} % for typesetting units
\usepackage[T1]{fontenc} % needed for less than symbol in connection with siunitx package
\usepackage{relsize} % for \smaller command
\usepackage{tabularx} % for tabularx environment

\ifthenelse{\not\equal{\paperversion}{\extendedversion}}{%
    \usepackage[available,functional,reproduced]{ieeebadges}
}

\setabbreviationstyle[acronym]{long-short-sc}

\newboolean{show_full_ssm_agent_protocol}
\newboolean{show_full_soundness_appendix}
\newboolean{show_composition_soundness_proofs}
\newboolean{show_meta_review}

\ifisconferenceversion{%
    \setboolean{show_meta_review}{true}
    \setboolean{show_full_soundness_appendix}{false}
    \setboolean{show_composition_soundness_proofs}{false}
    \setboolean{show_full_ssm_agent_protocol}{false}
}
\ifisauthorversion{%
    \setboolean{show_meta_review}{false}
    \setboolean{show_full_soundness_appendix}{false}
    \setboolean{show_composition_soundness_proofs}{false}
    \setboolean{show_full_ssm_agent_protocol}{false}
}
\ifisextendedversion{%
    \setboolean{show_meta_review}{false}
    \setboolean{show_full_soundness_appendix}{true}
    \setboolean{show_composition_soundness_proofs}{true}
    \setboolean{show_full_ssm_agent_protocol}{true}
}

\hypersetup{
    colorlinks=true,
    linkcolor=blue,
    filecolor=magenta,      
    urlcolor=cyan,
    pdftitle={Diodon},
    pdfpagemode=FullScreen,
}

\usepackage{xspace}

\newcommand*{\tool}[1]{\textsc{#1}\xspace}
\newcommand*{\core}{\tool{Core}}
\newcommand*{\app}{\tool{Application}}
\newcommand*{\approach}{\tool{Diodon}}
\newcommand*{\ioindependence}{\gls{io-independence}\xspace}
\newcommand*{\ioIndependence}{\glstitle{io-independence}\xspace}

\newcommand*{\argot}{\tool{Argot}}
\newcommand*{\gobra}{\tool{Gobra}}

\newcommand*{\tamarin}{\tool{\gls*{tamarin}}} % starred version disables the hyperlink
\newcommand*{\proverif}{\tool{ProVerif}}

\newcommand*{\verifast}{\tool{VeriFast}}
\newcommand*{\vst}{\tool{VST}}
\newcommand*{\nagini}{\tool{Nagini}}
\newcommand*{\prusti}{\tool{Prusti}}

\newcommand*{\fstar}{\tool{F$^{\star}$}}

\newcommand*{\owlc}{\tool{OwlC}}

\newcommand*{\capslock}{\tool{Capslock}}

\newcommand*{\astree}{\tool{Astr\'ee}}

\ifthenelse{\boolean{our_draft}}{%our_draft
    \newcommand{\todo}[1]{{\color{red}\textbf{(TODO:} #1\textbf{)}}}
    \newcommand{\fix}[1]{\footnote{#1}}

    \newenvironment{newtext}{\color{blue}}{}

    \newcommand{\remove}[1]{\textcolor{red}{#1}}
    
}{
    \ifthenelse{\boolean{colored}}{%colored
      \newenvironment{newtext}{\color{blue}}{}
      
    }{

    }
    \newcommand{\todo}[1]{}
    \newcommand{\fix}[1]{}

    \excludecomment{removetext}
    \newcommand{\remove}[1]{}
    
}

\usepackage{suffix}
\newcommand*{\Ie}{I.e.,\xspace}
\newcommand*{\ie}{i.e.,\xspace}
\newcommand*{\Eg}{E.g.,\xspace}
\newcommand*{\eg}{e.g.,\xspace}

\newcommand*{\cf}{cf.\xspace}

\newcommand*{\wrt}{w.r.t.\xspace}

\newcommand*{\etal}{et~al.\xspace}
\newcommand{\quotes}[1]{``#1''}

\newcommand*{\symb}[1]{\ensuremath{\mathit{#1}}}

\newcommand{\mypar}[1]{\smallskip\noindent\textbf{#1.}}
\newcommand*{\Figref}[1]{Fig.~\ref{fig:#1}}
\newcommand*{\figref}{\Figref}

\newcommand*{\Secref}[1]{Sec.~\ref{sec:#1}}
\newcommand*{\secref}{\Secref}
\newcommand{\Appprefprefix}{App.}

\newcommand{\appprefprefix}{\Appprefprefix}
\newcommand*{\appref}[1]{\appprefprefix~\ref{app:#1}}
\newcommand*{\Lineref}[1]{Line~\ref{line:#1}}
\newcommand*{\lineref}[1]{line~\ref{line:#1}}
\newcommand*{\Linerange}[2]{Lines~\ref{line:#1}--\ref{line:#2}}
\newcommand*{\linerange}[2]{lines~\ref{line:#1}--\ref{line:#2}}
\newcommand*{\Thmref}[1]{Thm.~\ref{thm:#1}}
\newcommand*{\thmref}{\Thmref}
\newcommand*{\Lemref}[1]{Lemma~\ref{lem:#1}}
\newcommand*{\lemref}{\Lemref}

\newcommand*{\Defref}[1]{Def.~\ref{def:#1}}
\newcommand*{\defref}{\Defref}
\newcommand*{\Asmref}[1]{Asm.~\ref{asm:#1}}
\newcommand*{\asmref}{\Asmref}
\newcommand*{\Corref}[1]{Cor.~\ref{cor:#1}}
\newcommand*{\corref}{\Corref}
\WithSuffix{\newcommand*}\Figref*[1]{Fig.~#1}
\WithSuffix{\newcommand*}\figref*{\Figref*}
\WithSuffix{\newcommand*}\Secref*[1]{Sec.~#1}
\WithSuffix{\newcommand*}\secref*{\Secref*}
\WithSuffix{\newcommand*}\Lineref*[1]{Line~#1}
\WithSuffix{\newcommand*}\lineref*[1]{line~#1}
\WithSuffix{\newcommand*}\Lines*[1]{Lines~{#1}}
\WithSuffix{\newcommand*}\lines*[1]{lines~{#1}}
\WithSuffix{\newcommand*}\Linerange*[1]{Lines~{#1}}
\WithSuffix{\newcommand*}\linerange*[1]{lines~{#1}}

\newcommand*{\citeauthors}[2]{#1~\etal~\cite{#2}}

\newenvironment{myitemize}{\begin{itemize}[align=left, leftmargin=*, label={\ding{228}}]}{\end{itemize}}

\usepackage{listings}
\usepackage{newfloat}
\DeclareFloatingEnvironment[
    listname={List of Protocols},
    name=Protocol,
    placement=!ht,
]{protocol}

\DeclareFloatingEnvironment[
    listname={List of Attacks},
    name=Attack,
    placement=!ht,
    within=section,
]{attack}

\usepackage{amsthm}
\newtheorem{definition}{Definition}

\newtheorem{lemma}{Lemma}
\newtheorem{theorem}{Theorem}

\newtheorem{assumption}{Assumption}

\newcommand*{\proofsketchname}{Proof sketch}
\newenvironment{proofsketch}{\begin{proof}[\proofsketchname]}{\end{proof}}

\definecolor{gobracommentcolor}{HTML}{747678}

\lstdefinelanguage{gobra}{
  language=go,
  sensitive=true,
  morecomment=[l]{//},
  morecomment=[s]{/*}{*/},
  morekeywords=[1]{ %% Keywords of the programming language subset
    pred, implements, ghost, set, match, foreach, of
  },
  morekeywords=[2]{ %% Keywords of the specification language subset
    requires, ensures, invariant, req, ens, pres, pure, unfolding, in, forall, acc, let
  },
  morekeywords=[3]{ %% Keywords of the proof language subset
    fold, unfold,
    assume, assert, inhale, exhale
  },
  basicstyle={\ttfamily\footnotesize},
  commentstyle={\color{gobracommentcolor}\textit},
  keywordstyle={[1]\color[HTML]{0005FF}},
  keywordstyle={[2]\color[HTML]{CC5500}},
  keywordstyle={[3]\color[HTML]{EC008C}},
  mathescape=true,
  moredelim=**[is][\normalfont\itshape]{'}{'}
}

\lstnewenvironment{gobraenv}[1][]{
  \lstset{
    language=gobra,
    floatplacement={tbp},
    captionpos=b,
    frame=lines,
    numbers=left,
    numberblanklines=false,
    xleftmargin=8pt,
    xrightmargin=8pt,
    numberstyle=\scriptsize,
    breaklines=true,
    #1
  }
}{}

\def\code{%
    \lstinline[language=gobra,basicstyle=\ttfamily]}

\def\codebw{%
    \lstinline[language=gobra,basicstyle=\ttfamily\color{black},keywordstyle=\color{black},keywordstyle={[1]\color{black}},keywordstyle={[2]\color{black}},keywordstyle={[3]\color{black}}]}

\lstdefinelanguage{tamarin}{
  language=go,
  sensitive=true,
  morecomment=[l]{//},
  morecomment=[s]{/*}{*/},
  morekeywords=[1]{
    aenc, sdec, senc, sdec, sign, verify, hashing, multiset, revealSign, revealVerify, getMessage, true
  },
  morekeywords=[2]{
    axiom, begin, builtins, end, equations, functions, heuristic, in, let, options, predicate, predicates, property, protocol, restriction, section, subsection, text, theory, verdictfunction
  },
  morekeywords=[3]{
    new, in, out, lookup, as, in, else, if, lock, unlock, event, insert, delete, then, account, accounts, for, parties, otherwise
  },
  morekeywords=[4]{
    use_induction, sources, reuse, hide_lemma, left, right
  },
  morekeywords=[5]{
    F, T, All, Ex
  },
  basicstyle={\ttfamily\footnotesize},
  commentstyle={\color{gobracommentcolor}\textit},
  keywordstyle={[1]\color[HTML]{0005FF}},
  keywordstyle={[2]\color[HTML]{CC5500}},
  keywordstyle={[3]\color[HTML]{EC008C}},
  mathescape=true
}

\lstnewenvironment{tamarinenv}[1][]{
  \lstset{
    language=tamarin,
    floatplacement={tbp},
    captionpos=b,
    frame=lines,
    numbers=left,
    numberblanklines=true,
    xleftmargin=8pt,
    xrightmargin=8pt,
    numberstyle=\scriptsize,
    breaklines=true,
    #1
  }
}{}

\def\codetamarin{%
    \lstinline[language=tamarin,basicstyle=\ttfamily]}

\makeatletter
\lst@AddToHook{OnEmptyLine}{\vspace{-0.4\baselineskip}}
\makeatother

\newcommand{\condition}[1]{(\conditiondef{#1})}
\newcommand{\conditiondef}[1]{C#1}

\newcommand*{\fullversionsecref}[1]{\cite[#1]{full-version}\xspace}

\newboolean{display_both_references_for_checking}
\setboolean{display_both_references_for_checking}{false}

\newcommand*{\soundnessproofappendixletter}{A}
\newcommand*{\fullsoundnessref}[2]{%
    \ifthenelse{\boolean{show_full_soundness_appendix}}{%
        \ifthenelse{\boolean{display_both_references_for_checking}}{%
            \textcolor{orange}{#1~(\fullversionsecref{#2})}% show both references for easier checking
        }{%
            #1% directly reference the appendix
        }%
    }{%
        \fullversionsecref{#2}%
    }%
}
\newcommand*{\refsoundnessproof}{%
    \fullsoundnessref
        {\appref{diodon-soundness-proof}}
        {\Appprefprefix~\soundnessproofappendixletter}}

\newcommand*{\refsoundnessioindependence}{%
    \fullsoundnessref
        {\appref{diodon-soundness-io-independence}}
        {\Appprefprefix~\soundnessproofappendixletter.1}}

\newcommand*{\refsoundnessioindependencesoundness}{%
    \fullsoundnessref
        {\thmref{diodon-ioindependence-soundness}}
        {Thm.~3}}

\newcommand*{\compositionsoundnessappendixnumber}{\soundnessproofappendixletter.2}
\newcommand*{\refcompositionsoundness}{%
    \fullsoundnessref
        {\appref{diodon-composition-soundness}}
        {\Appprefprefix~\compositionsoundnessappendixnumber}}
\newcommand*{\refproofrules}{%
    \fullsoundnessref
        {\appref{diodon-proof-rules}}
        {\Appprefprefix~\compositionsoundnessappendixnumber.2}}
\newcommand*{\refsoundnessstaticanalyses}{%
    \fullsoundnessref
        {\appref{diodon-soundness-static-analyses}}
        {\Appprefprefix~\compositionsoundnessappendixnumber.3}}
\newcommand*{\refsideconditionshold}{%
    \fullsoundnessref
        {\lemref{side-conditions-hold}}
        {Lemma~10}}
\newcommand*{\refcrashfreedomassumption}{%
    \fullsoundnessref
        {\asmref{crash-freedom}}
        {Asm.~2}}
\newcommand*{\refdataracefreedomassumption}{%
    \fullsoundnessref
        {\asmref{data-race-freedom}}
        {Asm.~3}}
\newcommand*{\refsyntacticrestrictionsassumption}{%
    \fullsoundnessref
        {\asmref{syntactic-restrictions}}
        {Asm.~4}}

\newcommand*{\fullprotocolappendixletter}{B}
\newcommand*{\fullprotocolref}[2]{%
    \ifthenelse{\boolean{show_full_ssm_agent_protocol}}{%
        \ifthenelse{\boolean{display_both_references_for_checking}}{%
            \textcolor{orange}{#1~(\fullversionsecref{#2})}% show both references for easier checking
        }{%
            #1% directly reference the appendix
        }%
    }{%
        \fullversionsecref{#2}%
    }%
}
\newcommand*{\reffullprotocol}{%
    \fullprotocolref
        {\appref{full-protocol}}
        {\Appprefprefix~\fullprotocolappendixletter}}
\newcommand*{\reffullprotocolfig}{%
    \fullprotocolref
        {\figref{secure-sessions-full}}
        {Fig.~36}}

\usepackage{eqparbox}

\newcommand*{\true}{\ensuremath{\mathsf{true}}}
\newcommand*{\false}{\ensuremath{\mathsf{false}}}
\newcommand*{\hoaretriple}{\simpleHoare}

\newcommand{\id}[1]{\mathit{#1}}

\newcommand{\udis}{\uplus}

\newcommand{\interl}{\mathrel{|||}}
\newcommand{\sync}[1]{\mathrel{\parallel_{#1}}}

\newcommand{\sigsub}[1]{\Sigma_{\mathsf{#1}}}
\newcommand{\siglin}{\sigsub{lin}}
\newcommand{\sigper}{\sigsub{per}}

\newcommand{\sigin}{\sigsub{in}}
\newcommand{\sigout}{\sigsub{out}}
\newcommand{\msrsys}{\mathcal{R}}
\newcommand{\RR}{\ensuremath{\msrsys}}
\newcommand{\msrsub}[1]{\msrsys_\mathsf{#1}}
\newcommand{\Renv}{\msrsub{env}}
\newcommand{\Rbuf}{\msrsub{io}}

\newcommand{\Rrole}[1]{\msrsub{role}^{#1}}
\newcommand{\Rind}{\msrsub{ind}}
\newcommand{\MD}{\mathit{MD}}
\newcommand{\MDInd}{\ensuremath{\MD^\mathsf{ind}_\Sigma}}

\newcommand{\fact}[1]{\mathsf{#1}}

\newcommand{\Frfact}{\fact{Fr}}
\newcommand{\Infact}{\fact{in}}
\newcommand{\Outfact}{\fact{out}}
\newcommand{\Indfact}{\fact{ind}}
\newcommand{\Outindfact}{\fact{out}_\fact{ind}}
\newcommand{\Inindfact}{\fact{in}_\fact{ind}}
\newcommand{\Frindfact}{\fact{Fr}_\fact{ind}}
\newcommand{\K}{\mathsf{K}}		% no bang!
\newcommand{\knowsf}{\fact{K}}

\newcommand{\tracePre}{\mathrel{\preccurlyeq}}
\newcommand{\tamtracePre}{\mathrel{\tracePre_{\mathrm{t}}}}

\newcommand{\EE}{\mathscr{E}}

\newcommand{\realEnv}{\ensuremath{\EE}}
\newcommand{\CC}{\ensuremath{\mathscr{C}}}
\newcommand{\concretelts}[2]{\ensuremath{\mathscr{C}_#1(#2)}}

\newcommand{\E}{\mathcal{E}}

\newcommand{\freshtype}{\mathit{fresh}}
\newcommand{\pubtype}{\mathit{pub}}
\newcommand{\rid}{\mathit{rid}}
\newcommand{\ind}{\mathit{ind}}

\newcommand{\prog}{\ensuremath{c}}

\newcommand{\rew}[1]{\xrightarrow{#1}}
\newcommand{\reweq}[2][]{\xrightarrow{\eqmakebox[#1]{\ensuremath{#2}}}}

\newcommand{\msetopify}[1]{#1^\mathsf{m}}
\newcommand{\enumM}[1]{[#1]}
\newcommand{\emptyM}{\enumM{}}
\newcommand{\cupM}{\msetopify{\cup}}

\newcommand{\multileft}{\{\hspace{-0.2em}|}
\newcommand{\multiright}{|\hspace{-0.2em}\}}

\newcommand{\R}{\mathscr{R}}

\newcommand{\asyncFn}{\chi}      % abstract
\newcommand{\csyncFn}{\chi^+}    % concrete
\newcommand{\csyncrelabeledFn}{\chi'}    % relabeled concrete

\newcommand{\taintanalysis}{\ensuremath{\mathbb{T}}}
\newcommand{\taintanalysisconfig}{\ensuremath{s}}

\newcommand*{\simplifiedrefinementstmt}{\left(\mathrel{||}_\rid \prog{}(\rid)\right) \mathrel{||} \mathscr{O} \tamtracePre \msrsys}

\input{macros_composition_soundness}
\newcommand{\gacskipin}{\cskip}
\newcommand{\gacheapallocin}{\cheapalloc{x}}
\newcommand{\gacheapreadin}{\cheapread{x}{e}}
\newcommand{\gacheapwritein}{\cheapwrite{x}{e}}
\newcommand{\gaccoreallocin}{\ccorealloc{c}{\bar{e}}}
\newcommand{\gaccorecallin}{\ccorecall{k}{c}{\bar{e}}{\bar{r}}}
\newcommand{\gacseqin}{s_1 \cseq s_2}
\newcommand{\gacforkin}{\cfork{\bar{x}}{s}}

\newcommand{\gacskipout}{\cskip}
\newcommand{\gacheapallocout}{%
    \cheapalloc{x} \cseq
    \csetadd{\setlocalheap}{\var{x}}}
\newcommand{\gacheapreadoutlocal}{%
    \csetrem{\setlocalheap}{e} \cseq
    \cheapread{x}{e} \cseq
    \csetadd{\setlocalheap}{e}}
\newcommand{\gacheapreadoutglobal}{%
    \catomic{%
        \csetrem{\ederef{\setglobalheap}}{e} \cseq
        \cheapread{x}{e} \cseq
        \csetadd{\ederef{\setglobalheap}}{e}}}
\newcommand{\gacheapwriteoutlocal}{%
    \csetrem{\setlocalheap}{\var{x}} \cseq
    \cheapwrite{x}{e} \cseq
    \csetadd{\setlocalheap}{\var{x}}}
\newcommand{\gacheapwriteoutgloballineone}{%
    \csetrem{\ederef{\setglobalheap}}{\var{x}} \cseq
    \csetremmult{\setlocalheap}{(\reachable{\var{e}} \cap \setlocalheap)}}
\newcommand{\gacheapwriteoutgloballinetwo}{%
    \cheapwrite{x}{e} \cseq
    \csetaddmult{\ederef{\setglobalheap}}{\setliteral{\var{x}} \cup (\reachable{\var{e}} \cap \setlocalheap)}}

\newcommand{\gaccoreallocout}{%
    \catomic{\cassign{\ederef{\flagiospec}}{\true}} \cseq
    \csetremmult{\setlocalheap}{\bar{e}} \cseq
    \ccorealloc{c}{\bar{e}} \cseq
    \csetaddmult{\setlocalheap}{\bar{e}} \cseq
    \csetadd{\setinvariantheap}{\var{c}}}
\newcommand{\gaccorecallout}{%
    \csetrem{\setinvariantheap}{\var{c}} \cseq
    \csetremmult{\setlocalheap}{\bar{e}} \cseq
    \ccorecall{k}{c}{\bar{e}}{\bar{r}} \cseq
    \csetaddmult{\setlocalheap}{\bar{e} \cup \bar{r}} \cseq
    \csetadd{\setinvariantheap}{\var{c}}}
\newcommand{\gacseqout}{%
    \ghostalgorithm(s_1) \cseq
    \ghostalgorithm(s_2)}
\newcommand{\gacforkout}{%
    \csetremmult{\setlocalheap}{(\reachable{\bar{x}} \cap \setlocalheap)} \cseq
    \csetaddmult{\ederef{\setglobalheap}}{(\reachable{\bar{x}} \cap \setlocalheap)} \cseq
    \cfork{\bar{x}}{%
        \cassign{\setlocalheap}{\emptyset} \cseq
        \cassign{\setinvariantheap}{\emptyset} \cseq
        \ghostalgorithm(s)}}

\newacronym{SMT}{SMT}{satisfiability modulo theories}
\newacronym{KDF}{KDF}{key derivation function}
\newacronym{DH}{DH}{Diffie--Hellman}
\newacronym{NSL}{NSL}{Needham--Schroeder--Lowe}
\newacronym{DY}{DY}{Dolev--Yao}
\newacronym{KCI}{KCI}{key compromise impersonation}
\newacronym{AKC}{AKC}{actor key compromise}
\newacronym{ADT}{ADT}{algebraic data type}
\newacronym{RVL}{RVL}{reusable verification library}
\newacronym{CDS}{CDS}{concurrent data structure}
\newacronym{AWS}{AWS}{Amazon Web Services}
\newacronym{SSM Agent}{SSM Agent}{Systems Manager Agent}
\newacronym{KMS}{KMS}{Key Management Service}
\newacronym{pm}{pm}{person-month}
\newacronym{IH}{IH}{induction hypothesis}
\newacronym{SSA}{SSA}{static single assignment}
\newacronym{VPN}{VPN}{Virtual Private Network}
\newacronym{TLS}{TLS}{Transport Layer Security}
\newacronym{MLS}{MLS}{Messaging Layer Security}
\newacronym[
    shortplural={ADEM},
    description={The \glsfmtfull{ADEM} is an authentication mechanism for digital assets}
]{ADEM}{ADEM}{Authentic Digital EMblem}
\newacronym[
    shortplural={LoC},
    longplural={lines of code}
]{LOC}{LoC}{line of code}
\newacronym[
    shortplural={LoS},
    longplural={lines of specification}
]{LOS}{LoS}{line of specification}
\newacronym[
    shortplural={ALoC},
    longplural={auxiliary lines of code}
]{ALOC}{ALoC}{auxiliary line of code}
\newacronym{MSR}{MSR}{multiset rewriting}
\newacronym{LTS}{LTS}{labeled transition system}
\newacronym{DoS}{DoS}{denial of service}
\newacronym{AEAD}{AEAD}{authenticated encryption with associated data}
\newacronym{MAC}{MAC}{message authentication code}
\newacronym[
    longplural={certificate authorities}
]{CA}{CA}{certificate authority}
\newacronym{CT}{CT}{Certificate Transparency}
\newacronym{ICRC}{ICRC}{International Committee
of the Red Cross}
\newacronym{IHL}{IHL}{International Humanitarian Law}
\newacronym{PKI}{PKI}{public key infrastructure}
\newacronym{ITree}{ITree}{interaction tree}
\newacronym{API}{API}{application programming interface}
\newacronym{SDK}{SDK}{software development kit}
\newacronym{IP}{IP}{Internet Protocol}
\newacronym{TCP}{TCP}{Transmission Control Protocol}
\newacronym{MITM}{MITM}{Mallory-in-the-middle}
\newacronym{AC}{AC}{associativity and commutativity}
\newacronym{SSH}{SSH}{Secure Shell}

\newglossaryentry{separation-logic}
{
    name=separation logic,
    description={A logic for reasoning about programs with mutable state},
    text=separation logic,
    nonumberlist
}
\newglossaryentry{io-separation-logic}
{
    name=I/O separation logic,
    description={A logic for reasoning about I/O behavior of programs},
}
\newglossaryentry{protected-party}{
    name=protected party,
    description={A party that is protected under \actitle{IHL} in the context of \acs{ADEM}. Examples include humanitarian organisations like UNICEF and Médecins Sans Frontières},
}
\newglossaryentry{attacker-completeness}{
    name=attacker completeness,
    description={A property of trace invariants stating that a trace invariant remains valid under the full set of operations available to the attacker, as defined by the attacker model. This property is also known as \emph{robust safety}~\cite{DBLP:conf/csfw/GordonJ01} or \emph{attacker typability}~\cite{DBLP:conf/eurosp/BhargavanBDHKSW21}}
}
\newglossaryentry{non-injective-agreement}{
    name=non-injective agreement,
    description={An authentication property of security protocols stating that two parties agree on their identities and certain values. This property is typically expressed as a correspondence between events on a trace. See \secref{traces-authentication} for details}
}
\newglossaryentry{injective-agreement}{
    name=injective agreement,
    description={An authentication property of security protocols stating that two parties \emph{injectively} agree on their identities and certain values. In contrast to \gls{non-injective-agreement}, this property additionally rules out replay attacks. See \secref{traces-authentication} for details}
}
\newglossaryentry{forward-secrecy}{
    name=forward secrecy,
    description={A property of security protocols stating that the compromise of long-term keys does not compromise past session keys. See \secref{traces-wireguard} for details}
}
\newglossaryentry{post-compromise-security}{
    name=post-compromise security,
    description={A property of security protocols stating that the compromise of long-term keys does not compromise future session keys}
}
\newglossaryentry{goroutine}{
    name=goroutine,
    description={A lightweight thread in the Go programming language}
}
\newglossaryentry{fractional-permissions}{
    name=fractional permissions,
    description={A common permission model in \gls{separation-logic} readily supported by many program verifiers. One or a full~permission grants read and write access to a heap location, which is created at allocation time of this heap location. Any fraction strictly larger than zero grants read-only access. Separating conjunction adds up fractions of permissions to the same heap location}
}
\newglossaryentry{permission-accounting}{
    name=permission accounting,
    description={An alternative permission model to \gls{fractional-permissions} in \gls{separation-logic}. This model is based on a factory resource from which so-called \emph{shares} (another \gls{separation-logic} resource) can be split off. The factory resource keeps track of how many shares have been split off and allows to recollect shares}
}
\newglossaryentry{safety}{
    name=safety,
    description={A property of programs guaranteeing that a program does cause neither runtime exceptions nor undefined behavior. In particular, it covers the absence of memory errors, which include buffer overflows, data races, use-after-free errors, and reading from uninitialized memory}
}
\newglossaryentry{functional-property}{
    name=functional property,
    plural=functional properties,
    description={An implementation-specific property of a program describing its desired behavior, \eg that a sorting algorithm's result is a sorted permutation of the input}
}
\newglossaryentry{io-independence}{
    name=I/O independence,
    description={A property of I/O operations stating that an I/O operation that is classified as protocol-irrelevant does not depend on any secret data involved in the protocol. In \approach, we prove this property by performing a static taint analysis}
}
\newglossaryentry{equational-theory}{
    name=equational theory,
    plural=equational theories,
    description={A set of equations that is used to axiomatize the behavior of otherwise uninterpreted functions}
}
\newglossaryentry{ghost-code}{
    name=ghost code,
    description={Auxiliary program code asserting intermediate properties and manipulating auxiliary program state to facilitate reasoning about a program. Ghost code does not affect a program's runtime behavior and is erased before compilation. See Filli{\^{a}}tre~\etal~\cite{DBLP:conf/cav/FilliatreGP14} for details}
}

\newglossaryentry{tamarin}{
    name=Tamarin,
    description={A security protocol model verifier based on \actitle{MSR}},
    nonumberlist
}

\newcommand*{\ac}[1]{\gls{#1}}
\newcommand*{\acp}[1]{\glspl{#1}} % plural
\newcommand*{\acf}[1]{\glsreset{#1}\gls{#1}} % force full version and record its use
\newcommand*{\acfp}[1]{\glsreset{#1}\glspl{#1}} % force full plural version and record its use
\newcommand*{\acs}[1]{\glsfmtshort{#1}} % force short version and leaves the used status unchanged
\newcommand*{\acsp}[1]{\glsfmtshortpl{#1}} % force short plural version and leaves the used status unchanged
\newcommand*{\actitle}[1]{\glsfmtfull{#1}} % force full version for titles without recording its use. Use `\glsadd{#1}` immediately after the section title to ensure that the acronym appears in the glossary (also without marking it used).
\newcommand*{\acptitle}[1]{\glsfmtfullpl{#1}} % force full, plural version for titles without recording its use. Use `\glsadd{#1}` immediately after the section title to ensure that the acronym appears in the glossary (also without marking it used).
\newcommand*{\glstitle}[1]{\glsentrytitlecase{#1}{text}} % title case for a glossary entry (using the value in its `text` field)
\newcommand{\sign}[2]{\ensuremath{\symb{sign}(#1, #2)}}
\newcommand{\aenc}[2]{\ensuremath{\symb{aenc}(#1, #2)}}
\newcommand{\senc}[2]{\ensuremath{\symb{senc}(#1, #2)}}
\newcommand{\agentid}{\ensuremath{\symb{Id}_{A}}}
\newcommand{\clientid}{\ensuremath{\symb{Id}_{B}}}
\newcommand{\readerid}{\ensuremath{\symb{Id}_{M}}}
\newcommand{\agentltkey}{\symb{sk}_A}
\newcommand{\agentltkeyid}{\ensuremath{\symb{Id}_\symb{skA}}}
\newcommand{\clientltkey}{\symb{sk}_B}
\newcommand{\clientltkeyid}{\ensuremath{\symb{Id}_\symb{skB}}}
\newcommand{\readerpk}{\ensuremath{\symb{pk}_M}}
\newcommand{\sigX}{\symb{sig}_x}
\newcommand{\sigXfull}{\sign{\langle g^x, \readerid, \clientid \rangle}{\agentltkey}}
\newcommand{\sigY}{\symb{sig}_y}
\newcommand{\sigYfull}{\sign{\langle g^y, \agentid \rangle}{\clientltkey}}
\newcommand{\sharedsecret}{\ensuremath{g^{x*y}}}
\newcommand{\sesshash}{\ensuremath{h(\sharedsecret)}}
\newcommand{\kdfssone}{\ensuremath{\symb{kdf1}(\sharedsecret)}}
\newcommand{\kdfsstwo}{\ensuremath{\symb{kdf2}(\sharedsecret)}}
\newcommand{\sessciphertext}{c_\symb{ss}}
\newcommand{\sessciphertextfull}{\aenc{\langle \kdfssone, \kdfsstwo \rangle}{\readerpk}}
\newcommand{\sigSS}{\symb{sig}_\symb{ss}}
\newcommand{\sigSSfull}{\sign{\langle \sessciphertext, \clientid \rangle}{\agentltkey}}

\newboolean{complete_protocol}
\newcommand{\externalsign}[1]{\ifthenelse{\boolean{complete_protocol}}{#1}{}}

\newcounter{msgcounter}

\begin{document}
%-------------------------------------------------------------------------------

% IEEE title apparently does not support a different title at the top of each page
\ifthenelse{\equal{\paperversion}{\extendedversion}}{%
    \title{The Secrets Must Not Flow:\\Scaling Security Verification to Large Codebases (extended version)}
}{%
    \title{The Secrets Must Not Flow: Scaling Security Verification to Large Codebases}
}

% we redefine the IEEEauthorrefmark command to use \dagger and \ddagger by removing "*\or":
% original command:
% \DeclareRobustCommand*{\IEEEauthorrefmark}[1]{\raisebox{0pt}[0pt][0pt]{\textsuperscript{\footnotesize\ensuremath{\ifcase#1\or *\or \dagger\or \ddagger\or%
%     \mathsection\or \mathparagraph\or \|\or **\or \dagger\dagger%
%     \or \ddagger\ddagger \else\textsuperscript{\expandafter\romannumeral#1}\fi}}}}
\DeclareRobustCommand*{\IEEEauthorrefmark}[1]{\raisebox{0pt}[0pt][0pt]{\textsuperscript{\footnotesize\ensuremath{\ifcase#1\or \dagger\or \ddagger\or%
    \mathsection\or \mathparagraph\or \|\or **\or \dagger\dagger%
    \or \ddagger\ddagger \else\textsuperscript{\expandafter\romannumeral#1}\fi}}}}

\author{
    % put all authors into a single `\IEEEauthorblockN` as recommended for 4+ authors by IEEEtran documentation:
    \IEEEauthorblockN{%
        Linard Arquint\IEEEauthorrefmark{1} \orcidlink{0000-0002-6230-8014},
        Samarth Kishor\IEEEauthorrefmark{2} \orcidlink{0009-0005-3795-3117},
        Jason R. Koenig\IEEEauthorrefmark{2} \orcidlink{0000-0002-5611-4408},\\
        Joey Dodds\IEEEauthorrefmark{2} \orcidlink{0009-0004-1534-6968},
        Daniel Kroening\IEEEauthorrefmark{2} \orcidlink{0000-0002-6681-5283}, and
        Peter M{\"{u}}ller\IEEEauthorrefmark{1} \orcidlink{0000-0001-7001-2566}
    }
    \IEEEauthorblockA{\IEEEauthorrefmark{1}\textit{Department of Computer Science, ETH Zurich, Switzerland}}
    \IEEEauthorblockA{\IEEEauthorrefmark{2}\textit{Amazon Web Services, USA}}
}

\maketitle

\thispagestyle{plain}
\pagestyle{plain}

\ifthenelse{\not\equal{\paperversion}{\extendedversion}}{%
    \nocite{full-version} % reserve [1] for full-version
}

\begin{abstract}
Existing program verifiers can prove advanced properties about security protocol implementations, but are difficult to scale to large codebases because of the manual effort required.
We develop a novel methodology called \approach that addresses this challenge by splitting the codebase into the protocol implementation (the \core) and the remainder (the \app).
This split allows us to apply powerful semi-automated verification techniques to the security-critical \core, while fully-automatic static analyses scale the verification to the entire codebase by ensuring that the \app cannot invalidate the security properties proved for the \core.
The static analyses achieve that by proving \emph{\ioindependence}, \ie that the I/O operations within the \app are independent of the \core's security-relevant data (such as keys), and that the \app meets the \core's requirements.
We have proved \approach sound by first showing that we can safely allow the \app to perform I/O independent of the security protocol, and second that manual verification and static analyses soundly compose.
We evaluate \approach on two~case studies: an implementation of the signed Diffie--Hellman key exchange and a large (100k+ LoC) production Go codebase implementing a key exchange protocol for which we obtained secrecy and \gls{injective-agreement} guarantees by verifying a \core of about \qty{1}{\percent} of the code with the auto-active program verifier \gobra in less than three~person months.
\end{abstract}

\IEEEpeerreviewmaketitle

\section{Introduction}
Security protocols such as TLS or Signal ensure security and privacy for browsing the web, sending private messages, and using cloud services.
It is, thus, crucial that these ubiquitous and critical protocols are designed \emph{and} implemented correctly.

Automatic protocol verifier tools such as
\tamarin~\cite{DBLP:conf/csfw/SchmidtMCB12,DBLP:conf/cav/MeierSCB13} and
\proverif~\cite{DBLP:conf/csfw/Blanchet01} make it viable to formally verify protocol \emph{models}.
Their applications to TLS~\cite{DBLP:conf/sp/BhargavanBK17},
EMV~\cite{DBLP:conf/sp/BasinST21}, Signal~\cite{DBLP:conf/eurosp/KobeissiBB17}, and
5G~\cite{DBLP:conf/ccs/BasinDHRSS18,DBLP:conf/ndss/CremersD19} demonstrate that they can handle realistic protocols. However, proving protocol \emph{models} secure does not
result in secure \emph{implementations} on its own. Coding errors such as omitted protocol steps (as in the Matrix SDK~\cite{CVEMatrix21})
or ignored errors (\eg returned by a TLS
library~\cite{CVETLStormPacketReassembly,CVETLStormAuthenticationBypass}) may invalidate all security
properties proven for the corresponding models.

Verifying security properties for protocol \emph{implementations} is possible as well~\cite{DBLP:conf/csfw/DupressoirGJN11, DBLP:conf/sp/ArquintWLSSWBM23, DBLP:conf/ccs/ArquintSM023}. For instance, \citeauthors{Arquint}{DBLP:conf/sp/ArquintWLSSWBM23} first verify security properties for a \tamarin model of the protocol in the presence of a \ac{DY} attacker~\cite{DBLP:journals/tit/DolevY83} fully controlling the network. Then, they prove that the protocol implementation refines this model, \ie that the model justifies every I/O operation performed by the implementation. Refinement guarantees that the implementation inherits the security properties proven for the model.

Existing approaches to verifying protocol implementations are sound \emph{only} if they are applied to the \emph{entire} implementation.
Verifying only a subset of the codebase is unsound, and would fail to prevent, \eg code seemingly unrelated to a security protocol accidentally logging key material~\cite{CVETerraformLogSecrets24, CVEGitHubEnterpriseLogSecrets23}.
However, the required expertise and annotation overhead make it infeasible to verify entire \emph{production} codebases, which often consist of hundreds of thousands of lines of code.

\mypar{This work}
We present \approach\footnote{Diodon is a genus of fish known for their inflation capabilities. Erecting spines and scaling their volume by a multiple provide security, like our verification methodology.}, a proved-sound methodology that scales verification of security properties to large production codebases.
\approach works with codebases where a small, syntactically-isolated component implements a security protocol, whose security argument can be made separately from the rest of the code.
Our methodology decomposes the overall codebase into this protocol implementation~(the \core) and the remainder~(the \app).

\begin{figure}[t]
    \centering
    \includegraphics[width=0.475\textwidth]{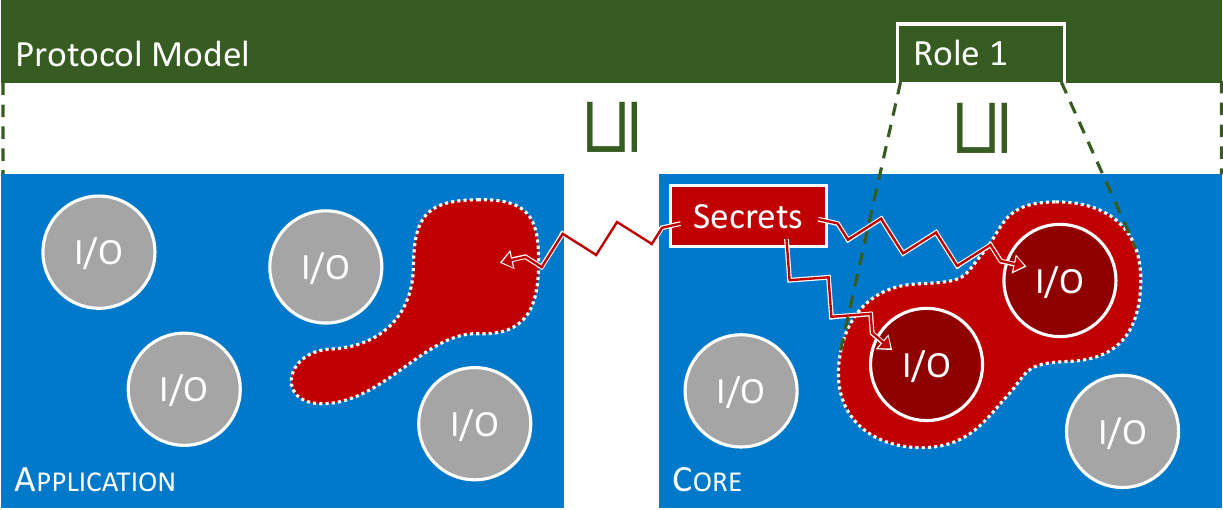}
    \caption{%
        The \approach methodology.
        We partition the codebase~(blue) into the module implementing a protocol~(\core) and the remaining codebase~(\app).
        We prove that the \core refines~(trace inclusion on the right) a particular role of the verified protocol model~(green) by auto-active verification.
        We apply static analyses to the entire codebase to enforce that secrets~(red) do not influence~(red arrows) the I/O operations~(gray circles) of the \app and to ensure that the \app cannot invalidate the security properties proved for the \core.
        Consequently, \approach guarantees that the entire codebase refines the protocol model~(trace inclusion in the middle) and, thus, enjoys all security properties proved for that model.
    }
    \label{fig:proof-decomposition}
    \ifisextendedversion{%
        % we slightly reduce the spacing below the figure in the paper's extended version such that the introduction section fits onto the first two pages:
        \vspace{-.6em}
    }
\end{figure}

This decomposition allows us to apply different verification techniques to the two~parts.
We verify the \core using Arquint~\etal's approach to show refinement \wrt a verified \tamarin model, which requires precise reasoning about, \eg the payloads of I/O operations.
Instead of applying the same annotation-heavy approach to the \app, we use automatic static analyses to ensure that security-relevant data of the \core (in particular, secrets such as keys) does not influence any I/O operation within the \app.
If this \emph{\ioindependence} holds, the \app cannot perform any I/O operations that could interfere with the protocol and invalidate its proven security.
Additionally, we use static analyses to prove that the \app satisfies the assumptions made for the proof of the \core, in particular, that the preconditions of \core functions hold when called from the \app and that the \app does not violate any invariants of \core data structures. These checks ensure that the proofs of the \core and the \app compose soundly.
Consequently, the entire codebase refines the protocol model and enjoys all security properties proved for the model.
\approach significantly reduces the proof effort of verifying software that contains protocol implementations.
\figref{proof-decomposition} illustrates our methodology.

% I/O independence
We prove \ioindependence for the \app by executing an automatic taint analysis on the entire codebase to identify I/O operations that are possibly affected by secrets~(also implicitly via control flow) and checking that all such operations are within the \core, which shows that the codebase's decomposition is valid and the \core is sufficiently large.
It would be too restrictive to enforce that all secrets are confined within the \core.
In most implementations, secrets exist outside the \core,
\eg the \app might have access to secrets either via program inputs or the \core's state (red~area within the \app in \figref{proof-decomposition}).
It is therefore essential to ensure (via a whole-program analysis) that the \app does not \emph{use} these secrets to violate the security properties of the \tamarin model.

Most I/O operations within the \core correspond to a protocol step and are relevant for proving refinement \wrt a protocol model.
In production code, however, the \core might also contain operations irrelevant to the protocol, such as logging a protocol step.
To reduce the verification effort further, we also check \ioindependence \emph{within} the \core to classify each I/O operation based on whether it depends on secrets occurring in a protocol run (dark red circles in \figref{proof-decomposition}) or not (gray circles).
The former need to be considered during the refinement proof, while the latter can safely be ignored.
This classification simplifies the refinement proof and shortens the abstract protocol model.

% refinement proof via auto-active verification
\looseness=-1
We prove refinement of the \core \wrt a protocol model using an \emph{auto-active} program verifier~\cite{krml212}.
These program verifiers take as input an implementation annotated with specifications such as pre- and postconditions and loop invariants, and attempt to verify the implementation automatically using a \ac{SMT} solver.

Auto-active verification is generally sound only if it is applied to the entire codebase because \emph{all} callers of a function must establish its precondition and \emph{all} functions must preserve data structure invariants.
To ensure that our methodology is sound while avoiding this requirement for the \app, we design our methodology such that static analyses automatically discharge the proof obligations on the \app.
Nevertheless, our methodology is flexible enough to permit complex interactions between the \core and \app, \eg through concurrency and callbacks.
Some assumptions remain, in particular, the absence of data races and undefined behavior; we discuss those in \secref{diodon-threat-model-assumptions}.

We prove \approach sound, providing a blueprint for combining the distinct formalisms of auto-active verifiers and static analyses.
First, we prove that a \ac{DY} attacker can simulate all secret-\emph{independent} I/O operations.
Consequently, if a \tamarin model permits every secret-\emph{dependent} I/O operation in a codebase, then this codebase refines the model.
Second, we show that \approach reasons about these secret-\emph{dependent} I/O operations \emph{without} verifying the entire codebase.
\Ie we construct the corresponding proof for the entire codebase by starting from the proof for the \core, which we obtain from auto-active verification, and discharging the remaining proof obligations using our static analyses.

% evaluation
We evaluate \approach on two~Go implementations, a signed \ac{DH} key exchange and a fork of the \ac{AWS} \ac{SSM Agent}~\cite{SSMAgent}, a large (100k+ LoC) codebase.
Part of the latter codebase implements an experimental protocol for encrypted shell sessions. We prove secrecy for and \gls{injective-agreement} on the session keys established by both protocols.
For the \ac{SSM Agent} codebase, \approach allowed us to limit auto-active verification to only about 1\% of the entire codebase, which took less than three~person months. This demonstrates that \approach enables, for the first time, the verification of strong security properties at the scale of production codebases.
Our static analyses and case studies are open-source~\cite{paper-artifact-zenodo, paper-artifact-github}.

\mypar{Contributions}
We make the following contributions:
\begin{myitemize}

\looseness=-1
\item We present a scalable verification methodology for implementations of security protocols within large codebases, which applies to any codebase with a clear distinction between the protocol core and the rest of the code.

\item We identify I/O independence, enabling concise protocol models for complex implementations.

\item We show how to use static analyses to automatically discharge the \core's proof obligations, enabling \approach to scale to large codebases.

\item We prove the soundness of I/O independence \wrt a \ac{DY} attacker, and the soundness of \approach's combination of auto-active verification and static analyses.

\item We evaluate our methodology on two~case studies, an implementation of the signed \ac{DH} key exchange and an \ac{AWS} Systems Manager Agent fork, to
demonstrate that \approach scales to large, production codebases.
\end{myitemize}

%%% Local Variables:
%%% mode: LaTeX
%%% TeX-master: "../main"
%%% End:

\section{Running Example of \approach}
\label{sec:diodon-running-example}
% # Example
We demonstrate the core ideas of \approach on a sample program in the Go programming language, which implements a simple \acf{MAC} protocol that sends and receives signed messages using a pre-shared key.
In the remainder of this section, \emph{we} refers to a user of \approach.
First, we manually split the codebase into \core and \app following function boundaries.
We make the \core as small as possible to reduce auto-active verification efforts while making sure that the entire protocol implementation is contained therein and that we can define an invariant for the \core's \ac{API} with which the \app interacts.

We model the protocol and prove security properties with the \tamarin protocol verifier~\cite{DBLP:conf/csfw/SchmidtMCB12,DBLP:conf/cav/MeierSCB13}.
The goal is to prove that the entire program, \ie the composition of the \core and \app, refines the \tamarin model and, thus, satisfies the same security properties as the protocol model.
We auto-actively verify the \core using \gobra~\cite{DBLP:conf/cav/WolfACOPM21} and apply the automatic \argot~\cite{argot} static analyses to the entire codebase.

\begin{figure}[t]
\begin{gobraenv}
package core

type Chan struct {
  psk []byte
  cb  Cb
}

type Cb = func(msg []byte)

//@ req  acc(msg, 1) $\label{line:cb-spec-start}$
//@ func CbSpec(msg []byte) $\label{line:cb-spec-end}$

//@ req  cb != nil ==> cb implements CbSpec{} $\label{line:closure-spec}$
//@ pres psk != nil ==> acc(psk, 1)
//@ ens  Inv(c) $\label{line:InitChannel-inv}$
func InitChannel(psk []byte, cb Cb) (c *Chan) {
  //@ inhale AliceIOPermissions() $\label{line:inhale-io-permissions}$
  c = &Chan{append([]byte(nil), psk...), cb} $\label{line:alloc-chan}$
  go continuousRecv(c) $\label{line:start-goroutine}$
  return c
}

//@ pres c != nil ==> Inv(c) $\label{line:send-pre-inv}$
//@ pres msg != nil ==> acc(msg, 1) $\label{line:send-pre-msg}$
func Send(c *Chan, msg []byte) { $\label{line:send-decl}$
  if c == nil || msg == nil { return }
  fmt.Printf("Send %x\n", msg) $\label{line:send-log-msg}$
  packet := append(msg, HMAC(msg, c.psk)...)
  sendToNetwork(packet)
}

/*@ pred Inv(c *Chan) { $\label{line:invariant}$
  c != nil && acc(c, 1/2) &&$\label{line:chan-struct-permissions}$
  acc(c.psk, 1/2) && AliceIOPermissions() &&$\label{line:io-permissions}$
  (c.cb != nil ==> c.cb implements CbSpec{})
} @*/
\end{gobraenv}
\vspace{-0.7em}
\caption{%
  \looseness=-1
  Sample \core for a simple \acs{MAC} communication.
  In Go, function definitions take a list of input parameters and may have a second list for outputs.
  We omit the \code{continuousRecv}~\gls{goroutine}'s implementation that invokes the \code{c.cb}~closure (if non-\nil{}) whenever a message has been received.
  We simplify the representation of I/O permissions, which describe permitted protocol-relevant I/O operations, and omit proof-related statements.
}
\label{fig:sample-core}
\end{figure}

\begin{figure}[t]
\begin{gobraenv}
package main

import . "core"

func main(psk []byte) {
  cb := func(m []byte) {fmt.Printf("%x\n", m)} $\label{line:main-log-msg}$
  c := InitChannel(psk, cb)
  Send(c, []byte("hello world"))
  fmt.Printf("Log: message sent.\n") 
  // fmt.Printf("%v\n", c) $\label{line:log-struct}$
}
\end{gobraenv}
\vspace{-0.7em}
\caption{%
  Sample \app that is a client of \figref{sample-core}.
  We omit parsing of command line arguments for presentation purposes and, thus, assume that \code{psk} stores the parsed pre-shared key.
}
\label{fig:sample-app}
\vspace{0.7em}
\begin{tamarinenv}
rule Alice_Send:
    let packet = <msg, sign(msg, psk)> in
    [ Alice_1(rid, A, B, psk), In(msg) ] $\label{line:alice-in-fact}$
  --->
    [ Alice_1(rid, A, B, psk), Out(packet) ]
rule Alice_Recv:
    let packet = <msg, sign(msg, psk)> in
    [ Alice_1(rid, A, B, psk), In(packet) ]
  --->
    [ Alice_1(rid, A, B, psk), Out(msg) ] $\label{line:alice-out-fact}$
\end{tamarinenv}
\vspace{-0.7em}
\caption{%
  \tamarin model excerpt for the MAC protocol implemented in \figref{sample-core}.
}
\label{fig:sample-model}
\end{figure}

\mypar{\core}
\looseness=-1
The \core~(\figref{sample-core}) consists of a struct definition, two~\ac{API} functions, \code{InitChannel} and \code{Send}, which access this struct, and a predicate~\codebw{Inv} that represents the \gls{separation-logic}~\cite{DBLP:conf/lics/Reynolds02} invariant used to verify the functions.
The definition of \codebw{Inv} includes permissions to access the struct fields and the pre-shared key's bytes.
\Gls{separation-logic} controls heap access with these permissions to reason about
side effects and to prove data-race freedom, as detailed in
\secref{diodon-code-level-verification}.
Accessibility predicates~(\codebw{acc})
represent permissions in specifications.  Their first argument indicates the
heap location and the second argument characterizes the permitted access: a
value of 1 provides exclusive read and write access, and any value strictly
between 0 and 1 provides read-only access that might be shared.
For instance, \codebw{acc(msg,1)} on
\lineref{send-pre-msg} passes full permission to write the contents of \code{msg} (if it is non-\nil)
from a caller to function \code{Send}, and back to the caller when the function returns.
Pre- and postconditions start with the keyword \codebw{req} and \codebw{ens}, respectively, and we use \codebw{pres} as syntactic sugar for properties that are preserved, that is, act as pre- \emph{and} postconditions.

To receive incoming packets, the \core spawns a \gls{goroutine} (lightweight thread) on
\lineref{start-goroutine} executing the function~\code{continuousRecv}.
We omit its implementation in the
figure for space reasons.
The \gls{goroutine} repeatedly calls a blocking receive
operation, checks the \ac{MAC}'s validity, and on success calls the closure that
is stored in the struct field~\code{cb} as a callback.
If the callback is non-\nil, it delivers the resulting message to the \app.

We verify the \core for any callback closure that satisfies
the specification \codebw{CbSpec} (\cf
\lineref{closure-spec} \& \ref{line:cb-spec-start}--\ref{line:cb-spec-end}),
which states that a caller must pass permission for modifying the message to
the closure when invoking it and that the closure does not have to return
any permissions.
On \lineref{alloc-chan}, we duplicate the pre-shared key (which the \app obtains as a program input) to keep the \core's memory footprint separated from the \app.
Thus, we can pass half of the permissions for
accessing the struct fields to the \gls{goroutine} spawned on \lineref{start-goroutine} and store the remaining
permissions in the invariant~\codebw{Inv}, which is then returned to the
caller of \code{InitChannel}.

\mypar{\app}
\looseness=-1
The \app~(\figref{sample-app}) consists of a single function that creates a closure that will print any incoming message, initializes the \core with the pre-shared key~\code{psk} and this closure, and then sends a message by invoking the \code{Send} function of the \core.
In realistic programs, the \app might have thousands of lines of code, making auto-active verification prohibitively expensive.
\approach allows us to apply automatic static analyses instead, as detailed below.

\mypar{Protocol model}
\figref{sample-model} excerpts our abstract protocol model as a multiset rewriting system in \tamarin (\cf \secref{diodon-protocol-model-verification}) with two~protocol roles, Alice and Bob, each starting off with a pre-shared key~\codetamarin{psk}.
Both roles can send and receive unboundedly many packets, each of which are the composition of a message plus the appropriate \ac{MAC}\@.
To make zero assumptions about the messages themselves, we treat them as being attacker-controlled, \ie the sending role obtains a message from the attacker-controlled network via an \codetamarin{In}~fact, as shown on \lineref{alice-in-fact}.
For this protocol model, we prove that all received messages were previously sent by either Alice or Bob, unless the attacker obtains the pre-shared key, which \tamarin proves automatically.

In order to prove that our program is actually a refinement of this model
and, thus, inherits all proven properties, \approach combines auto-active verification and static analyses to
obtain provably-sound guarantees.

\mypar{Verification}
\looseness=-1
Our goal is to ensure that the composition of the \app and \core refines the abstract \tamarin model, \ie the program's I/O behavior is contained in the model's I/O behavior.
This refinement implies that any trace-based safety property proven in \tamarin also holds for the program because the program performs the same or fewer I/O operations than the protocol model.
\approach splits the refinement proof into three~steps:
We prove that (1)~non-protocol I/O is independent of protocol secrets, (2)~all remaining I/O refines a protocol role, and (3)~the proof steps soundly compose.

First, we manually identify protocol-relevant calls to I/O operations within the \core. In our example, these are the \code{sendToNetwork} call and the corresponding network receive operation.
We then perform an automatic taint analysis on the entire codebase to prove \ioindependence for all other calls to I/O operations (in our example, the calls to \code{Printf}), \ie we check that they do not use tainted data.
Uncommenting \lineref{log-struct} in \figref{sample-app} would result in printing all struct fields of variable~\code{c} including the pre-shared key~\code{psk}, which is the only secret.
\ioindependence would correctly fail for this modified program, resulting
in an error message indicating the flow of secret data to the print statement.
In general, we treat data as a secret (\ie tainted) if the protocol model's attacker might not know this data.
Checking \ioindependence ensures that we do not miss any protocol-relevant I/O operations and that the chosen \core is sufficiently large.

The \core may execute protocol-relevant operations not only by performing I/O operations, but also by communicating with the \app. For example, Alice's protocol step of taking an arbitrary message from the environment (before signing and sending it), is implemented by the \core obtaining \code{msg} from the \app (\lineref{send-decl} in \figref{sample-core}). Similarly, Alice may (after receiving a packet and checking its signature) release its payload to the environment, which is implemented as passing the payload to the \app when invoking the closure \code{c.cb} (not shown in \figref{sample-core}). 
To handle such protocol-relevant operations uniformly, we treat them as \emph{virtual} protocol-relevant I/O operations.
This allows us to enforce or assume constraints on the arguments' taint status while creating the necessary proof obligations in the next step of the refinement proof. Here, the fact that releasing the payload is permitted by the protocol model (\lineref{alice-out-fact} in \figref{sample-model}) informs the taint analysis that the callback's argument may be considered as untainted, which allows printing it on \lineref{main-log-msg} in \figref{sample-app}.

\looseness=-1
Second, we prove the \core using the auto-active \gobra verifier.
This proof includes showing that the protocol model permits every protocol-relevant I/O operation, including virtual I/O\@. Note that step~(1) ensures that these operations must all be in the \core.
We use an I/O specification for each protocol role describing the permitted protocol-relevant I/O operations (\cf \secref{diodon-code-level-refinement}).
In our example, Alice obtains the permissions to perform these operations during the initialization of the \core (\lineref{inhale-io-permissions}) and maintains them as part of the invariant (\lineref{io-permissions}), where \codebw{inhale} adds the specified permissions to the current program state by assumption.
When performing a protocol-relevant I/O operation, like \code{sendToNetwork}, \gobra proves that the I/O specification permits this operation with the specific arguments.
Otherwise, \gobra reports a verification failure.

Third, since the \gobra proof for the \core assumes that callers respect the functions' preconditions, \approach restricts the class of supported pre- and postconditions such that static analyses are able to prove that the \app satisfies them.
For example, the precondition of \code{Send} requires exclusive access for the argument \code{msg}; \approach enforces this condition using a combination of static pointer, escape, and pass-through analyses to ensure that no other \gls{goroutine} accesses the memory pointed to by \code{msg}. \code{Send}'s other precondition requires the \core's invariant to hold, which is established by \code{InitChannel}. The \app could in principle violate this precondition, for example, by creating a \code{Chan} instance without calling \code{InitChannel}, or by invalidating the invariant of a \code{Chan} instance through field updates or concurrency.
We apply this combination of static analyses to prevent all such cases~(\cf \secref{diodon-approach-discharging-assumptions}).

Together, these three~proof steps ensure that the program refines the abstract \tamarin model and inherits the security properties proved for the model.

\section{Background}
\looseness=-1
In this section, we provide the necessary background on the verification techniques that we reference in the rest of the paper.
We detail verification of abstract protocol models~(\secref{diodon-protocol-model-verification}), verification of implementations~(\secref{diodon-code-level-verification}), and code-level refinement~(\secref{diodon-code-level-refinement}), which transfers security properties from a protocol model to implementations.

\subsection{Protocol Model Verification}
\label{sec:diodon-protocol-model-verification}
\looseness=-1
We model a security protocol and prove security properties about it using \tamarin, an automated protocol model verifier.
A protocol model consists of protocol roles and a \ac{DY} attacker that are expressed as multiset rewrite rules. Each rule has the shape \codetamarin{$L$ --[ $A$ ]-> $R$}, where $L$ and $R$ are multisets of facts and $A$ is a set of actions.
The system's state~$S$ is a multiset of facts, which is initially empty, and a rule can be applied if $L$ is (multiset) included in S, \ie $L \subseteq_m S$.
Applying a rule removes the facts in $L$ from and adds those in $R$ to the system state, \ie results in a new state $S \setminus_m L \cup_m R$.
While most facts are user-defined to represent the state of a protocol role, \tamarin uses certain predefined facts. In particular, \codetamarin{In(x)} and \codetamarin{Out(x)}~facts represent receiving and sending a message~\codetamarin{x} from and to the attacker-controlled network, respectively.
Since \tamarin uses \glspl{equational-theory} to describe otherwise uninterpreted functions, such as $\symb{dec}(\symb{enc}(p, \symb{pk}(\symb{sk})), \symb{sk}) = p$ to describe asymmetric decryption (\symb{dec}) \wrt asymmetric encryption (\symb{enc}), \tamarin performs multiset inclusion modulo \glspl{equational-theory}.
A possible sequence of rule applications forms a trace that consists of each rule application's set of actions~($A$).
\tamarin symbolically explores all possible traces involving unboundedly many instances of protocol roles to prove a security property, or, if the proof fails, provides a trace of an attack as a counterexample to the security property.

\tamarin's symbolic \ac{DY} attacker fully controls the network and can construct new messages by applying symbolic operations to terms it has observed.
However, a symbolic attacker can neither perform arbitrary computations, nor can it exploit algorithmic weaknesses or side channels such as timing.
Nevertheless, verification using \tamarin guarantees the absence of many relevant vulnerabilities and has proven effective in applications to 5G~\cite{DBLP:conf/ccs/BasinDHRSS18}, EMV card protocols~\cite{DBLP:conf/fm/BasinHST24}, and the Noise protocol family~\cite{DBLP:conf/uss/GirolHSJCB20}.

\subsection{Code-Level Verification}
\label{sec:diodon-code-level-verification}
We prove \gls{safety} and \glspl{functional-property} of an implementation by reasoning about all possible executions statically, without any runtime checks.
In this context, the term \emph{\gls{safety}} expresses that an implementation neither causes runtime exceptions nor undefined behavior.
In particular, it covers the absence of memory errors, buffer overflows, and data races.
\emph{\Glspl{functional-property}} are implementation-specific and express the desired behavior, \eg that a sorting algorithm's result is a sorted permutation of the input.

We perform \emph{modular} verification, \ie we reason about each function in an implementation in isolation.
To do this, we equip a function with a \emph{specification} that consists of a pre- and postcondition.
A function's precondition is a logical formula specifying all valid program states in which this function can be called, and a function's postcondition specifies properties that hold for all
valid program states after executing the function's body.

To reason about heap-manipulating programs, we use \emph{\gls{separation-logic}}~\cite{DBLP:conf/lics/Reynolds02}.
\Gls{separation-logic} allows us to express precisely which heap fragment $f$ a function operates on and provides connectives that split and combine heap fragments, \eg when calling another function that operates only on a subfragment $f' \subseteq f$, we know that the function does not modify any heap location in the \emph{frame}~$f \setminus f'$.

\Gls{separation-logic} associates a \emph{permission} with each heap location.
A permission represents ownership of a particular heap location and is required for each access.
We use fractional permissions~\cite{DBLP:conf/sas/Boyland03} to distinguish between exclusive and shared ownership, which permits multiple threads in a concurrent program to simultaneously share ownership of a heap location.
In specifications, we express permission to a heap location~\code{l} with fraction~\codebw{p} as \codebw{acc(l,p)}, \cf \secref{diodon-running-example}.

The \emph{separating conjunction}~$*$ is akin to regular conjunction, but requires the \emph{sum} of the permissions in both conjuncts.
For instance, \codebw{acc(l1,1)${}*{}$acc(l2,1/2)} specifies full and half permissions for heap locations~\code{l1} and \code{l2}, respectively.
Additionally, this example implicitly specifies that the heap locations are disjoint, \ie \code{l1} $\neq$ \code{l2}.
Otherwise, if \code{l1} and \code{l2} were aliased, the permission amounts would add up to $\sfrac{3}{2}$ contradicting \gls{separation-logic}'s invariant that at most a full permission exists for a heap location.
In our code listings, we overload~\codebw{&&} to denote separating conjunction.

We use \gls{separation-logic} \emph{predicates}~\cite{DBLP:conf/popl/ParkinsonB05} to abstract over individual permissions to heap locations as demonstrated by the \core{}'s invariant in our running example.
Conceptually, we can treat a predicate instance such as \codebw{Inv(c)} in \figref{sample-core} as a shorthand notation for the predicate's body.

\looseness=-1
A \glsdisp{separation-logic}{separation-logic} proof guarantees \gls{safety}, in part by requiring a proof that each function has sufficient permissions for each heap access.
\Eg a buffer overflow corresponds to accessing an array element out of bounds; this is prevented since allocating an array  creates permissions only for in-bound elements.
Similarly, data races are prevented since two~threads simultaneously writing the same heap location would require that each thread has write permission for this heap location, which is impossible as there is only a single write permission for any given heap location.

Various \gls{separation-logic}-based program verifiers exist, including \verifast~\cite{DBLP:conf/nfm/JacobsSPVPP11}
and \vst~\cite{DBLP:journals/jar/CaoBGDA18} for C, \gobra~\cite{DBLP:conf/cav/WolfACOPM21} for Go, \nagini~\cite{DBLP:conf/cav/Eilers018} for Python, and \prusti~\cite{DBLP:journals/pacmpl/Astrauskas0PS19} for Rust.
These verifiers are \emph{auto-active}, \ie use manual annotations and proof automation to prove properties about programs.

\subsection{Code-Level Refinement}
\label{sec:diodon-code-level-refinement}
\citeauthors{Arquint}{DBLP:conf/sp/ArquintWLSSWBM23} split verification of security protocol implementations into two~steps:
proving security properties for an abstract protocol model using \tamarin, and using an auto-active program verifier to prove that an implementation refines this model.
This approach disentangles proving global security properties from local reasoning about implementations, while exploiting each tool's automation.

To connect these two~steps, \tamarin automatically generates an \emph{I/O specification} for each protocol role in a given abstract protocol model.
A protocol role's I/O specification describes the permitted I/O operations including their sequential ordering and arguments.
By verifying that an implementation executes at most the operations permitted by the I/O specification, we prove that all executions of this implementation result in a trace of I/O operations that is included in the set of traces considered by \tamarin when verifying the security properties.
Thus, the implementation satisfies the same security properties as the model.

The I/O specification is expressed in \gls{io-separation-logic}~\cite{DBLP:conf/esop/Penninckx0P15}, a dialect of \gls{separation-logic}, which is readily supported by \gls{separation-logic}-based verifiers.
\gls{io-separation-logic} extends the use of permissions beyond guarding heap accesses to guarding I/O operations by associating an \emph{I/O permission} with each I/O operation.
We equip library functions performing I/O operations with a specification that checks and consumes the corresponding I/O permission.

Successful code-level verification against the library functions equipped with I/O permissions guarantees that an implementation performs at most the I/O operations specified in the \tamarin model and respects their sequential ordering.
Otherwise, verification fails because a prohibited I/O operation would require an unavailable I/O permission.

\section{\approach}
\label{sec:diodon-approach}

Our methodology, \approach, proves security properties for implementations by refinement and scales to large codebases by significantly reducing verification effort.
\approach enables more concise protocol models than previous approaches and leverages fully automatic analyses on most of the implementation to discharge proof obligations.

\begin{figure}[t]
    \centering
    \includegraphics[width=0.47\textwidth]{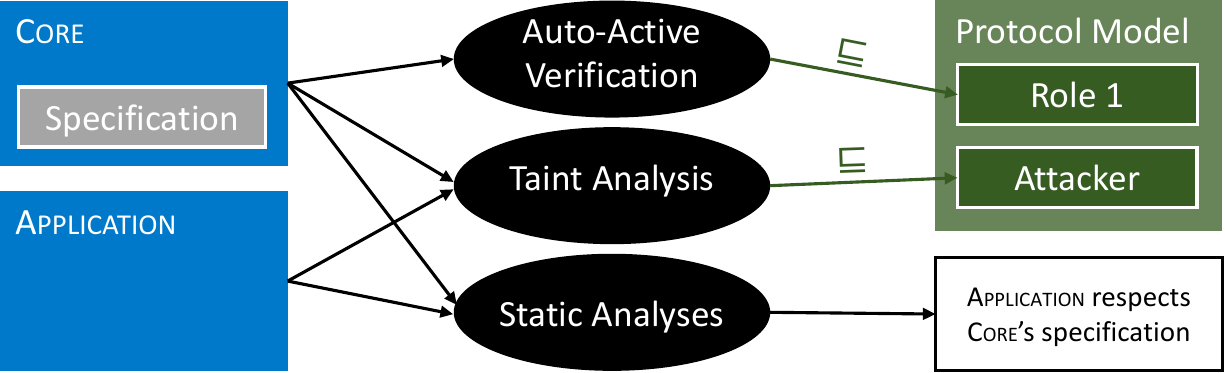}
    \caption{%
        \looseness=-1
        \approach proves that the entire codebase~(blue) refines a protocol model~(green) by soundly composing auto-active verification with automatic static analyses.
        We auto-actively verify the \core based on its specification to show that the protocol-relevant I/O operations refine a protocol role~(upper trace inclusion).
        This specification is partially generated from the protocol model, which is omitted.
        The static taint analysis proves that all other I/O operations within the entire codebase refine our attacker model~(lower trace inclusion).
        Lastly, we discharge the \core's assumptions by applying automatic static analyses, proving that the \app satisfies the calling restrictions expressed in the \core's specification.
    }
    \label{fig:approach-tool-diagram}
\end{figure}

We manually decompose a codebase into two~syntactically isolated components, the \core implementing a security protocol, and the \app consisting of the remaining code.
Typically, this decomposition is natural and follows module boundaries as a protocol's implementation is localized.
As illustrated in \figref{approach-tool-diagram}, this decomposition allows us to split the proof that the entire codebase refines a protocol model into three~steps and uses the best-suited tool for each step.
We explain in \secref{diodon-approach-io-independence} how \approach identifies which I/O operations are protocol-relevant by performing a static taint analysis on the entire codebase.
\secref{diodon-approach-role-refinement} covers the \core's auto-active verification using \gobra proving that protocol-relevant I/O operations refine a particular protocol role.
Finally, \secref{diodon-approach-discharging-assumptions} explains how we discharge the assumptions made when auto-actively verifying the \core by performing static analyses on the \app.

\subsection{\texorpdfstring{\ioIndependence}{I/O Independence}}
\label{sec:diodon-approach-io-independence}
% protocol model / Tamarin
One of our key insights is to distinguish between I/O operations that are relevant for a security protocol from those that are not (\eg sending log messages to a remote server). This distinction has two main benefits. First, protocol-irrelevant operations do not have to be reflected in the abstract protocol model, which makes the model concise, more general, and easier to maintain, review, and prove secure.
Second, by ensuring that protocol-irrelevant I/O operations cannot possibly invalidate the security properties proven for the protocol model, we do not have to consider them during the laborious auto-active refinement proof and instead can check simpler properties using automatic static analyses.
We classify all calls to I/O operations as either protocol-relevant or protocol-irrelevant. In the \core, an I/O operation is protocol-irrelevant if and only if its specification requires no I/O permissions. In contrast, all I/O operations in the \app are implicitly considered protocol-irrelevant.

\looseness=-1
To ensure that I/O operations classified as protocol-irrelevant indeed do not interfere with the protocol or invalidate proven security properties of the protocol, we check that they do not use any secret data (such as key material);
more precisely, we check \emph{non-interference} between protocol secrets and the parameters of these I/O operations. We call this important property of an I/O operation \emph{\ioindependence}.
It guarantees that an I/O operation cannot possibly invalidate the protocol's proven security properties: any I/O operation that uses only non-secret data could also have been performed by the \ac{DY} attacker and, thus, was already considered by \tamarin during the protocol verification.
In other words, proving that all protocol-irrelevant I/O operations satisfy \ioindependence guarantees that they refine our \ac{DY}~attacker~(\cf \secref{diodon-soundness-proof-overview-io-independence}).

From a cryptographic perspective, \ioindependence allows us to reduce the security of an entire codebase to the security of its \core. This reduction is valid because most of the \app can be treated as part of the attacker, while the parts of the \app that manipulate secrets (\eg code that loads long-term keys from disk) are shown not to perform I/O, and thus can conceptually be considered part of the \core, without introducing violations of the I/O specification of its protocol role.

We prove \ioindependence by performing an automatic static taint analysis on the entire codebase that includes implicit information flows from control flow.
A taint analysis checks for a set of sources and sinks whether there are any flows of information from a source to a sink.
The analysis starts at each source, \ie{} a function which produces secret data, and explores how secret information propagates through the program by keeping track of program locations storing a \emph{tainted} value, \ie a value that is influenced by a source.
We disallow branching on tainted data to avoid information flows via control flow.

We configure the taint analysis to consider all calls to key-generation functions within the \core and long-term secrets that are passed as program inputs, like the pre-shared key in our running example, as sources because the \ac{DY}~attacker does not have access to them.
This set of initial sources taints all protocol secrets including session keys.
\Eg if the \core implements a \ac{DH} key exchange, the analysis correctly considers the generated secret key and the resulting shared key tainted because the shared key is computed from the secret key and the other party's public key.
We then configure the taint analysis to treat all I/O operations in the \app as well as all protocol-irrelevant I/O operations in the \core as sinks.
We use Capslock~\cite{capslock} to identify such I/O performing functions in the Go standard library.
We consider all functions with at least one of the following capabilities as a sink:
write to the file system or network, modify the state of the operating system (\eg \code{os.Setenv}), perform a system call, and execute external programs (\eg \code{(*os/exec.Cmd).Run}). 

We run the taint analysis on the entire codebase.
If taint reaches a sink, verification fails because a secret reached a supposedly protocol-irrelevant I/O operation.
Otherwise, we have correctly identified the protocol-relevant I/O operations (and thereby confirmed that we have correctly delimited the \core);
it remains to reason about those I/O operations, as we discuss next.

\subsection{\core Refinement}
\label{sec:diodon-approach-role-refinement}
\looseness=-1
We auto-actively verify the entire \core, which allows us to state and prove (besides \gls{safety} and \glsdisp{functional-property}{functional correctness}) precise constraints about protocol-relevant calls to I/O functions and their arguments.
We prove that the implementation uses the payload for each I/O operation specified in the protocol model.
The corresponding verification effort is feasible since, in industrial codebases like our main case study, the \core comprises only a small fraction of the codebase.

\looseness=-1
We prove that the \core refines a protocol role by building on the approach explained in \secref{diodon-code-level-refinement}.
In particular, we equip each protocol-relevant I/O operation with a specification that requires an I/O permission for executing this operation with the provided arguments.
Since we provide exactly the I/O permissions justified by the protocol role's model to the \core during its initialization, successful verification with \gobra implies that the \core executes at most the protocol-relevant I/O operations permitted by the model and, thus, refines this protocol role.
This approach is inspired by \citeauthors{Arquint}{DBLP:conf/sp/ArquintWLSSWBM23}, but differs in three~significant ways.

\looseness=-1
First, we do not auto-actively verify the entire codebase and, instead, verify only the \core. As we will discuss in \secref{diodon-approach-discharging-assumptions}, we syntactically restrict the preconditions of the \core functions so that we can apply automatic static analyses to check that each call from the \app satisfies them, which is necessary for soundness.

\looseness=-1
Second, our approach supports codebases that use multiple instances of the \core, \eg to run multiple roles of the protocol or to run the protocol multiple times. Since \tamarin considers unboundedly many role instantiations, we can soundly create the required I/O permissions for executing a role instance whenever we create a new \core instance.
These I/O permissions are bound to an instance's unique identifier such that each \core instance has its own set of I/O permissions for executing the security protocol once.

\looseness=-1
Third, to reflect that interactions in the model between the protocol and the environment may manifest as interactions between the \core and the \app in the implementation, we treat the boundary between them as a virtual network interface and enforce I/O permissions for the corresponding virtual I/O operations, as we illustrated in \secref{diodon-running-example}.

\subsection{Analyzing the \app}
\label{sec:diodon-approach-discharging-assumptions}
\looseness=-1
We now show how to scale auto-active verification to the entire codebase.
Applying auto-active verification to an entire codebase is typically not feasible within the resource constraints of industrial projects. A key insight of \approach is that this is not necessary: we can use static analyses to automatically discharge \gls{separation-logic} proof obligations arising in the \app to obtain, together with the verified \core, a proof in \gls{separation-logic} for the \emph{entire} codebase. 

The refinement proof for the \core is valid in the context of the entire application if (1)~each call to a \core function from the \app satisfies the function precondition, and (2)~the \app respects permissions on memory accesses.
Our soundness proof for \approach (\cf \secref{diodon-soundness-proof-overview-composition}) ensures that these proof obligations are sufficient and that our novel combination of static analyses can soundly discharge them.
We illustrate these proof obligations and how we discharge them by considering the exemplary \core function in \figref{core-api-spec}, taking two~integer pointers as input and returning an integer pointer.
This function maintains the \core invariant (if \code{c} is non-\nil), needs full permissions for both inputs, and returns full permissions for the input and output parameters (if they are non-\nil).
Thus, we cannot allow, \eg the \app to pass two~aliased arguments (\cf \secref{diodon-code-level-verification}) to this function or to concurrently access heap locations pointed to by these arguments as this would violate the precondition, \ie the permissions specified therein.

\begin{figure}[t]
\begin{gobraenv}
//@ pres c$\color{gobracommentcolor} {}\neq{}$nil$\color{gobracommentcolor} {}\implies{}$inv(c)
//@ pres a0$\color{gobracommentcolor} {}\neq{}$nil$\color{gobracommentcolor} {}\implies{}$acc(a0)
//@ pres a1$\color{gobracommentcolor} {}\neq{}$nil$\color{gobracommentcolor} {}\implies{}$acc(a1)
//@ ens  r$\color{gobracommentcolor} {}\neq{}$nil$\color{gobracommentcolor} {}\implies{}$acc(r)
func (c *Core) ApiFn(a0, a1 *int) (r *int)
\end{gobraenv}
\vspace{-0.7em}
\caption{%
  Example of a signature and specification of a \core \acs{API} function.
}
\label{fig:core-api-spec}
\end{figure}

\mypar{Implicit annotations}
To construct a proof for the entire codebase, we enrich the \app with a \emph{hypothetical program instrumentation} that connects the \app to the necessary proof obligations imposed by the proof of the \core. 
These \emph{implicit annotations} track the permissions that the \app owns by using sets of heap locations and a \emph{program invariant} specifying permissions for the heap locations in these sets.
More precisely, each thread has a set~\setlocalheap{}~(short for \quotes{local heap set}) for thread-local objects such as buffers, and a set~\setinvariantheap{}~(short for \quotes{invariant heap set}) for \core instances.
Similarly, a global set~\setglobalheap{}~(\quotes{global heap set}) keeps track of objects that might be shared between threads, which becomes relevant later.
These sets are mutable and, thus, their content depends on a particular program point.
The sets~\setlocalheap{} and \setinvariantheap{} allow us to state the following local program invariant that must hold at every program point in the \app.
\newcommand*{\localproginvexpanded}{\ensuremath{\bigl(\forallstar_{l \in \setlocalheap} \acc{l}\bigr) \star \bigl(\forallstar_{i \in \setinvariantheap} \inv{i}\bigr)}}
\begin{equation*}
\localproginv \triangleq \; \localproginvexpanded
\end{equation*}
Here, the iterated separating conjunction~$\forallstar_{e \in s} a(e)$ conjoins the assertions~$a(e)$ using separating conjunction for all elements~$e$ in set~$s$. 
$\localproginv$ states that a thread holds full permissions for all objects in \setlocalheap{} and the \core invariant for all instances in \setinvariantheap{}.
In addition, these permissions are disjoint allowing the \app to write to heap locations in \setlocalheap{} without breaking the \core invariant. When a thread obtains or gives up permissions, our implicit annotations adjust \setlocalheap{} and \setinvariantheap{} to maintain the program invariant.

\looseness=-1
\figref{transformed-core-api-call} shows these \emph{implicit annotations} for calls to \code{ApiFn}.
To highlight that each statement in the \app maintains the program invariant, we assert $\localproginv$ on \lines*{\ref{line:local-prog-inv-start} and~\ref{line:local-prog-inv-end}}. For each permission required by the callee's precondition, we remove the corresponding heap location from one of the sets  to reflect that ownership is being passed to the callee. Assuming (for now) that the location was originally in the set, this removal extracts the corresponding permission from $\localproginv$, as illustrated by the intermediate assert statements starting on \lines*{\ref{line:assert-after-removal-1-start}, \ref{line:assert-after-removal-2-start}, and \ref{line:assert-precondition-start}} for the three~arguments of the call. After the call, we conversely add those heap locations to the sets for which the  callee's postcondition provides permissions.

For each permission in the precondition, if the corresponding heap location was contained in one of the sets before the removal operation, then we have effectively proved that the precondition holds (syntactic restrictions ensure that the preconditions cannot contain constraints other than permission requirements, see below).
In the rest of this subsection, we explain how we use static analyses to check this set containment. Then, we explain the proof obligations for memory accesses within the \app.

\begin{figure}[t]
\begin{gobraenv}
//@ assert $\color{gobracommentcolor} \localproginvexpanded$ $\label{line:local-prog-inv-start}$
//@ $\color{gobracommentcolor} \setinvariantheap := \setinvariantheap \setminus \setliteral{\progvar{c}}$ $\label{line:c-removal}$
//@ assert $\color{gobracommentcolor} \localproginvexpanded \star {}$ $\label{line:assert-after-removal-1-start}$
//@        $\color{gobracommentcolor} (\progvar{c} \neq \nil \implies \inv{\progvar{c}})$
//@ $\color{gobracommentcolor} \setlocalheap := \setlocalheap \setminus \setliteral{\progvar{a0}}$ $\label{line:a0-removal}$
//@ assert $\color{gobracommentcolor} \localproginvexpanded \star {}$ $\label{line:assert-after-removal-2-start}$
//@        $\color{gobracommentcolor} (\progvar{c} \neq \nil \implies \inv{\progvar{c}}) \star {}$
//@        $\color{gobracommentcolor} (\progvar{a0} \neq \nil \implies \acc{\progvar{a0}})$
//@ $\color{gobracommentcolor} \setlocalheap := \setlocalheap \setminus \setliteral{\progvar{a1}}$ $\label{line:a1-removal}$
//@ assert $\color{gobracommentcolor} \localproginvexpanded \star {}$ $\label{line:assert-precondition-start}$
//@        $\color{gobracommentcolor} (\progvar{c} \neq \nil \implies \inv{\progvar{c}}) \star {}$
//@        $\color{gobracommentcolor} (\progvar{a0} \neq \nil \implies \acc{\progvar{a0}}) \star {}$
//@        $\color{gobracommentcolor} (\progvar{a1} \neq \nil \implies \acc{\progvar{a1}})$ $\label{line:assert-precondition-end}$
r := c.ApiFn(a0, a1) $\label{line:core-api-call}$
//@ assert $\color{gobracommentcolor} \localproginvexpanded \star {}$ $\label{line:assert-postcondition-start}$
//@        $\color{gobracommentcolor} (\progvar{c} \neq \nil \implies \inv{\progvar{c}}) \star {}$
//@        $\color{gobracommentcolor} (\progvar{a0} \neq \nil \implies \acc{\progvar{a0}}) \star {}$
//@        $\color{gobracommentcolor} (\progvar{a1} \neq \nil \implies \acc{\progvar{a1}}) \star {}$
//@        $\color{gobracommentcolor} (\progvar{r} \neq \nil \implies \acc{\progvar{r}})$ $\label{line:assert-postcondition-end}$
//@ $\color{gobracommentcolor} \setlocalheap := \setlocalheap \cup (\setliteral{\progvar{a0}, \progvar{a1}, \progvar{r}} \setminus \nil)$ $\label{line:param-insertion}$
//@ $\color{gobracommentcolor} \setinvariantheap := \setinvariantheap \cup (\setliteral{\progvar{c}} \setminus \nil)$ $\label{line:c-insertion}$
//@ assert $\color{gobracommentcolor} \localproginvexpanded$ $\label{line:local-prog-inv-end}$
\end{gobraenv}
\vspace{-0.7em}
\caption{%
  \emph{Conceptually} inserted implicit annotations for a \core \acs{API} call \code{r := c.ApiFn(a0, a1)} in the \app.
  The assert statements solely illustrate our deductions and, thus, can be omitted.
}
\label{fig:transformed-core-api-call}
\end{figure}

\mypar{Guaranteeing permissions for parameters}
For the arguments~\code{a0} and \code{a1} (we will discuss the core instance~\code{c} below), we need to prove that 
(1)~$\setliteral{\progvar{a0}, \progvar{a1}} \subseteq \setlocalheap$ holds before the call to \code{ApiFn}  and (2)~\code{a0} and \code{a1} do not alias.
If (2) was violated, \code{a1} would no longer be in \setlocalheap{} after removing \code{a0} on \lineref{a0-removal} in \figref{transformed-core-api-call}, \ie we would obtain only \acc{\progvar{a0}} instead of $\acc{\progvar{a0}} \star \acc{\progvar{a1}}$.

\looseness=-1
We discharge these two~proof obligations by checking the conditions~\condition{6} and~\condition{7} in \figref{static-conditions}, resp., using static analyses.
We check \condition{6} by using a thread escape analysis, which delivers judgments $\text{local}(x)$ for a particular program point expressing that $*x$ is definitely not accessible by any other thread.
We show in \secref{diodon-soundness-proof-overview-composition} that \condition{6} suffices to discharge $\setliteral{\progvar{a0}, \progvar{a1}} \subseteq \setlocalheap$ (if the arguments are non-\nil) by proving a lemma that relates $\text{local}(x)$ for a program point~$p$ with $x \in \setlocalheap$.
We obtain \condition{7} by applying a pointer analysis, which computes may-alias information, \ie \pointstoset{x} for a pointer~$x$, where $a \in \pointstoset{x}$ denotes that $*x$ may-alias any location allocated at site~$a$.
More precisely, we check for each pair of arguments that the sets of locations they may-alias are disjoint, which is sufficient as we restrict parameters to be shallow. 

\begin{figure}[t]
\small\centering
\begin{tabular*}{\columnwidth}{@{}l@{\hskip 5pt}l@{\hskip 6pt}p{5.25cm}@{}}
\hline
& Condition & Details\\\hline
\conditiondef{1} & \core init & \core instances are created in a function ensuring the invariant in its postcondition\\
\conditiondef{2} & No modification & \app does \emph{not} write to \core instances' internal state, even through an alias\\
\conditiondef{3} & \core preservation & \core instances are  passed only to \core functions that preserve the invariant\\
\conditiondef{4} & \core locality & \core instances are used only in the thread they are created in\\
\conditiondef{5} & \core callback & \core \acsp{API} are not invoked in \app callbacks\\\hline
\conditiondef{6} & Parameter locality & Parameters to \core \acsp{API} are local\\
\conditiondef{7} & Disjoint parameters & Parameters to the same \core \acs{API} call do not alias one another\\\hline
\conditiondef{8} & \app access & Reads and writes in the \app occur to memory allocated in the \app or transferred from the \core
\end{tabular*}
\caption{%
  Sufficient conditions checked by our static analyses, grouped into those involving \core instances, other parameters to \core functions, and memory accesses in the \app.
}
\label{fig:static-conditions}
\end{figure}

\mypar{Guaranteeing the \core invariant}
Similarly to parameters, we have to prove that $\progvar{c} \in \setinvariantheap$ holds such that removing \progvar{c} from \setinvariantheap{} on \lineref{c-removal} grants us the \core invariant~$\inv{\progvar{c}}$, if \progvar{c} is non-\nil.
In \secref{diodon-soundness-proof-overview-composition}, we prove that $\progvar{c} \in \setinvariantheap$ if the following premises hold.
(1)~The \core instance~\progvar{c} must have been returned as a result from a \core \ac{API} function initially establishing the \core invariant, \eg \code{InitChannel} in our running example.
(2)~All heap modifications in the \app must not modify the internal state of the \core instance, even through an alias, since this could invalidate the \core invariant.

In a single-threaded program without callbacks from the \core to the \app, the above premises are sufficient.
However, in the presence of these two~features, we need to ensure that the \app  does not call two~\core functions on the same \core instance simultaneously, which would effectively duplicate permissions and, thus, make reasoning unsound:
(3)~The \app must not pass the same \core reference to more than one~thread, and
(4)~the \app must not call a \core function in a callback on the same instance that is invoking the callback.

We establish the four~premises by checking the conditions~\condition{1} to~\condition{5} in \figref{static-conditions}.
Conditions~\condition{1} and~\condition{3} can be enforced by checking that the \app calls only \core functions that establish or preserve the invariant.
While the postconditions provide this information for \core instances that are passed as arguments or results, our analyses need to prevent a subtle loophole:
We need to prevent \core functions from allocating a \core instance \emph{without} establishing its invariant and letting the \app access it via global variables or shared memory.
We implemented a pass-through analysis computing $\text{pass}_f(x, r)$ for a function~$f$ stating that outside of calls to $f$, $*x$ definitely passed through return parameter~$r$.
We use this pass-through analysis to ensure that all references to \core instances in the \app are obtained exclusively through the return parameter, such that the postcondition establishes the invariant.

\looseness=-1
To establish \condition{2}, we use a pointer analysis to ensure that all reads and writes in the \app never access a \core instance's internal state.
In particular, we ensure that the \app accesses \emph{only} heap locations that must-not-alias locations reachable from \setinvariantheap{}, \ie internal state of \core instances.
Since we use a sound pointer analysis, this check conservatively over-approximates the heap locations about which the \core invariant states properties.
While it is possible to access \core memory without breaking the invariant, we could not treat the \core invariant as an opaque \gls{separation-logic} resource when analyzing the \app, which would require a static analysis capable of reasoning about \gls{fractional-permissions} and arbitrary \glspl{functional-property}.

For \condition{4}, we use the thread escape analysis to ensure that each \core instance does not escape its thread (we show $\text{local}(\progvar{c})$ for each call to \core instance~\progvar{c}), guaranteeing that each thread operates on a disjoint set of \core instances~\setinvariantheap{}.
While it is possible to safely pass \core instances between threads, this would require a significantly more sophisticated static analysis that can reason about the ordering of concurrent executions.
Condition~\condition{5} is enforced by checking that the call graph does not contain \core functions invoked transitively from \app callbacks. Allowing such calls would require proving that the same instance is not used in the inner call, which requires a more precise pointer analysis.

Our explanations generalize from the exemplary \core function in \figref{core-api-spec} to arbitrary \core \ac{API} functions as long as they satisfy the following restrictions on pre- and postconditions.
We support an arbitrary number of input and output parameters with arbitrary value and pointer types.
Our restrictions mandate that \core \ac{API} functions preserve the \core invariant and full permissions for each parameter of pointer type, both only under the condition that the receiver and parameters are non-\nil.
Additionally, the postcondition specifies full permissions for each return parameter if it is non-\nil and of pointer type.
These restrictions ensure that preconditions do not specify \glspl{functional-property}, such as require an input array to have a certain length, which we cannot check using our static analyses.
As seen with our example in \figref{core-api-spec}, we cannot rule out that the \app passes \nil as an argument because there is no sound nilness analysis for Go to the best of our knowledge and, thus, we account for this possibility in our restrictions.

\mypar{\app memory access}
\looseness=-1
Finally, we need to ensure that the \app accesses only memory to which it has permissions.
While we have already established that the \app does not write to internal state of \core instances~(C2), we need to particularly consider the case where memory is transferred after its allocation between the \core and the \app.
The other case, namely the \core or \app allocating memory without transferring it, is straightforward.
\Ie if \core-allocated memory is never transferred to the \app then the \app cannot access it.
Similarly, if \app-allocated memory is not transferred to the \core then the \app retains the corresponding permission.

\looseness=-1
Checking condition~\condition{8} is sufficient.
If \app-allocated memory is transferred to the \core, our syntactic restrictions guarantee that the \core only temporarily borrows the corresponding permissions until the \core \ac{API} call returns.
If the \core allocates memory and transfers it to the \app, the \core must also transfer the corresponding permissions, which we enforce via our pass-through analysis checking that this transfer happens via a return parameter as our syntactic restrictions guarantee that the postcondition specifies permissions for this return parameter.
Using \condition{8}, we prove that each memory access in the \app is to a location in either \setlocalheap{} or \setglobalheap{} (\cf \secref{diodon-soundness-proof-overview-composition}).
In the latter case, we need to reason about concurrent access. We assume that the \app is free from data races: if two~accesses race, then the program is invalid according to the Go specification. If there are no races, then there is some total order in which the threads can atomically pull permission from \setglobalheap{}, perform the access, and then return permissions to \setglobalheap{} before the next thread needs to access the same location.

\subsection{Threat Model, Assumptions, and Limitations}
\label{sec:diodon-threat-model-assumptions}
The \approach methodology provides strong guarantees for large codebases, namely that a codebase satisfies the same trace-based safety properties as the abstract protocol model.
Like other verification techniques, \approach relies on assumptions about the codebase, execution environment, and the employed tools.

\approach considers an arbitrary number of potentially concurrent protocol sessions, allowing the \ac{DY} attacker to, \eg replay messages across sessions or apply cryptographic operations thereto to construct messages of unbounded size.
As is standard for symbolic cryptography, we assume cryptographic operations such as signing are perfectly secure, \eg the attacker can create valid signatures only if it possesses the correct signing key.
The attacker can obtain such keys only by observing or constructing them, never by guessing.

Our methodology allows us to prove that each implementation individually refines a particular role of an abstract protocol model.
Since the security properties we prove about an abstract model are typically global, they hold only if each involved implementation refines one of the protocol roles.
Next, we discuss limitations of this refinement proof, grouped by limitations of the methodology itself and limitations of our instantiation in Go.

The \approach methodology requires a partitioning of a codebase into \core and \app, while satisfying the syntactic restrictions for the \core \ac{API} specifications.
This partitioning limits applicability, not soundness as the taint analysis checking \ioindependence guides correct partitioning and fails otherwise.
Additionally, \approach requires the absence of undefined behavior in the codebase, which we prove for the \core.
However, this remains an assumption for the \app, which could be mitigated by performing an additional static analysis establishing this property. \Eg we could use \astree~\cite{astree} for a subset of C and C++. 
Finally, we inherit the \emph{pattern requirement} from \citeauthors{Arquint}{DBLP:conf/sp/ArquintWLSSWBM23}, which allows multiple terms to have the same byte-level representation in general, but requires a unique representation for terms corresponding to protocol messages.

Our instantiation of \approach in Go uses several tools to discharge proof obligations, and we rely on the soundness of each tool:
the abstract protocol model verifier, the auto-active program verifier, and the static analyses.
The risk that any of these tools is unsound can be mitigated by choosing mature tools such as \tamarin and \gobra.

More specifically, the \core's auto-active verification relies on trusted specifications for libraries, such as the I/O or cryptographic libraries that, \eg consume I/O permissions or specify the cryptographic relations between input and output parameters.
\approach could be combined with verified libraries like EverCrypt~\cite{DBLP:conf/sp/ProtzenkoPFHPBB20} to reduce this trust assumption.

Furthermore, our taint analysis relies on the correct specification of secrets and I/O operations (we use an existing tool~\cite{capslock} to identify I/O operations).
\Eg not treating the pre-shared key in the running example as a secret would allow us to perform I/O operations in the \app that depend on this key.

The employed static analyses assume that the entire codebase is free of data races and, thus, exhibits defined behavior only~\cite{go-memory-model}.
While we auto-actively prove race freedom for the \core, this remains an assumption for the \app.
Our implicit annotations clearly indicate where in the \app we rely on this assumption.
Additionally, the static analyses do not soundly handle certain hard-to-analyze features such as the \code{unsafe} package (\eg allowing arbitrary pointer arithmetic), \code{cgo} (\ie the ability to invoke C functions), or reflection. We rely on the codebase not using them in a way that would invalidate the analysis results.
\approach could be extended by additional static analyses to reduce these assumptions, \eg by performing a data race analysis and checking for uses of the \code{unsafe} and \code{cgo} packages and reflection.
As such, these assumptions are not an inherent limitation of \approach itself.
We report case-studies-related limitations of the static analyses in \secref{diodon-evaluation}.

%%% Local Variables:
%%% mode: LaTeX
%%% TeX-master: "../main"
%%% End:

\section{Formalization and Soundness}
\label{sec:diodon-soundness-proof-overview}
\setboolean{mainbody}{true}
\looseness=-1
\ifthenelse{\boolean{show_full_soundness_appendix}}{%
    We provide an overview of \approach's soundness proof by highlighting its key steps and main ideas.
    Readers interested in full details may refer directly to \appref{diodon-soundness-proof}, which subsumes this section.
}{%
    We provide an overview of \approach's soundness proof by highlighting its key steps and main ideas, and refer to the paper's extended version~\refsoundnessproof for details.
}%
We split the soundness proof into two~parts.
First, we prove that we can soundly allow protocol-independent I/O operations in a codebase while assuming that we auto-actively verify an entire codebase.
Second, we relax the requirement of verifying an entire codebase by showing that we can still construct a proof for the entire codebase in \gls{separation-logic} even though only the \core is verified when certain side conditions are satisfied, most of which can be discharged by static analyses.

\subsection{\texorpdfstring{\ioIndependence}{I/O Independence}}
\label{sec:diodon-soundness-proof-overview-io-independence}
As explained in \secref{diodon-approach-io-independence}, we execute a taint analysis to identify protocol-independent I/O operations in a codebase.
Furthermore, we assume that we have a proof that a codebase~\prog{} satisfies the Hoare triple \hoaretriple{\phi}{\prog}{\true}, where $\phi$ is an I/O specification providing I/O permissions to execute protocol-\emph{dependent} I/O operations.
Protocol-independent I/O operations do not require an I/O permission and, thus, the codebase~\prog{} may contain arbitrarily many protocol-independent I/O operations.

We prove in \refsoundnessioindependence that these independent I/O operations do not violate the security properties proven for the abstract protocol model by showing that the \ac{DY} attacker can simulate these I/O operations.
Hence, \tamarin considers the existence of these I/O operations when proving security properties for a protocol model.
More specifically, we extend the soundness proof by \citeauthors{Arquint}{DBLP:journals/corr/abs-2212-04171} by an additional refinement step.
This step defines a more refined protocol model by augmenting a given protocol model with rules enabling the protocol roles to perform protocol-independent I/O operations.
We then prove that this refined protocol model refines the original protocol model by establishing a refinement relation that simulates protocol-independent I/O operations by actions of the \ac{DY} attacker.

\looseness=-1
The key insight of the \ioindependence proof is that we split the state of every protocol role instance into two~parts.
The first~part is involved in the protocol's execution, \eg keeping track of progress therein.
The second~part is involved only in the protocol-independent I/O operations.
Our refinement relation leaves the former part unchanged while refining the latter to a corresponding state of the \ac{DY} attacker.
We map a transition in the augmented protocol model executing a protocol-independent I/O operation to a transition of the \ac{DY} attacker performing the same I/O operation.

Therefore, we prove that the codebase~\prog{} refines the abstract protocol model~$\msrsys$ even though the codebase contains \emph{more} I/O operations than specified by the model for this protocol role, the difference being all protocol-independent I/O operations in \prog{}.
We prove $\simplifiedrefinementstmt$ (simplified here, see \refsoundnessioindependencesoundness), where we consider unboundedly many instances of the codebase~\prog{}, parameterized by a run identifier~$\rid$, $\mathscr{O}$ denotes instances of all other verified protocol role implementations and the environment including the \ac{DY} attacker, $\mathrel{||}$ represents parallel composition, and $\tamtracePre$ expresses trace inclusion.

\subsection{Combining Auto-Active Verification and Static Analyses}
\label{sec:diodon-soundness-proof-overview-composition}
\begin{figure*}[t]
% we use a minipage to make the equation span the full width too
\footnotesize % a bit smaller than "\small"
\begin{minipage}{\linewidth}
    \begin{align*}
        % skip
        \ghostalgorithm(\gacskipin) &\leadsto \gacskipout\\
        % heap alloc
        \ghostalgorithm(\gacheapallocin) &\leadsto \gacheapallocout\\
        % heap read
        \ghostalgorithm(\gacheapreadin) &\leadsto
            \begin{cases}
                \gacheapreadoutlocal & \text{if $e \in \setlocalheap$}\\
                \gacheapreadoutglobal & \text{otherwise}\\
            \end{cases}\\
        % heap write
        \ghostalgorithm(\gacheapwritein) &\leadsto
            \begin{cases}
                \gacheapwriteoutlocal & \text{if $\var{x} \in \setlocalheap$}\\
                \begin{aligned}
                    \catomic{
                        & \gacheapwriteoutgloballineone \cseq\\
                        & \gacheapwriteoutgloballinetwo
                    }
                \end{aligned}
                & \text{otherwise}
            \end{cases}\\
        % core alloc
        \ghostalgorithm(\gaccoreallocin) &\leadsto \gaccoreallocout\\
        % core api call
        \ghostalgorithm(\gaccorecallin) &\leadsto \gaccorecallout\\
        % seq
        \ghostalgorithm(\gacseqin) &\leadsto \gacseqout\\
        % fork
        \ghostalgorithm(\gacforkin) &\leadsto \gacforkout
    \end{align*}
\end{minipage}
\caption{%
    Algorithm~\ghostalgorithm{} transforms a codebase by inserting \glsdisp{ghost-code}{ghost statements}.
    We define this algorithm by cases, \ie describe how \ghostalgorithm{} transforms each statement~$s$ to a statement~$s'$, written as $\ghostalgorithm(s) \leadsto s'$.
    \reachable{e} computes the set of transitively reachable heap locations from expression~$e$.
    The set union operation ignores \nil{}, as variables might be \nil{}, \ie $S_1 \cupnil S_2 \triangleq \; (S_1 \cup S_2) \setminus \nil$.
    This ensures that \nil{} is never contained in any \glsdisp{ghost-code}{ghost set}.
}
\ifthenelse{\boolean{mainbody}}{%
    \label{fig:ghostalgorithm-mainbody}
}{%
    \label{fig:ghostalgorithm-appendix}
}
\end{figure*}

In the second~part of the soundness proof~\refcompositionsoundness, we want to prove the Hoare triple \hoaretriple{\phi}{\prog}{\true} for an entire codebase~\prog{}, such that we can apply the soundness proof's first~part, while auto-actively verifying only a small part of \prog{}, namely the \core.

At a high-level, we prove a Hoare triple for each function~$f$ in the \core using an auto-active verifier, \ie \hoaretriple{P}{f}{C}, where $P$ and $Q$ are $f$'s pre- and postconditions.
Since the \app calls multiple \core functions and performs arbitrary I/O and memory operations in between, we have to prove that the \app establishes each invoked \core function's precondition and has sufficient permissions to execute its memory operations in order that we obtain a proof in concurrent \gls{separation-logic}~\cite{DBLP:conf/lics/Reynolds02, DBLP:journals/entcs/Vafeiadis11} for the entire codebase~\prog{}.
While not all codebases~\prog{} have such a proof, we prove its existence under certain side conditions that can be discharged using static analyses.

The proof sketch proceeds as follows.
After presenting a simple programming language allowing us to focus on the main ideas, we present an algorithm inserting the implicit annotations for every statement in our language (\cf \secref{diodon-approach-discharging-assumptions}).
The implicit annotations then allow us to define a program invariant that each statement in our language maintains, under some side conditions.
By constructing derivation trees, we prove that each statement maintains the program invariant and make all side conditions apparent.
We then prove lemmata showing that all these side conditions hold if our static analyses succeed on the codebase~\prog{}.
Finally, we compose the individual proof rules to construct a proof for the entire codebase~\prog{}, \ie \hoaretriple{\phi}{\prog}{\true}, and discuss extensions of our soundness proof to support features commonly found in programming languages.

\mypar{Programming language}
We consider an imperative, concurrent, and heap-manipulating programming language in which the codebase~\prog{} is written.
To focus on the main ideas, we keep this language simple by restricting the codebase to a single run of a protocol role and omitting function boundaries, complex control flow, and callbacks;
we separately discuss extensions lifting these restrictions at the end.
Thus, we make calls to \core functions first-class statements in our language.
\Ie we consider \ccorealloc{c}{\bar{e}} to correspond to calling the \core function allocating and initializing a new \core instance (\cf \code{InitChannel} in the running example) and \ccorecall{k}{c}{\bar{e}}{\bar{r}} to correspond to calling any other \core function (indexed by $k$) on a \core instance~$c$ with arguments~$\bar{e}$ and results~$\bar{r}$.

\mypar{Implicit annotations}
As explained in \secref{diodon-approach-discharging-assumptions}, we conceptually introduce \glsdisp{ghost-code}{ghost sets} to track the permissions for heap locations owned by the \app.
We recall that each thread has a set~\setlocalheap{} for thread-local objects such as buffers and a set~\setinvariantheap{} for \core instances.
In addition, a set~\setglobalheap{} contains heap locations that are shared between threads, and a flag~\flagiospec{} tracks whether the codebase has already used the I/O specification~$\phi$ to create an instance of the \core and, thus, a run of the corresponding protocol role.

While \figref{transformed-core-api-call} demonstrates the implicit annotations to manipulate the mentioned \glsdisp{ghost-code}{ghost sets} for a \core \ac{API} call, \figref{ghostalgorithm-mainbody} presents the general algorithm~\ghostalgorithm{} for inserting these implicit annotations for every statement in our language.

Writing to a heap location~(\cheapwrite{x}{e}) in the case that \var{x} does not point to a thread-local heap location, \ie $\var{x} \not\in \setlocalheap$, and our fork statement are of particular interest.
For the former statement, the implicit annotations wrap the heap access in an atomic block to synchronize this access.
This synchronization is sound as we assume that the \app is free of data races, and, thus, a linearization of accesses to a particular heap location exists (we assume that the underlying memory model is sequentially-consistent, like Go's memory model).
Furthermore, we not only temporarily remove \var{x} from \setglobalheap{} but also move all heap locations that are transitively reachable from $e$ from \setlocalheap{} to \setglobalheap{} because the write operation makes these heap locations become potentially accessible to other threads (via the heap location to which \var{x} points).
Since our fork statement~(\cfork{\bar{x}}{s}) passes the arguments~$\bar{x}$ to a newly forked thread executing the statement~$s$, we similarly move all heap locations that are transitively reachable from $\bar{x}$ from \setlocalheap{} to \setglobalheap{}.
The newly forked thread then starts with its own, initially empty \setlocalheap{} and \setinvariantheap{} as no thread-local heap locations and \core instances exist yet.

\mypar{Program invariant}
With \glsdisp{ghost-code}{ghost sets} and algorithm~\ghostalgorithm{} defined, we can define a program invariant that holds at every original program point and that each statement maintains.
More specifically, this program invariant is split into a thread-local~(\localproginv{}) and global~(\globalproginv{})~part.
\begin{align*}
    \localproginv \triangleq \;
        & \left(\forallstar_{l \in \setlocalheap} \acc{l}\right) \star \left(\forallstar_{i \in \setinvariantheap} \inv{i}\right)\\
    \globalproginv \triangleq \;
        & \acc{\setglobalheap} \star \left(\forallstar_{g \in \ederef{\setglobalheap}} \acc{g}\right) \star {}\\
        & \acc{\flagiospec} \star (\neg(\ederef{\flagiospec}) \implies \phi)
\end{align*}

\looseness=-1
Since \secref{diodon-approach-discharging-assumptions} already explains \localproginv{}, we briefly explain \globalproginv{} here.
As \setglobalheap{} and \flagiospec{} are shared between threads, we treat them as pointers and \globalproginv{} specifies permissions for the corresponding heap locations.
Furthermore, \globalproginv{} specifies full permissions for each heap location in \ederef{\setglobalheap} and the I/O specification~$\phi$, unless $\phi$ has already been used to create a \core instance, \ie \ederef{\flagiospec} is set to \true{}.
The last separating conjunct is sufficient for programs that create at most one core instance.
We later discuss how this conjunct can be adapted to provide a family of I/O permissions, enabling arbitrarily many \core instances.

% expects a boolean flag `mainbody` to be set if this file is included in the main body of the paper (otherwise, it is included in the appendix).
\begin{figure*}[t]
\footnotesize % a bit smaller than "\small"
% reduce spacing above and below:
\setlength{\abovedisplayskip}{0pt}%
\setlength{\belowdisplayskip}{0pt}%
\begin{flalign*}
    % simple commands
    \Inf[\rnsimple]
        {\sidecondition{\omega(s_\text{simple})}}
        {\ctxhoare
            {\globalproginv}
            {\localproginv}
            {\ghostalgorithm(s_\text{simple})}
            {\localproginv}}
    &&
    %
    % seq
    \Inf[\rnseq]
        {\ctxhoare{\globalproginv}{\localproginv}{\ghostalgorithm(s_1)}{\localproginv}}
        {\ctxhoare{\globalproginv}{\localproginv}{\ghostalgorithm(s_2)}{\localproginv}}
        {\ctxhoare{\globalproginv}{\localproginv}{\ghostalgorithm(s_1 \cseq s_2)}{\localproginv}}
    &&
    %
    % fork
    \Inf[\rnfork]
        {\ctxhoare{\globalproginv}{\localproginv}{\ghostalgorithm(s)}{\localproginv}}
        {\ctxhoare{\globalproginv}{\localproginv}{\ghostalgorithm(\cfork{\bar{x}}{s})}{\localproginv}}
\end{flalign*}
\ifthenelse{\boolean{mainbody}}{%
    \caption{%
        \looseness=-1
        Proof rules. $s_\text{simple}$ ranges over all \emph{simple} statements; $s$, $s_1$, and $s_2$ range over all statements.
        $\omega$ denotes a statement's side conditions (\cf \figref{side-conditions-mainbody}).
    }
    \label{fig:proof-rules-mainbody}
}{%
    \caption{%
        \looseness=-1
        Proof rules. $s_\text{simple}$ ranges over all \emph{simple} statements; $s$, $s_1$, and $s_2$ range over all statements.
        $\omega$ denotes a statement's side conditions (\cf \figref{side-conditions-appendix}).
    }
    \label{fig:proof-rules-appendix}
}
\end{figure*}

\begin{figure}[t]
% reduce spacing above and below:
\setlength{\abovedisplayskip}{0pt}%
\setlength{\belowdisplayskip}{0pt}%
\begin{align*}
    % \app{} does not invalidate \core{} invariant. \Ie for each \cheapwrite{x}{e}, $*x$ is not transitively reachable from a \core{} instance.
    \omega(\cheapread{x}{e}) \triangleq \;
        &e \in \setlocalheap \cup \ederef{\setglobalheap}\\
    \omega(\cheapwrite{x}{e}) \triangleq \;
        &x \in \setlocalheap \cup \ederef{\setglobalheap}\\
    % \ccoreallocshort{\bar{e}} returns a single \core{} instance (for which the invariant holds)
    \omega(\ccorealloc{c}{\bar{e}}) \triangleq \;
        &\ederef{\flagiospec} = \false \land {}\\
        &(\set{\bar{e}} \setminus \nil) \subseteq \setlocalheap \land {}\\
        &\disjoint{\bar{e}}\\
    % \core{} instances do not escape (relevant for \cheapwrite{x}{e} and \cfork{\bar{x}}{C})
    \omega(\ccorecall{k}{c}{\bar{e}}{\bar{r}}) \triangleq \;
        &(\set{\bar{e}} \setminus \nil) \subseteq \setlocalheap \land {}\\
        &\disjoint{\bar{e}} \land {}\\
        &(c \in \setinvariantheap \lor c = \nil)
\end{align*}
\caption{%
    Side conditions for our statements, which are amenable to static analyses.
    $\omega$ evaluates to \true{} for all statements not listed above and \set{l} returns the set of elements in list~$l$.
    We implicitly refer to variables' values, \eg $v \in S$ denotes that the value of variable~$v$ is contained in set stored in variable~$S$ as opposed to the variables' syntactical representation.
}
\ifthenelse{\boolean{mainbody}}{%
    \label{fig:side-conditions-mainbody}
}{%
    \label{fig:side-conditions-appendix}
}
\end{figure}

\mypar{Proof rules}
As shown in \figref{proof-rules-mainbody}, we use algorithm~\ghostalgorithm{} and the program invariant to define proof rules that have structurally identical conclusions, namely \ctxhoare{\globalproginv}{\localproginv}{\ghostalgorithm(s)}{\localproginv} for each statement~$s$.
This Hoare triple expresses that executing the statement~$s$, transformed by algorithm~\ghostalgorithm, starting in a context satisfying \globalproginv{} and state satisfying \localproginv{}, maintains this context and results in a state satisfying \localproginv{}.
We prove these rules sound in \refproofrules by constructing a proof tree in concurrent \gls{separation-logic}~\cite{DBLP:journals/entcs/Vafeiadis11}.
Our proof relies on the side conditions~$\omega$ (\cf \figref{side-conditions-mainbody}) for statements other than sequential composition and fork.
We require that the specification of \core functions satisfies our syntactic restrictions as mentioned in \secref{diodon-approach-discharging-assumptions}.

\mypar{Discharging side conditions}
Since the side conditions~$\omega$ refer to containment of heap locations in certain \glsdisp{ghost-code}{ghost sets} and disjointness of heap locations, we prove several lemmata in \refsoundnessstaticanalyses to bridge the gap between these side conditions and the properties that a successful execution of our static analyses provides.

Our pointer analysis computes judgments~$\pointstoset{\var{x}}$ for a pointer~\var{x}, where $a \in \pointstoset{\var{x}}$ denotes that $\ederef{\var{x}}$ may-alias any location allocated at site~$a$.
Applying our escape analysis delivers $\local{x}{p}$ guaranteeing that \var{x} points at program point~$p$ to a heap location that is accessible only by the current thread and not by any other thread.
Finally, our pass-through analysis computes the judgments
\passthroughcoretrace{a}{\tau}{p} and \passthroughreturntrace{a}{\tau}{p} denoting that a heap location allocated at allocation site~$a$ passed through ($\mathit{pt}$) the return argument of a \ccorealloc{c}{\bar{e}} statement and through one of the return arguments~$\bar{r}$ of a \ccorecall{k}{c}{\bar{e}}{\bar{r}} statement, respectively, between site~$a$ and program point~$p$ on trace~$\tau$.

More specifically, we use our static analyses to obtain the judgments for each statement in the codebase as shown in \defref{diodon-judgements-mainbody} and prove in \refsideconditionshold that these judgments are sufficient to discharge the side conditions~$\omega$.
While the static analyses run on the original program, before applying algorithm~\ghostalgorithm{}, the runtime (non-ghost) behavior is identical and so the judgments apply equally to the codebase after transformation.
This proof relies on our assumptions, \ie the \app's data race freedom, the syntactic restrictions for \core functions, and the soundness of our static analyses.

\begin{definition}[Static analyses for \approach]
    \label{def:diodon-judgements-mainbody}
    In \approach, we execute the static analyses on a codebase to obtain the following judgments for every statement~$s$ at label~$\ell$ therein, denoted as $\judgements{s^\ell}$.
    \begin{align*}
        \judgements{\cheapread{x}{e}} \triangleq \;
            \forall a, \tau \ldotp a \in &\:\pointstoset{e} \implies \appmanaged{a}{\tau}{\prestatesuperscript{\ell}}\\
        \judgements{\cheapwrite{x}{e}} \triangleq \;
        \forall a, \tau \ldotp a \in &\:\pointstoset{x} \implies \appmanaged{a}{\tau}{\prestatesuperscript{\ell}}\\
        \judgements{\ccorealloc{c}{\bar{e}}} \triangleq \;
            &\disjointallocationsites{\bar{e}} \land \localappmanaged{\bar{e}}{\ell}\\
        \judgements{\ccorecall{k}{c}{\bar{e}}{\bar{r}}} \triangleq \;
            &\disjointallocationsites{\bar{e}} \land \localappmanaged{\bar{e}}{\ell}\\
            &\land \localcore{c}{\ell} \land \localreturn{\bar{r}}{\ell}
    \end{align*}
    where
    \begin{align*}
        \disjointallocationsites{\bar{e}} \triangleq \;
            &\forall i, j \ldotp 0 \leq i < j < \len{\bar{e}} \implies{}\\
            &\phantom{\forall i, j} \pointstoset{\at{\bar{e}}{i}} \cap \pointstoset{\at{\bar{e}}{j}} = \emptyset\\
        \localappmanaged{\bar{e}}{\ell} \triangleq \;
            &\forall e, h, \tau \ldotp e \in \set{\bar{e}} \land {}\\
            &\phantom{\forall e, h, \tau \ldotp } \allocationsite{h}{\tau} \in \pointstoset{e} \implies{}\\
            &\phantom{\forall e, h} \local{e}{\poststatesuperscript{\ell}} \land \appmanaged{h}{\tau}{\prestatesuperscript{\ell}}\\
        \localcore{c}{\ell} \triangleq \;
            &\forall h, \tau \ldotp \allocationsite{h}{\tau} \in \pointstoset{c} \implies{}\\
            &\phantom{\forall h, \tau} \local{c}{\prestatesuperscript{\ell}} \land \passthroughcoretrace{h}{\tau}{\prestatesuperscript{\ell}}\\
        \localreturn{\bar{r}}{\ell} \triangleq \;
            &\forall r, \tau \ldotp r \in \set{\bar{r}} \implies \local{r}{\poststatesuperscript{\ell}}
    \end{align*}
    \prestatesuperscript{\ell} and \poststatesuperscript{\ell} refer to the program points immediately preceding and following the statement at label~$\ell$, respectively.
    \allocationsite{h}{\tau} returns heap location~$h$'s allocation site for a particular program trace~$\tau$.
    $\appmanaged{a}{\tau}{\prestatesuperscript{\ell}}$ denotes that an allocation site~$a$ is \app-managed on trace~$\tau$ at the program point~\prestatesuperscript{\ell}.
    An allocation site~$a$ is \app-managed if $a$ is either within the \app or the \core passed the corresponding heap location as a return argument to the \app.
    We decide whether $a$ is \app-managed by checking its location in the program text and running our pass-through analysis.
\end{definition}

\mypar{Proof construction}
While we showed that we can compose the proof rules in \figref{proof-rules-mainbody} and discharge their side conditions~$\omega$, it remains to show that we initially establish the global context~\globalproginv{} and the local program invariant~\localproginv{}, such that we obtain a proof for the entire codebase~\prog{}.
We close this gap in \corref{proof-construction-mainbody}.

\begin{corollary}[Proof construction]
    \ifthenelse{\boolean{mainbody}}{%
        \label{cor:proof-construction-mainbody}
    }{%
        \label{cor:proof-construction-appendix}
    }
    Successfully executing \approach's static analyses on a codebase~$c$ and the \core{}'s auto-active verification combined with our assumptions allow us to construct a \gls{separation-logic} proof for $c$.
    \begin{align*}
        \text{If }\;& \forall s, k \ldotp s \in c \land \judgements{s} \land{}\\
            &\phantom{ \forall } \Bigl(s = \ccorealloc{c}{\bar{e}} \implies\\
            &\phantom{ \forall s,} \ctxhoare{\globalproginv}{\precorealloc{\bar{e}}}{s}{\postcorealloc{c}{\bar{e}}}\Bigr) \land {}\\
            &\phantom{ \forall } \Bigl(s = \ccorecall{k}{c}{\bar{e}}{\bar{r}} \implies\\
            &\phantom{ \forall s,} \ctxhoare{\globalproginv}{\precorecall{k}{c}{\bar{e}}}{s}{\postcorecall{k}{c}{\bar{e}}{\bar{r}}}\Bigr)\text{,}\\
        \text{then }\;& \emp \vdash \simpleHoare{\phi}{s_\text{init} \cseq \ghostalgorithm(c)}{\true}
    \end{align*}
    where $s_\text{init}$ is \gls{ghost-code} creating and initializing the thread-local \glsdisp{ghost-code}{ghost sets} \setlocalheap{} and \setinvariantheap{} for the main thread, as well as the global \glsdisp{ghost-code}{ghost set} \ederef{\setglobalheap} and the \glsdisp{ghost-code}{ghost flag} \ederef{\flagiospec}.
\end{corollary}
% omit the proof sketch if we place this corollary in the main body:
\ifthenelse{\boolean{mainbody}}{}{%
    \begin{proofsketchcomposition}
    All our proof rules~(\cf \ifthenelse{\boolean{mainbody}}{\figref{proof-rules-mainbody}}{\figref{proof-rules-appendix}}) have the same shape, namely \ctxhoare{\globalproginv}{\localproginv}{\ghostalgorithm(s)}{\localproginv} for a statement~$s$.
    As shown by \lemref{side-conditions-hold}, the judgments obtained from the static analyses allow us to discharge the side conditions that are associated with each proof rule (\ifthenelse{\boolean{mainbody}}{\figref{side-conditions-mainbody}}{\figref{side-conditions-appendix}}).
    Therefore, left to show is that we initially establish \localproginv{} and \globalproginv{} such that we can compose the proof rules to form a proof for an entire codebase~$c$.
    The \glsdisp{ghost-code}{ghost statement}~$s_\text{init}$ creates and initializes the \glsdisp{ghost-code}{ghost sets} \setlocalheap{}, \setinvariantheap{}, and \ederef{\setglobalheap} as well as the \glsdisp{ghost-code}{ghost flag} \ederef{\flagiospec}.
    Thus, we can complete the proof tree as shown in \figref{whole-program-prooftree}.
    This constitutes a proof for $c$ as neither $s_\text{init}$ nor the statements added by \ghostalgorithm{} modify $c$'s runtime behavior.
    \end{proofsketchcomposition}
}

We show that we obtain the desired proof for the entire codebase, namely that the codebase satisfies the I/O specification~$\phi$ expressed as the Hoare triple $\emp \vdash \simpleHoare{\phi}{s_\text{init} \cseq \ghostalgorithm(\prog)}{\true}$.
This Hoare triple relies on $s_\text{init}$ that initializes \setlocalheap{}, \setinvariantheap{}, and \ederef{\setglobalheap} to empty sets, as well as sets the \glsdisp{ghost-code}{ghost flag} \ederef{\flagiospec} to \false.
$s_\text{init}$ is similar in spirit to the \glsdisp{ghost-code}{ghost statements} that algorithm~\ghostalgorithm{} inserts as these statements are necessary to construct a proof for the codebase~\prog{}.
\Corref{proof-construction-mainbody}'s premise states that our static analyses succeed on the codebase~\prog{}, such that we obtain \judgements{s} for each statement~$s$ therein, and that we prove a Hoare triple for each \core function satisfying the syntactic restrictions.

We combine the proof for the entire codebase that we obtain from \corref{proof-construction-mainbody} with the result of \secref{diodon-soundness-proof-overview-io-independence} to obtain \approach's overall soundness result.
This result states that successfully executing our static analyses on codebase~$\prog{}$ and auto-actively verifying its \core suffices to prove that the traces of executing $\prog{}$ together with other verified implementations and the environment are contained in the traces described by the abstract protocol model.

\begin{theorem}[Overall soundness]
    \label{thm:diodon-soundness-proof-overview-overall-soundness}
    Suppose \corref{proof-construction-mainbody}'s antecedent holds, and we established \ioindependence.
    Then, $\simplifiedrefinementstmt$ (simplified) holds.
\end{theorem}
\begin{proofsketch}
    By \corref{proof-construction-mainbody} and \refsoundnessioindependencesoundness (\cf \secref{diodon-soundness-proof-overview-io-independence}).
\end{proofsketch}

\subsubsection{Limitations}
\label{sec:diodon-soundness-limitations}
Our formalization defines a simple programming language to focus on the main ideas of our soundness proof and to show that successfully executing our static analyses discharges all side conditions.
We believe this language covers the most critical features like heap manipulations and concurrency as these features are relevant for the results of our static analyses.
In addition, we abstract each function making up the \core's \ac{API} to a dedicated statement in our language, and assume that the specification of each such function satisfies our syntactic restrictions~\refsyntacticrestrictionsassumption.
However, there is a slight risk that this language misses Go features that would be a threat to soundness such as function boundaries, complex control flow, and callbacks;
the former two features would be straightforward to add, and we discuss in \secref{diodon-soundness-extensions} how to add the latter.

To prove a Hoare triple for the entire codebase, we assume that the \app is free of crashes~\refcrashfreedomassumption and data races~\refdataracefreedomassumption.
While our soundness proof does not make any statement in the case that the program crashes, our compositional proof informally guarantees that the trace inclusion holds for the program's prefix up to the program point at which a crash occurs, such that the crash freedom assumption could be dropped, which we leave to future work.
However, data race freedom remains an assumption;
more generally, we assume the absence of undefined behavior for programming languages other than Go and our formalized one.
This assumption can be mitigated by performing additional static analyses.

\subsubsection{Extensions}
\label{sec:diodon-soundness-extensions}
Having covered the main soundness result, we discuss two~extensions to bridge the gap to realistic applications of \approach as used in our case studies.
We first lift the restriction of at most one~\core instance to allow a codebase to create unboundedly many \core instances.
Second, we allow the \core to invoke callbacks into the \app and discuss the side conditions that arise by this extension.

\mypar{Unboundedly many \core{}~instances}
So far, our global program invariant~\globalproginv{} contains the separating conjunct
\begin{equation*}
    \acc{\flagiospec} \star (\neg(\ederef{\flagiospec}) \implies \phi).
\end{equation*}
As explained in
\refsoundnessioindependence,
each execution of a protocol role is parameterized by a unique $\rid{}$.
\Ie $\phi$ and all I/O permissions that $\phi$ internally provides are parameterized by $\rid{}$ and, thus, are not interchangeable but specific to a particular $\rid{}$.
Hence, we can change the separating conjunct stated above to
\begin{equation*}
    \acc{\flagiospec} \star (\forall \rid \not\in \ederef{\flagiospec} \implies \phi(\rid))
\end{equation*}
providing a family of I/O permissions, where \flagiospec{} points to a \glsdisp{ghost-code}{ghost set} containing the $\rid{}$s that have already been used.
In addition, we adapt the entire program's precondition from $\phi$ to $\forall \rid \ldotp \phi(\rid)$ and change the translation~$\ghostalgorithm(\ccorealloc{c}{\bar{e}})$ to, first, pick a fresh $\rid'$ such that $\rid' \not\in \ederef{\flagiospec}$ and, second, adding $\rid'$ to \ederef{\flagiospec}.
Picking such a fresh $\rid'$ is always possible since $\rid$ ranges over $\naturals$.

\mypar{Adding callbacks to the \core}
So far, we have treated the statements \ccoreallocshort{\bar{e}} and \ccorecall{k}{c}{\bar{e}}{\bar{r}} as atomic statements in our language.
These two~statements are internally implemented as sequences of statements, which we hereafter call \core~statements.
As these statements constitute the \core, we auto-actively prove that a particular postcondition holds when control transfers back to the \app after fully executing these statements.

In the presence of callbacks, however, calling into the \core becomes non-atomic and control flow might transfer to the \app before reaching the post-state for which we know that the postcondition holds.
We can treat callbacks as temporarily pausing the execution of these auto-actively verified \core~statements to (sequentially) execute some statements belonging to the \app before eventually resuming execution of \core~statements.

\looseness=-1
With respect to algorithm~\ghostalgorithm{} and the \glsdisp{ghost-code}{ghost sets}, interrupting the execution of \core~statements to execute certain \app~statements~$s_\text{app}$ means that heap locations on which the \core~statements operate are missing from the \glsdisp{ghost-code}{ghost sets} while executing $s_\text{app}$ as we remove them from the \glsdisp{ghost-code}{ghost sets} before executing \core~statements and put them back only after the \core~statements' postcondition holds.
Missing permissions include both arguments~$\bar{e}$ and the \core instance~$c$.
Therefore, we have to make sure that $s_\text{app}$ neither accesses heap locations to which $\bar{e}$ points nor invokes \ac{API} calls on the \core instance~$c$ as the \core~invariant might not hold.

We can lift these restrictions by introducing additional proof obligations for the auto-active verification.
More specifically, if we auto-actively prove that the \core~statements satisfy a particular precondition for the callback, then we can update the \glsdisp{ghost-code}{ghost sets} accordingly.
\Eg such a precondition can specify permissions for heap locations passed to the callback or that the \core invariant holds.

In our \ac{SSM Agent} case study (\secref{diodon-ssm-agent}), we make use of these proof obligations for the callback delivering incoming messages to the \app as we specify that the \core transfers permission for the incoming message to the \app.
Conceptually, this allows us to add the corresponding heap location to \setlocalheap{} before executing the statements constituting the callback because the auto-active proof guarantees that no statement in the \core thereafter accesses this heap location.

For our case studies, it was not necessary to transfer permissions from a callback back to the \core via a callback's postcondition.
Extending \approach to allow such permission transfers would require an analysis of the callback showing that the \app possesses these permissions while executing the callback and that the corresponding heap locations do not get accessed by the \app after the callback returns.

%%% Local Variables:
%%% mode: LaTeX
%%% TeX-master: "../main"
%%% End:

\section{Case Studies}
\label{sec:diodon-evaluation}
To demonstrate that \approach scales to large codebases, we evaluate it on the \ac{AWS} \acf{SSM Agent}~\cite{SSMAgent}, a 100k+\thinspace \acp{LOC} production Go codebase.
Furthermore, we apply \approach to a small implementation of the signed \acf{DH} key exchange to showcase that our methodology applies to other implementations and coding styles.

\subsection{AWS Systems Manager Agent}
\label{sec:diodon-ssm-agent}
\looseness=-1
The \ac{AWS} \acf{SSM Agent}~\cite{SSMAgent} provides features for configuring, updating, and managing Amazon EC2~instances, and is widely used by \ac{AWS}~customers.
A fork of this codebase implements a novel protocol which enables encrypted interactive shell sessions with remote host machines, similar to the \ac{SSH} protocol, without needing to open inbound ports or manage \ac{SSH} keys.
This protocol establishes these shell sessions with a handshake protocol involving a signed elliptic-curve \ac{DH} key exchange to derive sessions keys that are subsequently used in the transport phase to encrypt the shell commands and their results.

\looseness=-1
We apply \approach by first modeling the protocol in \tamarin and proving secrecy and \gls{injective-agreement}~(\secref{diodon-eval-model}).
Second, we partition the codebase into the code implementing the protocol (the \core) and the remaining codebase (the \app), and prove \ioindependence~(\secref{diodon-eval-io-independence}).
Third, we auto-actively verify the \core using \gobra to prove that the \core refines the \ac{SSM Agent}'s role~(\secref{diodon-eval-core}).
Finally, we apply the automatic static analyses~\argot~\cite{argot} to discharge the assumptions within the \app on which the auto-active proof relies~(\secref{diodon-eval-discharging-assumptions}).

\begin{figure}[t]
\small\centering
\begin{tabular}{l c c c c c}
    \hline
    & Tool & Proof Effort & Execution Time \\
    \hline
    Protocol Model & \tamarin{} & \qty{< 2}{\text{\acsp{pm}}} & \qty{3.30}{\minute} \\
    \core{} Refinement & \gobra{} & \qty{< 3}{\text{\acsp{pm}}} & \qty{1.17}{\minute} \\
    I/O Independence & \argot{} & \qty{< 0.5}{\text{\acs{pm}}} & \qty{0.48}{\minute} \\
    \core Assumptions & \argot{} & \qty{< 1.5}{\text{\acsp{pm}}} & \qty{2.12}{\minute} \\ % invariant-proof + concurrency-proof + argument-alias-proof + pass-through-proof
    \hline
\end{tabular}
\caption{%
    Execution time for running each tool on the \acs{SSM Agent} codebase and approximate proof effort in \acptitle{pm} for creating a protocol model, adding specifications, and adapting the \argot analyses, respectively.
}
\label{fig:evaluation-times-ssm}
\end{figure}

\Figref{evaluation-times-ssm} overviews each tool's execution time, for which we use the \qty{10}{\percent} Winsorized mean of the wall-clock runtime across \num{10}~verification runs, measured on a 2023~Apple MacBook Pro with M3 Pro~processor and macOS~15.6.

\subsubsection{Protocol Model}
\label{sec:diodon-eval-model}
We model in \tamarin the security protocol for establishing a remote shell session between an \ac{SSM Agent} running on an EC2~instance and an \ac{AWS} customer.
The protocol offloads all signature operations to the \ac{AWS} \ac{KMS}~\cite{aws-kms} such that neither protocol role has to manage their own signing keys.
We model the connections to \ac{KMS} as secure channels.
Furthermore, the \ac{SSM Agent} sends the asymmetrically-encrypted session keys to a trusted third party to monitor the transmitted shell commands should this be necessary for regulatory reasons.
We provide the full description of the protocol in \ifthenelse{\boolean{show_full_ssm_agent_protocol}}{\appref{diodon-full-protocol}}{the extended version of this paper~\reffullprotocol}.

In \tamarin, we prove secrecy for the two~symmetric session keys, \ie the attacker does not learn these keys unless the \ac{SSM Agent}'s or customer's signing key or the monitor's secret key is corrupted.
Additionally, we prove that the \ac{SSM Agent} \glsdisp{injective-agreement}{injectively agrees} with the customer, and vice versa, on their identities and the session keys, unless one of the three~aforementioned corruption cases occurs.

The abstract protocol model amounts to \qty{319}{\acp{LOC}} and is automatically verified by \tamarin~1.10.0 in \qty{3.30}{\minute} using an auxiliary oracle consisting of \qty{75}{lines} of Python code.

\subsubsection{Proving \texorpdfstring{\ioIndependence}{I/O Independence}}
\label{sec:diodon-eval-io-independence}
We perform a taint analysis to prove \ioindependence.
We configure the taint analysis to consider all generated elliptic-curve secret keys as sources of protocol secrets.
We assume that only the \core uses the \ac{SSM Agent}'s signing keys and do not treat \ac{KMS} responses as taint sources because \ac{KMS} only sends us signatures and never key material.
As described in \secref{diodon-approach-io-independence}, we use \capslock's capability information to automatically configure the taint analysis' sinks.

\looseness=-1
We annotated some branching operations, instructing the taint analysis to ignore that the branch condition is tainted.
We identified two~classes of such branching operations.
The first class is justified by cryptography.
\Eg we allow branching on the success of decrypting a transport message because leakage is minimal.
The second class results from imprecisions of the taint analysis and corresponds to false positives, \ie the analysis deems a branch condition tainted although it is not.
To avoid another source of false positives, we configured the taint analysis to ignore taint escaping the current thread (which would otherwise always lead to errors).
Such cases could be handled precisely by marking certain struct fields as potentially storing concurrently-accessed, tainted data, such that the analysis can track the taint.

The taint analysis succeeds for the \ac{SSM Agent} codebase in \qty{29.0}{\second}, proving that there are no taint flows.

\subsubsection{\core Refinement}
\label{sec:diodon-eval-core}
The \ac{SSM Agent} contains a Go package called \code{datachannel} that implements the protocol.
More precisely, this package contains struct definitions that together store all necessary internal state.
Additionally, this package exposes publicly accessible functions to initialize the internal state, perform a handshake, and send a payload, which internally rely on several private functions.
We refer to these struct definitions and functions as the \core.
For backward compatibility, the \core also implements a legacy protocol; we assume that this legacy protocol is disabled.

\mypar{Implementation}
Each \core instance corresponds to one run of the protocol with a particular \ac{AWS} customer.
During initialization of a new \core instance, the \core starts a new thread,
responsible for receiving and processing incoming packets for this protocol run, similar to the running example.
If an incoming packet contains a transport phase payload, this payload is delivered by a callback to the \app.
Thus, the \core uses two different threads, one for sending messages and another one for receiving messages, which both operate on shared state.
This shared state keeps track of the progress within the protocol and the secret data involved in the protocol, such as the elliptic-curve \ac{DH} points and the resulting session keys.

Since the shared state is modified during the handshake, accesses must be synchronized to avoid data races.
Hence, the \core employs Go channels, \ie lightweight message passing, to signal a transfer of the shared state's ownership from one~thread to another.
During the handshake phase, exclusive ownership is transferred such that the threads have synchronized write access to the shared state.
Afterwards, the shared state, which includes the established session keys, is used in a read-only way permitting both threads to concurrently read the shared state while sending and receiving transport messages.

\mypar{Auto-active refinement proof}
We verify the \core using \gobra, which proves that the \core refines the \tamarin model's \ac{SSM Agent} role.
This proof encompasses \gls{safety}, \ie we prove that the \core does not crash and has sufficient permissions for every heap access, thus, guaranteeing absence of data races.
In particular, this forces us to reason precisely about the accesses to shared state that the two~threads within the \core perform.

Due to the intricate interplay of these threads, the resulting \gls{safety} proof is substantial and requires \gobra's expressivity.
We isolate and axiomatize operations that \gobra does not yet support such as simultaneously receiving on multiple channels and \glsdisp{functional-property}{functionally} reasoning about serialization and deserialization.
For the purpose of the proof, we treat the \core as a state machine consisting of 12~different states.
This allows us to refer to these states in the \core's invariant and precisely express for each state the permissions and progress \wrt the abstract protocol model.

Although the entire complexity of the proof is encapsulated in the \core's invariant, function calls to the \core must respect its state machine.
To avoid exposing the state machine in these functions' preconditions and imposing additional restrictions on callers, we slightly changed the implementation to perform a dynamic check consisting of a comparison with \nil and a single integer comparison ensuring that the state machine is in a correct state; otherwise, these \core functions return a descriptive error.
Thus, the \core functions' specifications are similar to those of our running example, \ie mention only the invariant and specify permissions for parameters without referring to the state machine.
While most parameters are of primitive type or shallow, there are a few non-shallow input parameters, which the \core treats as opaque.
Similarly, the callback from the \core to the \app delivers a non-shallow struct for which we ensure that the \core passes permissions for all transitively reachable heap locations to the \app.

We prove \gls{safety} and refinement of the \core in \qty{1.17}{\minute} for \qty{749}{lines} of code requiring \qty{3825}{lines} of specification and proof annotations; \num{1064}~thereof are related to the I/O specification and generated automatically by \tamarin.

\subsubsection{Analyzing the \app}
\label{sec:diodon-eval-discharging-assumptions}
The auto-active proof for the \core relies on callers satisfying the specified preconditions, which we establish using a combination of static analyses. We implemented automatic checks as described in \secref{diodon-approach-discharging-assumptions} for conditions~\condition{1}--\condition{4} and \condition{6}--\condition{8}. Condition~\condition{5} requires a more precise call graph than is currently available in our tool and is, thus, left as future work.

We implemented our analyses by forking and extending the existing \argot tool. Most of our analyses are obtained by interpreting the output of an existing analysis; \eg the parameter alias check uses the off-the-shelf pointer analysis to show parameters do not alias one another.

For some conditions, our static analyses were not able to validate the \app due to tool limitations.
For example, the escape analysis cannot reason about which fields are accessed after a struct escapes.
This can cause the tool to raise alarms when a struct stores a \core instance in a field. We found it was straightforward to rewrite the \core and \app to eliminate these failures. For example, the struct leakage can be fixed by moving the relevant field accesses before thread creation, so that the new thread has access only to the values of those fields and not the entire struct, and by extension the \core instance.

By running our escape analysis, we observed that \core instances escape the thread in which they are created because the \app creates a closure that closes over an object that points to a \core instance.
This capture is incidental in that the closure does not access the captured \core instance, which we verified by manual inspection.
This capture can be eliminated by rewriting the application to reference only the state necessary in this closure, rather than the full object. This change would result in a more defensive implementation by reducing the scope of possibly concurrent accesses.

Our pass-through analysis is a prototype that succeeds on our second~case study.
However, for the \ac{SSM Agent}, we obtain false positives due to allocations in functions called from both \core and \app, which could be addressed by adding calling context information.

\looseness=-1
Some \core functions take a pointer to a logger object as a parameter, which is internally thread-safe and shared between
threads. We can safely ignore escape errors due to these parameters because the \core does not access any memory of the logger object;
the pointer is just used as an opaque reference to invoke log functions that are part of the \app.

In summary, this case study demonstrates that \approach allows one to obtain strong security guarantees for a production codebase that was not designed with formal verification in mind. The remaining limitations (manual overrides of false positives in the static analyses, checking condition~\condition{5}, extremely lightweight dynamic checks enforcing non-nilness and
correct ordering of \ac{API} calls, and minor code changes) are modest compared to the complexity of the overall verification challenge and we conjecture that we can lift them by employing more precise static analyses.

\subsection{Signed \texorpdfstring{\actitle{DH}}{Diffie--Hellman} Key Exchange}
We also apply our approach to a codebase employing \emph{inverted I/O}, \ie has a \core that only produces and consumes byte arrays corresponding to protocol messages while the \app performs all I/O operations.
We adapted the \tamarin model and Go implementation of the signed \ac{DH} key exchange from \citeauthors{Arquint}{DBLP:conf/sp/ArquintWLSSWBM23} and extended both by a transport phase that uses the established session key to send and receive unboundedly many payloads.
\tamarin verifies the abstract model with \qty{177}{lines} of code in \qty{3.2}{\second} while \gobra verifies the \core consisting of \qty{178}{lines} of code in \qty{14.2}{\second} requiring \qty{1726}{\acp{LOS}}.
Executing all static analyses including the taint analysis takes \qty{9.7}{\second}.

This case study clearly exhibits the concept of virtual I/O\@.
The \core performs a virtual input operation for messages that the \app received from the network and forwarded to the \core.
Similarly, we perform a virtual output operation for every message that the \core produces before returning this message to the \app.
Therefore, we prove that the \tamarin model permits sending this message and in return, we sanitize the message from a taint analysis' perspective such that the \app can send the message without causing a false-positive taint flow.

\approach separates the justification of sending a particular message from the actual I/O operation.
This is important for tackling realistic codebases because identifying the actual send operation in a call stack is typically difficult as a message passes through several functions that, \eg add additional protocol headers before a message is handed to the network interface controller.

\subsection{Discussion}
\label{sec:diodon-evaluation-discussion}
Our evaluation demonstrates that \approach enables us to efficiently prove that an entire codebase refines a protocol model and therefore is secure.
To obtain the security properties as proven in \tamarin for a deployment of this protocol, we have to prove the implementations of all other protocol roles analogously against the same model using \approach.

As shown in \figref{evaluation-times-ssm}, the efforts for applying \approach to the \ac{SSM Agent} is manageable.
Thanks to \ioindependence, the \tamarin model is concise and can focus on the relevant interactions between the protocol roles.
In addition, \ioindependence allows us to apply automatic static analyses at the code-level to reason about all protocol-irrelevant I/O operations.
This contrasts existing approaches that would auto-actively verify the entire codebase and prove that every I/O operation is explicitly permitted by the model, which is completely impractical for this codebase.

To evaluate \approach's effectiveness at preventing security vulnerabilities, we deliberately introduce bugs in our case studies.
\Eg our taint analysis correctly fails if the \core's internal state, which includes the established session keys, is logged after the handshake.
Additionally, sending the \ac{DH} secret key in plaintext correctly results in \gobra failing to prove refinement \wrt the abstract protocol model.
The tools' execution time in the presence of these bugs remains comparable to that for the secure implementations.

By applying \approach we not only obtain security properties for the \ac{SSM Agent} codebase but we also discovered and fixed bugs along the way.
\tamarin allowed us to quickly locate and fix a \ac{MITM} attack in an earlier and unreleased version of the protocol, which is possible if the intended recipient's identity is omitted in the signatures ($\sigX{}$ and $\sigY{}$ in \reffullprotocolfig).
On the code level, we identified and fixed a potential data race in an earlier and unreleased version of the \core caused by insufficient synchronization between the two~threads that send and receive handshake messages.
We uncovered this data race because completing the \gls{safety} proof for the \core's earlier version is not possible as an additional synchronization point is necessary to transfer \gls{separation-logic} permissions between these threads.
This demonstrates the power of applying formal methods because detecting this data race with testing techniques would require to precisely time the reception of a handshake message such that the faulty memory access occurs and, thus, can be observed.

\section{Related Work}
\label{sec:diodon-related-work}
Much prior work on verifying security protocols exists and surveys~\cite{DBLP:conf/sp/BarbosaBBBCLP21, DBLP:journals/fac/AvallePS14, DBLP:conf/post/Blanchet12} provide an extensive overview.
Hence, we focus on approaches for verifying security properties for \emph{implementations} and their applicability to large and real-world codebases.
We end by comparing \approach to approaches based on dynamic verification.

\mypar{Implementation and model generation}
\looseness=-1
One approach to obtain verified protocol implementations generates secure-by-construction implementations from an abstract model, \eg \cite{DBLP:conf/aina/PozzaSD04, DBLP:conf/IEEEares/CadeB12, DBLP:conf/eurosp/BhargavanBDHKSW21, DBLP:conf/sp/GancherGSDP23, DBLP:conf/uss/SinghGP25}.
However, these implementations typically show subpar performance and optimizing them by hand or integrating them into a larger codebase forfeits proven security properties.
Thus, the abstract model has to cover the entire functionality (which we do not require).
While \owlc~\cite{DBLP:conf/uss/SinghGP25} enables embedding a generated implementation into a codebase, they rely on the Rust type system to shield secrets from the rest of the codebase.
We avoid this restriction by checking \ioindependence for the \app.
In addition, they use a Rust type to ensure that a previously established session key is used during a transport phase.
Instead, we use the \core invariant to maintain \gls{separation-logic} properties between \core \ac{API} calls, which is more expressive.

\looseness=-1
An alternative approach extracts an abstract model from an implementation, \eg \cite{DBLP:journals/toplas/BhargavanFGT08, OShea08, DBLP:conf/ccs/AizatulinGJ12, DBLP:conf/eurosp/KobeissiBB17, DBLP:conf/sp/BhargavanBK17}.
However, for this extraction to work, an implementation typically has to follow restrictive coding disciplines such that relevant protocol steps can be identified and extracted.
To achieve isolation between a verified component and potentially malicious code, 
\citeauthors{Kobeissi}{DBLP:conf/eurosp/KobeissiBB17} build on process isolation provided by operating systems and, thus, require verifying the entire critical process.
We cannot adopt this approach because it requires changing the codebase heavily to split it into several processes and results in an, for our use case, unacceptable overhead, since each process includes its copy of the Go runtime and the Go standard library.
\citeauthors{Bhargavan}{DBLP:conf/sp/BhargavanBK17} impose substantial restrictions on the \ac{API} of verified code, \eg disallowing state preservation between \ac{API} calls.
Codebases do not normally satisfy these restrictions, including all our case studies.
\Eg they use a session key for sending a transport message in one \ac{API} call that was established during the handshake, \ie a previous \ac{API} call.

\mypar{Existing implementations}
\citeauthors{Dupressoir}{DBLP:conf/csfw/DupressoirGJN11} embed a \emph{trace} storing relevant protocol operations as an auxiliary data structure for proof purposes into C~code implementing a security protocol.
This auxiliary data structure is removed before compilation and does not incur any runtime overhead while enabling reasoning about weak secrecy and \gls{non-injective-agreement}.
\citeauthors{Arquint}{DBLP:conf/ccs/ArquintSM023} generalize this approach to \gls{separation-logic}, making it applicable to a wide range of programming languages and supporting stronger security properties such as \gls{forward-secrecy} and \gls{injective-agreement}.
However, both approaches require a sufficiently strong invariant over this trace to prove security properties.
To avoid such a trace invariant, \citeauthors{Arquint}{DBLP:conf/sp/ArquintWLSSWBM23} prove security properties on the level of an abstract model using \tamarin's proof automation and prove that an implementation refines the abstract model.
All three~approaches require verifying the entire codebase using an auto-active verifier (which we do not).
We build on the latter approach and, to the best of our knowledge, are the first to relax this requirement to verifying just the \core and reason about the \app using lightweight static analyses.

\mypar{Dynamic verification}
Several approaches employ dynamic checks at runtime to allow for partially verified codebases.
\citeauthors{Agten}{DBLP:conf/popl/Agten0P15} target single-threaded C~code and generate runtime checks at the boundary between verified and unverified code to test that the verified code's specification holds.
To detect violations of properties expressed in \gls{separation-logic} such as ownership (via permissions) and aliasing, this approach tracks the heap locations accessed by the verified codebase at runtime and computes cryptographic hashes thereover.
It remains unclear whether these checks only at the boundary remain sufficient when targeting concurrent codebases or whether the runtime overhead increases further.
To avoid tracking heap locations at runtime, \citeauthors{Ho}{DBLP:conf/sp/HoPBB22} copy all heap data at this boundary to rule out aliasing.

Gradual verification (\eg \cite{DBLP:conf/vmcai/BaderAT18,DBLP:journals/pacmpl/WiseBWATS20}) combines auto-active verification with dynamic checks but aims at helping the proof developer by allowing incomplete specifications.
\Ie gradual verification enables incremental verification where each function's specification is extended over time to eventually obtain a fully specified and verified codebase.
However, as long as a codebase is not fully specified and verified, gradual verification requires tracking heap locations at runtime, which results in noticeable runtime overhead.

\looseness=-1
SCIO$^{\star}$~\cite{andrici24} is an \fstar transpiler that injects dynamic checks not only at the boundary between verified and unverified code but also at call sites of I/O operations.
While they can enforce access policies for I/O operations, it remains unclear how this approach extends to cryptographic message payloads.
To be applicable in our context, we would need to dynamically check whether a message sent by our \app is indeed protocol-irrelevant and, thus, does not contain any secrets from the \core{}\textemdash not even in encrypted form.
Like our work, SecRef$^{\star}$~\cite{DBLP:journals/corr/abs-2503-00404} considers the problem of verifying only a subset of a codebase due to the otherwise prohibitive proof effort.
While they also allow pre- and postconditions at the boundary between verified and unverified code, they rely on dynamic checks to enforce these conditions for heap locations accessible by unverified code.
For a verified component like our \core, this means that they check the entire invariant at runtime for each \ac{API} call (which we do not), as they treat the unverified code as potentially modifying a \core instance's entire state.
By targeting a single-threaded language~(\fstar), SecRef$^{\star}$ does not have to consider concurrent memory accesses (which we do).

By contrast, \approach performs only extremely lightweight dynamic checks enforcing non-nilness and correct ordering of \ac{API} calls, and checks all other constraints \emph{statically} to avoid runtime overhead while simultaneously requiring minimal code changes.

\section{Conclusions}
\looseness=-1
We present \approach, a novel methodology to scale verification of security protocol implementations to large existing codebases by symbiotically combining powerful auto-active verification of a relatively small part of the codebase with static analyses that scale to the entire codebase.
Since \approach is not inherently limited to Go, future work could apply it to codebases written in, \eg Rust and C. Adapting \approach to Rust would remove several static analyses due to the strong type system, and C has a variety of static analyses and auto-active verifiers that can be used.
Orthogonally, extending \approach with further static analyses, such as for nilness, would allow us to pass more guarantees from the \app to the \core.
We hope our work spurs both practical and theoretical understanding of how to soundly combine proof systems of different expressive power.

\section*{Acknowledgments}
We thank the Werner Siemens-Stiftung~(WSS) for their generous support of this project, Michael Hicks, K.\ Rustan M.\ Leino, and Margarida Ferreira for feedback on drafts of this paper, Christoph Sprenger and Joseph Lallemand for helpful discussions, and the anonymous reviewers for their insightful comments.
Parts of this work were conducted by the first author during internships at \ac{AWS}.
Some authors are developers of \gobra or \argot, which we disclose as potential non-financial interests.

%-------------------------------------------------------------------------------
\bibliographystyle{IEEEtran}
\balance
\bibliography{IEEEabrv, paper-bibliography, bibliography}

\begin{thebibliography}{10}
\providecommand{\url}[1]{#1}
\csname url@samestyle\endcsname
\providecommand{\newblock}{\relax}
\providecommand{\bibinfo}[2]{#2}
\providecommand{\BIBentrySTDinterwordspacing}{\spaceskip=0pt\relax}
\providecommand{\BIBentryALTinterwordstretchfactor}{4}
\providecommand{\BIBentryALTinterwordspacing}{\spaceskip=\fontdimen2\font plus
\BIBentryALTinterwordstretchfactor\fontdimen3\font minus
  \fontdimen4\font\relax}
\providecommand{\BIBforeignlanguage}[2]{{%
\expandafter\ifx\csname l@#1\endcsname\relax
\typeout{** WARNING: IEEEtran.bst: No hyphenation pattern has been}%
\typeout{** loaded for the language `#1'. Using the pattern for}%
\typeout{** the default language instead.}%
\else
\language=\csname l@#1\endcsname
\fi
#2}}
\providecommand{\BIBdecl}{\relax}
\BIBdecl

\bibitem{DBLP:conf/csfw/SchmidtMCB12}
B.~Schmidt, S.~Meier, C.~Cremers, and D.~A. Basin, ``Automated analysis of
  {Diffie--Hellman} protocols and advanced security properties,'' in
  \emph{{CSF}}.\hskip 1em plus 0.5em minus 0.4em\relax {IEEE}, 2012, pp.
  78--94.

\bibitem{DBLP:conf/cav/MeierSCB13}
S.~Meier, B.~Schmidt, C.~Cremers, and D.~A. Basin, ``The {TAMARIN} prover for
  the symbolic analysis of security protocols,'' in \emph{{CAV}}, ser. {LNCS},
  vol. 8044.\hskip 1em plus 0.5em minus 0.4em\relax Springer, 2013, pp.
  696--701.

\bibitem{DBLP:conf/csfw/Blanchet01}
B.~Blanchet, ``An efficient cryptographic protocol verifier based on {Prolog}
  rules,'' in \emph{{CSFW}}.\hskip 1em plus 0.5em minus 0.4em\relax {IEEE},
  2001, pp. 82--96.

\bibitem{DBLP:conf/sp/BhargavanBK17}
K.~Bhargavan, B.~Blanchet, and N.~Kobeissi, ``Verified models and reference
  implementations for the {TLS} 1.3 standard candidate,'' in
  \emph{{S{\&}P}}.\hskip 1em plus 0.5em minus 0.4em\relax {IEEE}, 2017, pp.
  483--502.

\bibitem{DBLP:conf/sp/BasinST21}
D.~A. Basin, R.~Sasse, and J.~Toro{-}Pozo, ``The {EMV} standard: {Break}, fix,
  verify,'' in \emph{{S{\&}P}}.\hskip 1em plus 0.5em minus 0.4em\relax {IEEE},
  2021, pp. 1766--1781.

\bibitem{DBLP:conf/eurosp/KobeissiBB17}
N.~Kobeissi, K.~Bhargavan, and B.~Blanchet, ``Automated verification for secure
  messaging protocols and their implementations: {A} symbolic and computational
  approach,'' in \emph{{EuroS{\&}P}}.\hskip 1em plus 0.5em minus 0.4em\relax
  {IEEE}, 2017, pp. 435--450.

\bibitem{DBLP:conf/ccs/BasinDHRSS18}
D.~A. Basin, J.~Dreier, L.~Hirschi, S.~Radomirovic, R.~Sasse, and V.~Stettler,
  ``A formal analysis of {5G} authentication,'' in \emph{{CCS}}.\hskip 1em plus
  0.5em minus 0.4em\relax {ACM}, 2018, pp. 1383--1396.

\bibitem{DBLP:conf/ndss/CremersD19}
C.~Cremers and M.~Dehnel{-}Wild, ``Component-based formal analysis of {5G-AKA:}
  {Channel} assumptions and session confusion,'' in \emph{{NDSS}}.\hskip 1em
  plus 0.5em minus 0.4em\relax The Internet Society, 2019.

\bibitem{CVEMatrix21}
\BIBentryALTinterwordspacing
{CVE}, ``{CVE}-2021-40823,'' 2021. [Online]. Available:
  \url{https://www.cve.org/CVERecord?id=CVE-2021-40823}
\BIBentrySTDinterwordspacing

\bibitem{CVETLStormPacketReassembly}
\BIBentryALTinterwordspacing
------, ``{CVE}-2022-22805,'' 2022. [Online]. Available:
  \url{https://www.cve.org/CVERecord?id=CVE-2022-22805}
\BIBentrySTDinterwordspacing

\bibitem{CVETLStormAuthenticationBypass}
\BIBentryALTinterwordspacing
------, ``{CVE}-2022-22806,'' 2022. [Online]. Available:
  \url{https://www.cve.org/CVERecord?id=CVE-2022-22806}
\BIBentrySTDinterwordspacing

\bibitem{DBLP:conf/csfw/DupressoirGJN11}
F.~Dupressoir, A.~D. Gordon, J.~J{\"{u}}rjens, and D.~A. Naumann, ``Guiding a
  general-purpose {C} verifier to prove cryptographic protocols,'' in
  \emph{{CSF}}.\hskip 1em plus 0.5em minus 0.4em\relax {IEEE}, 2011, pp. 3--17.

\bibitem{DBLP:conf/sp/ArquintWLSSWBM23}
L.~Arquint, F.~A. Wolf, J.~Lallemand, R.~Sasse, C.~Sprenger, S.~N. Wiesner,
  D.~A. Basin, and P.~M{\"{u}}ller, ``Sound verification of security protocols:
  {From} design to interoperable implementations,'' in \emph{{S{\&}P}}.\hskip
  1em plus 0.5em minus 0.4em\relax {IEEE}, 2023, pp. 1077--1093.

\bibitem{DBLP:conf/ccs/ArquintSM023}
L.~Arquint, M.~Schwerhoff, V.~Mehta, and P.~M{\"{u}}ller, ``A generic
  methodology for the modular verification of security protocol
  implementations,'' in \emph{{CCS}}.\hskip 1em plus 0.5em minus 0.4em\relax
  {ACM}, 2023, pp. 1377--1391.

\bibitem{DBLP:journals/tit/DolevY83}
D.~Dolev and A.~C. Yao, ``On the security of public key protocols,''
  \emph{{IEEE} Trans. Inf. Theory}, vol.~29, no.~2, pp. 198--207, 1983.

\bibitem{CVETerraformLogSecrets24}
\BIBentryALTinterwordspacing
{CVE}, ``{CVE-2024-47083},'' 2024. [Online]. Available:
  \url{https://www.cve.org/CVERecord?id=CVE-2024-47083}
\BIBentrySTDinterwordspacing

\bibitem{CVEGitHubEnterpriseLogSecrets23}
\BIBentryALTinterwordspacing
------, ``{CVE-2023-6746},'' 2023. [Online]. Available:
  \url{https://www.cve.org/CVERecord?id=CVE-2023-6746}
\BIBentrySTDinterwordspacing

\bibitem{krml212}
K.~R.~M. Leino and M.~Moskal, ``Usable auto-active verification,'' in
  \emph{Usable Verification Workshop}, 2010.

\bibitem{SSMAgent}
\BIBentryALTinterwordspacing
{Amazon Web Services, Inc.}, ``{Working with SSM Agent},'' 2023. [Online].
  Available:
  \url{https://docs.aws.amazon.com/systems-manager/latest/userguide/ssm-agent.html}
\BIBentrySTDinterwordspacing

\bibitem{paper-artifact-zenodo}
\BIBentryALTinterwordspacing
L.~Arquint, S.~Kishor, J.~R. Koenig, J.~Dodds, D.~Kroening, and
  P.~M{\"{u}}ller, ``The secrets must not flow: Scaling security verification
  to large codebases (artifact),'' Sep. 2025. [Online]. Available:
  \url{https://doi.org/10.5281/zenodo.17099763}
\BIBentrySTDinterwordspacing

\bibitem{paper-artifact-github}
\BIBentryALTinterwordspacing
------. (2025, Oct.) The secrets must not flow: Scaling security verification
  to large codebases. Artifact repository containing the protocol models, the
  forked SSM Agent's codebase, a DH implementation codebase, and the static
  analysis tools. [Online]. Available:
  \url{https://github.com/viperproject/diodon-artifact}
\BIBentrySTDinterwordspacing

\bibitem{DBLP:conf/cav/WolfACOPM21}
F.~A. Wolf, L.~Arquint, M.~Clochard, W.~Oortwijn, J.~C. Pereira, and
  P.~M{\"{u}}ller, ``{Gobra:} {Modular} specification and verification of {Go}
  programs,'' in \emph{{CAV}}, ser. {LNCS}, vol. 12759.\hskip 1em plus 0.5em
  minus 0.4em\relax Springer, 2021, pp. 367--379.

\bibitem{argot}
\BIBentryALTinterwordspacing
{AWS Labs}, ``{Argot},'' 2024. [Online]. Available:
  \url{https://github.com/awslabs/ar-go-tools}
\BIBentrySTDinterwordspacing

\bibitem{DBLP:conf/lics/Reynolds02}
J.~C. Reynolds, ``{Separation Logic:} {A} logic for shared mutable data
  structures,'' in \emph{{LICS}}.\hskip 1em plus 0.5em minus 0.4em\relax
  {IEEE}, 2002, pp. 55--74.

\bibitem{DBLP:conf/fm/BasinHST24}
D.~A. Basin, X.~Hofmeier, R.~Sasse, and J.~Toro{-}Pozo, ``Getting chip card
  payments right,'' in \emph{{FM} {(1)}}, ser. {LNCS}, vol. 14933.\hskip 1em
  plus 0.5em minus 0.4em\relax Springer, 2024, pp. 29--51.

\bibitem{DBLP:conf/uss/GirolHSJCB20}
G.~Girol, L.~Hirschi, R.~Sasse, D.~Jackson, C.~Cremers, and D.~A. Basin, ``A
  spectral analysis of {Noise:} {A} comprehensive, automated, formal analysis
  of {Diffie--Hellman} protocols,'' in \emph{{USENIX} Security
  Symposium}.\hskip 1em plus 0.5em minus 0.4em\relax {USENIX} Association,
  2020, pp. 1857--1874.

\bibitem{DBLP:conf/sas/Boyland03}
J.~Boyland, ``Checking interference with fractional permissions,'' in
  \emph{{SAS}}, ser. {LNCS}, vol. 2694.\hskip 1em plus 0.5em minus 0.4em\relax
  Springer, 2003, pp. 55--72.

\bibitem{DBLP:conf/popl/ParkinsonB05}
M.~J. Parkinson and G.~M. Bierman, ``{Separation Logic} and abstraction,'' in
  \emph{{POPL}}.\hskip 1em plus 0.5em minus 0.4em\relax {ACM}, 2005, pp.
  247--258.

\bibitem{DBLP:conf/nfm/JacobsSPVPP11}
B.~Jacobs, J.~Smans, P.~Philippaerts, F.~Vogels, W.~Penninckx, and F.~Piessens,
  ``{VeriFast:} {A} powerful, sound, predictable, fast verifier for {C} and
  {Java},'' in \emph{{NASA} Formal Methods}, ser. {LNCS}, vol. 6617.\hskip 1em
  plus 0.5em minus 0.4em\relax Springer, 2011, pp. 41--55.

\bibitem{DBLP:journals/jar/CaoBGDA18}
Q.~Cao, L.~Beringer, S.~Gruetter, J.~Dodds, and A.~W. Appel, ``{VST-Floyd:} {A}
  {Separation Logic} tool to verify correctness of {C} programs,'' \emph{J.
  Autom. Reason.}, vol.~61, no. 1-4, pp. 367--422, 2018.

\bibitem{DBLP:conf/cav/Eilers018}
M.~Eilers and P.~M{\"{u}}ller, ``{Nagini:} {A} static verifier for {Python},''
  in \emph{{CAV} {(1)}}, ser. {LNCS}, vol. 10981.\hskip 1em plus 0.5em minus
  0.4em\relax Springer, 2018, pp. 596--603.

\bibitem{DBLP:journals/pacmpl/Astrauskas0PS19}
V.~Astrauskas, P.~M{\"{u}}ller, F.~Poli, and A.~J. Summers, ``Leveraging {Rust}
  types for modular specification and verification,'' \emph{Proc. {ACM}
  Program. Lang.}, vol.~3, no. {OOPSLA}, pp. 147:1--147:30, 2019.

\bibitem{DBLP:conf/esop/Penninckx0P15}
W.~Penninckx, B.~Jacobs, and F.~Piessens, ``Sound, modular and compositional
  verification of the input/output behavior of programs,'' in \emph{{ESOP}},
  ser. {LNCS}, vol. 9032.\hskip 1em plus 0.5em minus 0.4em\relax Springer,
  2015, pp. 158--182.

\bibitem{capslock}
\BIBentryALTinterwordspacing
Google, ``Capslock,'' 2024. [Online]. Available:
  \url{https://github.com/google/capslock}
\BIBentrySTDinterwordspacing

\bibitem{astree}
\BIBentryALTinterwordspacing
{AbsInt Angewandte Informatik GmbH}, ``{Astr\'ee} static analyzer for {C} and
  {C++},'' 2025. [Online]. Available: \url{https://www.absint.com/astree}
\BIBentrySTDinterwordspacing

\bibitem{DBLP:conf/sp/ProtzenkoPFHPBB20}
J.~Protzenko, B.~Parno, A.~Fromherz, C.~Hawblitzel, M.~Polubelova,
  K.~Bhargavan, B.~Beurdouche, J.~Choi, A.~Delignat{-}Lavaud, C.~Fournet,
  N.~Kulatova, T.~Ramananandro, A.~Rastogi, N.~Swamy, C.~M. Wintersteiger, and
  S.~Z. B{\'{e}}guelin, ``{EverCrypt:} {A} fast, verified, cross-platform
  cryptographic provider,'' in \emph{{S{\&}P}}.\hskip 1em plus 0.5em minus
  0.4em\relax {IEEE}, 2020, pp. 983--1002.

\bibitem{go-memory-model}
\BIBentryALTinterwordspacing
{Go developers}, ``The {Go} memory model,'' 2022. [Online]. Available:
  \url{https://go.dev/ref/mem}
\BIBentrySTDinterwordspacing

\bibitem{DBLP:journals/corr/abs-2212-04171}
L.~Arquint, F.~A. Wolf, J.~Lallemand, R.~Sasse, C.~Sprenger, S.~N. Wiesner,
  D.~A. Basin, and P.~M{\"{u}}ller, ``Sound verification of security protocols:
  From design to interoperable implementations (extended version),''
  \emph{{CoRR}}, vol. abs/2212.04171, 2022.

\bibitem{DBLP:journals/entcs/Vafeiadis11}
V.~Vafeiadis, ``Concurrent {Separation Logic} and operational semantics,'' in
  \emph{{MFPS}}, ser. Electronic Notes in Theoretical Computer Science, vol.
  276.\hskip 1em plus 0.5em minus 0.4em\relax Elsevier, 2011, pp. 335--351.

\bibitem{aws-kms}
\BIBentryALTinterwordspacing
{Amazon Web Services, Inc.}, ``{AWS Key Management Service},'' 2024. [Online].
  Available: \url{https://aws.amazon.com/kms/}
\BIBentrySTDinterwordspacing

\bibitem{DBLP:conf/sp/BarbosaBBBCLP21}
M.~Barbosa, G.~Barthe, K.~Bhargavan, B.~Blanchet, C.~Cremers, K.~Liao, and
  B.~Parno, ``{SoK:} {Computer}-aided cryptography,'' in \emph{{S{\&}P}}.\hskip
  1em plus 0.5em minus 0.4em\relax {IEEE}, 2021, pp. 777--795.

\bibitem{DBLP:journals/fac/AvallePS14}
M.~Avalle, A.~Pironti, and R.~Sisto, ``Formal verification of security protocol
  implementations: a survey,'' \emph{Formal Aspects Comput.}, vol.~26, no.~1,
  pp. 99--123, 2014.

\bibitem{DBLP:conf/post/Blanchet12}
B.~Blanchet, ``Security protocol verification: Symbolic and computational
  models,'' in \emph{{POST}}, ser. {LNCS}, vol. 7215.\hskip 1em plus 0.5em
  minus 0.4em\relax Springer, 2012, pp. 3--29.

\bibitem{DBLP:conf/aina/PozzaSD04}
D.~Pozza, R.~Sisto, and L.~Durante, ``{Spi2Java:} {Automatic} cryptographic
  protocol {Java} code generation from {Spi} calculus,'' in
  \emph{{AINA}}.\hskip 1em plus 0.5em minus 0.4em\relax {IEEE}, 2004, pp.
  400--405.

\bibitem{DBLP:conf/IEEEares/CadeB12}
D.~Cad{\'{e}} and B.~Blanchet, ``From computationally-proved protocol
  specifications to implementations,'' in \emph{{ARES}}.\hskip 1em plus 0.5em
  minus 0.4em\relax {IEEE}, 2012, pp. 65--74.

\bibitem{DBLP:conf/eurosp/BhargavanBDHKSW21}
K.~Bhargavan, A.~Bichhawat, Q.~H. Do, P.~Hosseyni, R.~K{\"{u}}sters,
  G.~Schmitz, and T.~W{\"{u}}rtele, ``{DY*:} {A} modular symbolic verification
  framework for executable cryptographic protocol code,'' in
  \emph{{EuroS{\&}P}}.\hskip 1em plus 0.5em minus 0.4em\relax {IEEE}, 2021, pp.
  523--542.

\bibitem{DBLP:conf/sp/GancherGSDP23}
J.~Gancher, S.~Gibson, P.~Singh, S.~Dharanikota, and B.~Parno, ``{Owl:}
  {Compositional} verification of security protocols via an information-flow
  type system,'' in \emph{{S{\&}P}}.\hskip 1em plus 0.5em minus 0.4em\relax
  {IEEE}, 2023, pp. 1130--1147.

\bibitem{DBLP:conf/uss/SinghGP25}
P.~Singh, J.~Gancher, and B.~Parno, ``{OwlC:} {Compiling} security protocols to
  verified, secure, high-performance libraries,'' in \emph{{USENIX} Security
  Symposium}.\hskip 1em plus 0.5em minus 0.4em\relax {USENIX} Association,
  2025, pp. 5071--5090.

\bibitem{DBLP:journals/toplas/BhargavanFGT08}
K.~Bhargavan, C.~Fournet, A.~D. Gordon, and S.~Tse, ``Verified interoperable
  implementations of security protocols,'' \emph{{ACM} Trans. Program. Lang.
  Syst.}, vol.~31, no.~1, pp. 5:1--5:61, 2008.

\bibitem{OShea08}
N.~O'Shea, ``Using {Elyjah} to analyse {Java} implementations of cryptographic
  protocols,'' in \emph{{FCS-ARSPA-WITS-2008}}, 2008.

\bibitem{DBLP:conf/ccs/AizatulinGJ12}
M.~Aizatulin, A.~D. Gordon, and J.~J{\"{u}}rjens, ``Computational verification
  of {C} protocol implementations by symbolic execution,'' in
  \emph{{CCS}}.\hskip 1em plus 0.5em minus 0.4em\relax {ACM}, 2012, pp.
  712--723.

\bibitem{DBLP:conf/popl/Agten0P15}
P.~Agten, B.~Jacobs, and F.~Piessens, ``Sound modular verification of {C} code
  executing in an unverified context,'' in \emph{{POPL}}.\hskip 1em plus 0.5em
  minus 0.4em\relax {ACM}, 2015, pp. 581--594.

\bibitem{DBLP:conf/sp/HoPBB22}
S.~Ho, J.~Protzenko, A.~Bichhawat, and K.~Bhargavan, ``{Noise*:} {A} library of
  verified high-performance secure channel protocol implementations,'' in
  \emph{{S{\&}P}}.\hskip 1em plus 0.5em minus 0.4em\relax {IEEE}, 2022, pp.
  107--124.

\bibitem{DBLP:conf/vmcai/BaderAT18}
J.~Bader, J.~Aldrich, and {\'{E}}.~Tanter, ``Gradual program verification,'' in
  \emph{{VMCAI}}, ser. Lecture Notes in Computer Science, vol. 10747.\hskip 1em
  plus 0.5em minus 0.4em\relax Springer, 2018, pp. 25--46.

\bibitem{DBLP:journals/pacmpl/WiseBWATS20}
J.~Wise, J.~Bader, C.~Wong, J.~Aldrich, {\'{E}}.~Tanter, and J.~Sunshine,
  ``Gradual verification of recursive heap data structures,'' \emph{Proc. {ACM}
  Program. Lang.}, vol.~4, no. {OOPSLA}, pp. 228:1--228:28, 2020.

\bibitem{andrici24}
C.~Andrici, {\c{S}}.~Ciob{\^{a}}c{\u{a}}, C.~Hritcu, G.~Mart{\'{\i}}nez,
  E.~Rivas, {\'{E}}.~Tanter, and T.~Winterhalter, ``Securing verified {IO}
  programs against unverified code in {F*},'' \emph{Proc. {ACM} Program.
  Lang.}, vol.~8, no. {POPL}, pp. 2226--2259, 2024.

\bibitem{DBLP:journals/corr/abs-2503-00404}
C.~Andrici, D.~Ahman, C.~Hritcu, R.~Icleanu, G.~Mart{\'{\i}}nez, E.~Rivas, and
  T.~Winterhalter, ``{SecRef*:} {Securely} sharing mutable references between
  verified and unverified code in {F*},'' \emph{{CoRR}}, vol. abs/2503.00404,
  2025.

\bibitem{DBLP:journals/pacmpl/0001KEW0CB20}
C.~Sprenger, T.~Klenze, M.~Eilers, F.~A. Wolf, P.~M{\"{u}}ller, M.~Clochard,
  and D.~A. Basin, ``{Igloo:} {Soundly} linking compositional refinement and
  {Separation Logic} for distributed system verification,'' \emph{Proc. {ACM}
  Program. Lang.}, vol.~4, no. {OOPSLA}, pp. 152:1--152:31, 2020.

\end{thebibliography}

% clear page inserting a page break and flushing floats only for the full version:
\ifisextendedversion{%
    \clearpage
}
\appendices

\ifthenelse{\boolean{show_meta_review}}{%
    \input{sections/AA_meta_review}
}{}

\ifthenelse{\boolean{show_full_soundness_appendix}}{%
    \section{Soundness Proof Sketch}
\label{app:diodon-soundness-proof}
To prove \approach sound, we reason separately about allowing protocol-independent I/O operations in a codebase and combining auto-active verification with static analyses.

In \appref{diodon-soundness-io-independence}, we prove that a codebase~\prog{} containing protocol-dependent \emph{and} protocol-independent I/O operations refines a given \tamarin model if the I/O specification~$\phi$, corresponding to a protocol role in this \tamarin model, permits all protocol-dependent I/O operations in the codebase.
For this part of the soundness proof, we assume that the \emph{entire} codebase~\prog{} satisfies the Hoare triple \hoaretriple{\phi}{\prog}{\true}, where protocol-independent I/O operations do not consume an I/O permission and, thus, can be performed at arbitrary points within \prog{} and independently of the I/O specification~$\phi$.
Such a Hoare triple could be obtained by auto-actively verifying the \emph{entire} codebase~\prog{}, which \approach does not require.

In \appref{diodon-composition-soundness}, we show that we constructively obtain the Hoare triple \hoaretriple{\phi}{\prog}{\true} for an entire codebase~\prog{} using \approach by auto-actively verifying only parts thereof, namely the \core, and executing our static analyses on \prog{}, if we assume crash freedom and absence of undefined behavior for the parts of \prog{} that are not auto-actively verified.

\looseness=-1
By combining both results, we obtain that applying \approach to a codebase~\prog{} proves that \prog{} refines a \tamarin model, despite the presence of protocol-independent I/O operations, and auto-actively verifying the \core only, as long as we discharge the side conditions using our static analyses.

\subsection{\texorpdfstring{\ioIndependence}{I/O Independence}}
\label{app:diodon-soundness-io-independence}
We show that we can soundly allow protocol-independent I/O operations in a codebase by treating these I/O operations as a refinement of our attacker model.
For this purpose, we extend Arquint~\etal's soundness proof~\cite[App.~E]{DBLP:journals/corr/abs-2212-04171} to accommodate such I/O operations, and we adopt their notation for legibility.
More specifically, we add these I/O operations to the concrete system and show that this concrete system refines an abstract system that is composed of only protocol roles and our attacker, which corresponds to a protocol's Tamarin model.

Since we permit a codebase~\prog{} to perform independent I/O operations in addition to I/O operations permitted by an I/O specification~$\phi$, we adapt the verifier assumption~\cite[Asm.~1]{DBLP:journals/corr/abs-2212-04171} to account for both types of I/O operations.

\newcommand*{\independentiotraces}{\ensuremath{\delta}}
\begin{assumption}[Verifier assumption]
    \label{asm:diodon-verifier-assumption}
    \begin{equation*}
        \vdash_\alpha \hoaretriple{\phi}{\prog}{\true} \land \taintanalysis(\prog, \taintanalysisconfig) = \true \implies \alpha(\CC) \tracePre \phi \interl \independentiotraces.
    \end{equation*}
\end{assumption}

\Ie we assume that successfully verifying a program~\prog{} against an I/O specification~$\phi$ and successfully executing the taint analysis~\taintanalysis{} with some configuration~\taintanalysisconfig{} specifying sources and sinks of tainted data implies that the program's traces abstracted under a relabeling function~$\alpha$ are included in the parallel composition of the I/O specification's traces~$\phi$ and the traces of performing independent I/O operations~\independentiotraces{}.

\looseness=-1
We assume that the program's traces are described by the \acf{LTS} semantics~\CC{}, which is provided by the operational semantics of the programming 
language in which \prog{} is implemented\footnote{%
We leave the programming language intentionally unspecified to keep our soundness result general.
}.
$\alpha$ abstracts the program's traces, \eg referring to specific function names, to traces of operations that match the naming as used in $\phi$ and \independentiotraces{}.
\independentiotraces{} represents the set of traces resulting from generating fresh nonces and using received payloads as well as public constants to construct and send payloads, as will be made explicit in \defref{MDInd}.

Note that \asmref{diodon-verifier-assumption} expresses besides the trace inclusion itself that the states of $\phi$ and \independentiotraces{} are independent such that their parallel composition is possible.
We obtain this independence by successfully executing our taint analysis.
In particular, our taint analysis establishes that protocol-independent I/O operations do not operate on tainted data.
We configure the taint analysis such that long-term and short-term secrets known by a protocol role but not the attacker are a source of taint.
Hence, these I/O operations and all steps necessary to compute their data are either already part of \independentiotraces{} or the necessary computation steps can be replicated and added thereto, such that \independentiotraces{} is independent of $\phi$.

\looseness=-1
The other direction, namely that the I/O specification~$\phi$ is independent of from \independentiotraces{}, holds by construction of $\phi$.
Since we generate $\phi$ by a series of transformations from a protocol role's abstract model and use syntactically distinct elements to represent this protocol role's state and express the transitions that form \independentiotraces{}, as we shall see next, \independentiotraces{} cannot influence $\phi$.

For a set of function symbols~$\Sigma$ operating over terms, $\MD$ denotes the set of transition rules that the attacker can apply.
$\K(x)$ represents the fact that the attacker knows the term~$x$ and the fact symbols $\Outfact$ and $\Infact$ represent that a protocol participant sent and might receive a particular term, respectively.
Therefore, $\MD$ captures all operations available to the attacker, namely receiving a sent term, sending a term, adding a public constant to its knowledge, generating a fresh nonce, and applying a function~$f \in \Sigma$.

\begin{definition}[Attacker message deduction rules]
    As defined in \cite[Def.~9]{DBLP:journals/corr/abs-2212-04171}, $\MD_\Sigma$ denotes the set of message deduction rules representing our DY~attacker for $\Sigma$:
    \[\begin{array}{r@{\;}c@{\;}l}
    {[\Outfact(x)]}              &\rew{[]}          & {[\K(x)]}\\
    {[\K(x)]}                 &\rew{[\knowsf(x)]}& {[\Infact(x)]}\\
    {[]}                      &\rew{[]}          & {[\K(x\in\pubtype)]}\\
    {[\Frfact(x\in\freshtype)]}  &\rew{[]}          & {[\K(x)]}\\
    {[\K(x_1),\dots,\K(x_k)]} &\rew{[]}          & {[\K(f(x_1,\dots,x_k))]}\\ &&\text{for } f\in\Sigma \text{ with arity } k
    \end{array}
    \]
\end{definition}

Similarly, we define \MDInd{} in \defref{MDInd}, which consists of the transition rules a protocol-independent component can execute.
These transition rules represent sending known terms to the network and receiving terms from the network, using public constants, generating nonces, and applying functions to learn new terms.
We assume that these transition rules cover all operations that a protocol-independent component might perform.
In contrast to $\MD$, \MDInd{} operates on syntactically different, reserved fact symbols.
Avoiding these name clashes simplifies defining a simulation relation for proving \lemref{attacker-refinement}.

While $\Indfact$ represents knowledge of a particular term, $\Outindfact$ and $\Inindfact$ represent a term sent to and received from the network, respectively.
$\Indfact$, $\Outindfact$, and $\Inindfact$ are in the same class of fact symbols as their analogous counterparts $\K$, $\Outfact$, and $\Infact$, respectively.
\Ie $\K$ and $\Indfact$ are persistent fact symbols~$\sigper$ capturing the property that knowledge is monotonically increasing.
This means that applying a transition rule does not consume such facts and, thus, their multiplicity in the multiset comprising the state is irrelevant.
In contrast, $\Outfact$, $\Infact$, $\Outindfact$, and $\Inindfact$ are in the class of linear fact symbols~$\siglin$, meaning that applying a transition rule that states such a fact in its premise will remove this fact from the state while such a fact occurring in the rule's conclusion adds it to the state.
Additionally, $\Outindfact$ and $\Inindfact$ are in the class of output and input fact symbols~$\sigout$ and $\sigin$, respectively, as suggested by their intuitive semantics.

\begin{definition}[Protocol-independent message deduction rules]
    \label{def:MDInd}

    \[\begin{array}{r@{\;}c@{\;}l}
    {[\Indfact(x)]}              &\rew{[]}          & {[\Outindfact(x)]}\\
    {[\Inindfact(x)]}            &\rew{[]}          & {[\Indfact(x)]}\\
    {[]}                      &\rew{[]}          & {[\Indfact(x\in\pubtype)]}\\
    {[\Frfact(x\in\freshtype)]}  &\rew{[]}          & {[\Indfact(x)]}\\
    {[\Indfact(x_1),\dots,\Indfact(x_k)]} &\rew{[]}    & {[\Indfact(f(x_1,\dots,x_k))]}\\ &&\text{for } f\in\Sigma \text{ with arity } k
    \end{array}
    \]
    where $\Indfact$, $\Outindfact$, and $\Inindfact$ are reserved fact symbols and $\Indfact\in\sigper$, $\Outindfact\in\sigout\cap\siglin$, and $\Inindfact\in\sigin\cap\siglin$.
\end{definition}

Although we present \MDInd{} on the same abstraction level as the attacker deduction rules~$\MD_\Sigma$ to make them more legible, these deduction rules are \emph{not} part of the \ac{MSR} system~\RR.
Instead, we transform these rules according to \cite{DBLP:journals/corr/abs-2212-04171} and make them part of the component system as described next.

We introduce buffered versions for the $\Outindfact$ and $\Inindfact$ facts and split the rules in \MDInd involving I/O into two separate rules each, which we synchronize using transition labels.
This split allows us to assign half of the rules to the component executing protocol-independent operations~$\Rind(\rid)$ and assign the remaining rules~$\Rbuf^{+}$ to the environment forming $\Renv^{e+}$.
We use $\csyncFn$ to synchronize the execution of these now separated rules.

\begin{definition}[$\Rind(\rid)$]
    $\Rind(\rid)$ consists of the following multiset transition rules.
    \[\begin{array}{r@{\;}c@{\;}l}
    % begin send
    {[\Indfact(\rid,x)]} &
    \reweq[rind]{[\lambda^s_{\Outindfact}(\rid,x)]} &
    {[]}
    % end send
    \\
    % begin receive 
    {[]} &
    \reweq[rind]{[\lambda^s_{\Inindfact}(\rid,x)]} &
    {[\Indfact(\rid,x)]}
    % end receive
    \\
    % begin public
    {[]} &
    \reweq[rind]{[]} &
    {[\Indfact(\rid,x\in\pubtype)]}
    % end public
    \\
    % begin fresh
    {[]} &
    \reweq[rind]{[\lambda^s_{\Frindfact}(\rid,x)]} &
    {[\Indfact(\rid,x)]}
    % end fresh
    \\
    % begin function application
    {
        \begin{bmatrix}
            \Indfact(\rid,x_1),\\
            \dots,\\
            \Indfact(\rid,x_k)
        \end{bmatrix}
    } &
    \reweq[rind]{[]} &
    {
        \begin{aligned}
            &\\
            &[\Indfact(\rid,f(x_1,\dots,x_k))]\\
            &\text{for } f\in\Sigma \text{ with arity } k
        \end{aligned}
    }
    % end function application
    \end{array}
    \]
\end{definition}

\begin{definition}[$\Renv^{e+}$]
    $\Renv^{e+} = \Renv^{e} \udis \Rbuf^{+}$ where $\Renv^{e}$ is defined as in \cite[Sec.~3.2.2~(6)]{DBLP:journals/corr/abs-2212-04171} and $\Rbuf^{+}$ consists of the following multiset transition rules.
    \[\begin{array}{r@{\;}c@{\;}l}
    {[]}                        &\reweq[renv]{[\lambda^e_{\Outindfact}(\rid,x)]} & {[\Outindfact(x)]}\\
    {[\Inindfact(x)]}              &\reweq[renv]{[\lambda^e_{\Inindfact}(\rid,x)]}  & {[]}\\
    {[\Frfact(x\in\freshtype)]}    &\reweq[renv]{[\lambda^e_{\Frindfact}(\rid,x)]}     & {[]}\\
    \end{array}
    \]
\end{definition}

\begin{definition}[$\csyncFn$]
    We define the partial synchronization function~$\csyncFn : (\bigcup_{i,\rid} (\Rrole{i}(\rid) \cup \Rind(\rid))) \times \Renv^{e+} \rightarrow \E$ that synchronizes events of the two systems~$\interl_{i,\rid} \left( \Rrole{i}(\rid) \interl \Rind(\rid) \right)$ and $\Renv^{e+}$, \ie
    \begin{equation*}
        \csyncFn(e, e') =
        \begin{cases}
            \emptyM & \text{if $e = F^s(\rid,x)$ and}\\ & \text{\phantom{if }$e' = F^e(\rid,x)$}\\
            \chi(e, e') & \text{if $e \neq F^s(\rid,x)$ and}\\
            & \text{\phantom{if }$e' \neq F^e(\rid,x)$}\\
            \text{undefined} & \text{otherwise}
        \end{cases}
    \end{equation*}
    where $F \in \{ \lambda_{\Outindfact}, \lambda_{\Inindfact}, \lambda_{\Frindfact} \}$ and the partial function~$\chi$~\cite[App.~E.5]{DBLP:journals/corr/abs-2212-04171} synchronizes labels occurring in $\Rrole{i}$ and $\Renv^{e}$ and $\emptyM$ denotes the empty transition label.
\end{definition}

\begin{lemma}[Protocol-independent components refine the attacker]
\label{lem:attacker-refinement}
\begin{equation*}
\begin{split}
    & \left( \interl_{i,\rid} \left( \Rrole{i}(\rid) \interl \Rind(\rid) \right) \right) \sync{\csyncFn} \Renv^{e+} \\
    \tracePre
    & \left( \interl_{i,\rid} \Rrole{i}(\rid) \right) \sync{\asyncFn} \Renv^{e}
\end{split}
\end{equation*}
\end{lemma}

Given $\Rind(\rid)$ and $\Renv^{e+}$, \lemref{attacker-refinement} states that we can treat the system (on the first line) consisting of possibly unboundedly many instances of components executing a protocol role and executing protocol-independent operations as a refinement of the system on the second line that does not have components executing protocol-independent operations and uses an environment without the rules in $\Rbuf^{+}$.

The following proof proceeds by setting up a simulation relation that merges the states of components executing protocol-independent operations with the environment and renames certain fact symbols.
Using this simulation relation, we show that each transition in the concrete system can be simulated by the abstract system.
While this simulation is straightforward for transitions executed by components that are present in both, the concrete and abstract system, concrete transitions corresponding to protocol-independent operations are more insightful as we show that the abstract environment, namely our \ac{DY} attacker model, can simulate those transitions.

\begin{proof}

% macros for abstract system
\newcommand{\asys}{\EE}                         % abstract system
\newcommand{\astep}[1]{\rew{#1}_{\asys}}        % transition in the abstract system
\newcommand{\acstate}[1]{s_{#1}}                % component
\newcommand{\acjstate}[2]{s_{#1,#2}}            % indexed component
\newcommand{\acstep}[2]{\rew{#1}_{\Rrole{#2}(\rid)}} % step of component
\newcommand{\aestate}{s_e}                      % environment
\newcommand{\aejstate}[1]{s_{#1,e}}             % indexed environment
\newcommand{\aestep}[1]{\rew{#1}_{\Renv^{e}}}   % step of environment

% macros for concrete system
\newcommand{\csys}{\EE'}                        % concrete system
\newcommand{\cstep}[1]{\rew{#1}_{\csys}}        % transition in the concrete system
\newcommand{\ccstate}[1]{s'_{#1}}               % role component
\newcommand{\ccjstate}[2]{s'_{#1,#2}}           % indexed role component
\newcommand{\ccstep}[2]{\rew{#1}_{\Rrole{#2}(\rid)}} % step of role component
\newcommand{\cindstate}[1]{s'_{\ind,#1}}        % protocol-independent component
\newcommand{\cindjstate}[2]{s'_{#1,\ind,#2}}    % indexed protocol-independent component
\newcommand{\cindstep}[1]{\rew{#1}_{\Rind(\rid)}} % step of protocol-independent component
\newcommand{\cestate}{s'_e}                     % environment
\newcommand{\cejstate}[1]{s'_{#1,e}}            % indexed environment
\newcommand{\cestep}[1]{\rew{#1}_{\Renv^{e+}}}   % step of environment

\newcommand{\rnFn}{\ensuremath{r}} % function that renames facts used in simulation relation

\looseness=-1
We denote $\asys$ and $\csys$ the abstract and concrete systems, respectively, and prove this lemma by establishing refinement with a stuttering simulation relation~$\R$ between states of the abstract system~$\asys$ and states of the concrete system~$\csys$.
\Ie $\asys = \left( \interl_{i,\rid} \Rrole{i}(\rid) \right) \sync{\asyncFn} \Renv^{e}$ and $\csys = \left( \interl_{i,\rid} \left( \Rrole{i}(\rid) \interl \Rind(\rid) \right) \right) \sync{\csyncFn} \Renv^{e+}$.
Using a stuttering simulation relation in contrast to a standard simulation relation allows us to relate the abstract and concrete states even if the concrete system performs additional transitions that do not have a corresponding transition in the abstract system, \ie the abstract system can stutter as long as the observable behaviors of the two~systems remain the same.
Accordingly, we use $\astep{.}$ and $\cstep{.}$ to denote a transition step in the abstract system~$\asys$ and concrete system~$\csys$, respectively.
Additionally, we use $\acstep{.}{i}$ and $\aestep{.}$ for transitions performed by the individual components in the abstract system and, similarly, $\ccstep{.}{i}$, $\cindstep{.}$ and $\cestep{.}$ for the concrete system's components.

The abstract states are of the shape $((\acstate{i,\rid})_{1\leq i \leq n, \text{ for each }\rid},\aestate)$.
We use primed variables for referring to concrete states, which are of the shape $((\ccstate{i,\rid}, \cindstate{i,\rid})_{1\leq i \leq n, \text{ for each }\rid},\cestate)$, \ie they are composed of two multisets of facts for each $i$, $\rid$, and one for the environment.
Each multiset $\ccstate{i,\rid}$ corresponds to the state of instance~$\rid$ executing the protocol role~$i$, while $\cindstate{i,\rid}$ corresponds to the state of the component executing protocol-independent operations, which is conceptually co-located with $\ccstate{i,\rid}$ but guaranteed by our taint analysis to operate on distinct state.

We use a stuttering simulation relation~$\R$, such that $(s, s') \in \R$ iff
\begin{equation*}
s = ((\ccstate{i,\rid})_{1\leq i \leq n, \text{ for each }\rid}, \rnFn((\cupM_{i,\rid} \cindstate{i,\rid}) \cupM \cestate)),
\end{equation*}
where $s' = ((\ccstate{i,\rid}, \cindstate{i,\rid})_{1\leq i \leq n, \text{ for each }\rid},\cestate)$ and \rnFn{} is the identity function except for the cases specified below.
We lift \rnFn{} to operate on multiset of facts.
This lifted version removes duplicate $\K$~facts because $\K$ is a persistent fact symbol.
\begin{equation*}
\left.
\begin{aligned}
    \rnFn(\Indfact(\rid,x)) &= \K(x)\\
    \rnFn(\Outindfact(x)) &= \K(x)\\
    \rnFn(\Inindfact(x)) &= \K(x)
\end{aligned}
\right\}\text{for all $\rid$, $x$.}
\end{equation*}
\Ie to derive $\aestate$ from $s'$, we, first, combine all facts in the states of protocol-independent components~$\cindstate{i,\rid}$ with $\cestate$ by applying multiset union~$\cupM$ and, second, rename and deduplicate these facts according to the renaming function~\rnFn{}.

It is clear that the initial states are related, \ie $(s, s') \in \R$ with $s = ((\emptyset,\dots,\emptyset),\emptyset)$ and $s' = ((\emptyset,\emptyset),\dots,(\emptyset,\emptyset),\emptyset)$.
We now show that for all states $(s_1, s'_1) \in \R$ and for all concrete transition steps~$s'_1 \cstep{e} s'_2$ there exists an abstract transition $s_1 \astep{e} s_2$ such that $(s_2, s'_2) \in \R$.
We use the following naming convention to refer to individual multisets within the abstract and concrete states, respectively, for $j \in \{1, 2\}$:
\begin{equation*}
\begin{aligned}
    s_j &= ((\acjstate{j}{i,\rid})_{1\leq i \leq n, \text{ for each }\rid},\aejstate{j})\\
    s'_j &= ((\ccjstate{j}{i,\rid}, \cindjstate{j}{i,\rid})_{1\leq i \leq n, \text{ for each }\rid},\cejstate{j})\\
\end{aligned}
\end{equation*}

Based on the definition of the parallel and synchronizing composition, $\interl$ and $\sync{\csyncFn}$, resp., we distinguish the following two~cases for the transition step~$s'_1 \cstep{e} s'_2$:
\begin{itemize}
    % new synchronization label
    \item $e = \csyncFn(F^s(\rid,x), F^e(\rid,x))$ for $F \in \{ \lambda_{\Outindfact}, \lambda_{\Inindfact}, \lambda_{\Frindfact} \}$:\\
    Since $s'_1 \cstep{e} s'_2$, we have:
    \begin{equation*}
        \cindjstate{1}{i,\rid} \cindstep{F^s(\rid,x)} \cindjstate{2}{i,\rid}
    \end{equation*}
    \begin{equation*}
        \cejstate{1} \cestep{F^e(\rid,x)} \cejstate{2}
    \end{equation*}
    and all other component states remain unchanged, \ie
    \begin{equation*}
    \begin{aligned}
        \ccjstate{2}{i,\rid} &= \ccjstate{1}{i,\rid}\\
        \ccjstate{2}{j,\rid'} &= \ccjstate{1}{j,\rid'}\\
        \cindjstate{2}{j,\rid'} &= \cindjstate{1}{j,\rid'}
    \end{aligned}
    \end{equation*}
    for all $(j,\rid') \neq (i,\rid)$.
    We now need to distinguish the cases where $F = \lambda_{\Outindfact}$, $F = \lambda_{\Inindfact}$, and $F = \lambda_{\Frindfact}$.
    \begin{itemize}
        \item $F = \lambda_{\Outindfact}$:
        By definition of the transition rule~$F^s$, we have $\Indfact(\rid, x) \in \cindjstate{1}{i,\rid}$ and $\cindjstate{2}{i,\rid} = \cindjstate{1}{i,\rid} \setminus^m \multileft \Indfact(\rid, x) \multiright$.
        Similarly, by definition of $F^e$, we have $\cejstate{2} = \cejstate{1} \cup^m \multileft \Outindfact(x) \multiright$.
        By definition of $\csyncFn$, the transition label~$e$ is the empty label~$\emptyM$.
        We simulate this transition in $\asys$ by stuttering, \ie $s_2 = s_1$.
        Since \rnFn{} renames both facts $\Indfact(\rid, x)$ and $\Outindfact(x)$ to $\K(x)$ and $(s_1, s'_1) \in \R$, we have $\K(x) \in \aejstate{1}$.
        Additionally, the multiset minus and multiset union operations cancel out after applying \rnFn{} such that $\aejstate{2} = \aejstate{1}$.
        Therefore, $(s_2, s'_2) \in \R$.

        \item $F = \lambda_{\Inindfact}$:
        This case is analogous to $F = \lambda_{\Outindfact}$.
        
        \item$F = \lambda_{\Frindfact}$:
        By definition of $F^s$ and $F^e$, we have
        \begin{equation*}
            \Frfact(x \in \freshtype) \in \cejstate{1}\text{,}
        \end{equation*}
        \begin{equation*}
        \begin{aligned}
            \cejstate{2} &= \cejstate{1} \setminus^m \multileft \Frfact(x) \multiright\text{, and}\\
            \cindjstate{2}{i,\rid} &= \cindjstate{1}{i,\rid} \cup^m \multileft \Indfact(\rid, x) \multiright.
        \end{aligned}
        \end{equation*}
        Since $(s_1, s'_1) \in \R$, we obtain $\Frfact(x) \in \aejstate{1}$ enabling us to apply the attacker's message deduction rule (from $\MD_\Sigma$) ${[\Frfact(x\in\freshtype)]} \rew{[]} {[\K(x)]}$, which results in $\aejstate{2} = \aejstate{1} \setminus^m \multileft \Frfact(x) \multiright \cup^m \multileft \K(x) \multiright$.
        Due to the renaming function~\rnFn{} applied to $\cindjstate{2}{i,\rid}$, we obtain $(s_2, s'_2) \in \R$.
    \end{itemize}

    % old synchronization label
    \item $e = \chi(e, e')$:\\
    We consider the following four~subcases based on the definition of $\chi$:
    \begin{itemize}
        \item $e = \chi(\lambda^s_{F,i,\rid}(\bar{m}), \lambda^e_{F,i,\rid}(\bar{m}))$ for some $F$, $i$, $\rid$, $\bar{m}$:\\
        By definition, neither $\Rind(\rid)$ nor $\Rbuf^{+}$ contain any transition rule with a matching transition label.
        Hence, this transition step synchronizes a step in $\Rrole{i}$ and $\Renv^{e}$.
        By definition of our composition operators and since $s'_1 \cstep{e} s'_2$, we have
        \begin{equation*}
        \begin{aligned}
            \ccjstate{1}{i,\rid} & \ccstep{\lambda^s_{F,i,\rid}(\bar{m})}{i,\rid} && \ccjstate{2}{i,\rid}\\
            \cejstate{1} & \aestep{\lambda^e_{F,i,\rid}(\bar{m})} && \cejstate{2}\\
        \end{aligned}
        \end{equation*}
        and
        \begin{equation*}
        \begin{aligned}
            \ccjstate{2}{j,\rid'} &= \ccjstate{1}{j,\rid'}\\
            \cindjstate{2}{i,\rid} &= \cindjstate{1}{i,\rid}\\
            \cindjstate{2}{j,\rid'} &= \cindjstate{1}{j,\rid'}
        \end{aligned}
        \end{equation*}
        for all $(j,\rid') \neq (i,\rid)$.

        Since the renaming function~\rnFn{} behaves like the identity function for facts occurring in the premise and conclusion of rules $\lambda^s_{F,i,\rid}(\bar{m})$ and $\lambda^e_{F,i,\rid}(\bar{m})$, the same rules can be applied in the abstract states~$\acjstate{1}{i,\rid}$ and $\aejstate{1}$.
        \Ie we have
        \begin{equation*}
        \begin{aligned}
            \acjstate{1}{i,\rid} & \acstep{\lambda^s_{F,i,\rid}(\bar{m})}{i,\rid} && \acjstate{2}{i,\rid}\\
            \aejstate{1} & \aestep{\lambda^e_{F,i,\rid}(\bar{m})} && \aejstate{2}\\
        \end{aligned}
        \end{equation*}
        and $(s_2, s'_2) \in \R$.

        \item $e = \chi(e', skip)$ for some $e' \in \Rrole{i}(\rid)$:\\
        Then, $e' = e$ and by definition of our composition operators, we obtain $\ccjstate{1}{i,\rid} \ccstep{e}{i} \ccjstate{2}{i,\rid}$ and
        \begin{equation*}
        \begin{aligned}
            \ccjstate{2}{j,\rid'} &= \ccjstate{1}{j,\rid'}\\
            \cindjstate{2}{i,\rid} &= \cindjstate{1}{i,\rid}\\
            \cindjstate{2}{j,\rid'} &= \cindjstate{1}{j,\rid'}\\
            \cejstate{2} &= \cejstate{1}
        \end{aligned}
        \end{equation*}
        for all $(j,\rid') \neq (i,\rid)$.
        Since $(s_1, s'_1) \in \R$, we further have $\acjstate{1}{i,\rid} = \ccjstate{1}{i,\rid}$, $\acjstate{1}{i,\rid} \acstep{e}{i} \acjstate{2}{i,\rid}$, and, thus, $\acjstate{2}{i,\rid} = \ccjstate{2}{i,\rid}$.
        Therefore, $(s_2, s'_2) \in \R$.

        \item $e = \chi(e', skip)$ for some $e' \in \Rind(\rid)$:\\
        $e' \neq F^s(\rid,x)$ for $F \in \{ \lambda_{\Outindfact}, \lambda_{\Inindfact}, \lambda_{\Frindfact} \}$ by definition of $\csyncFn$.
        Therefore, $e'$ must be the transition adding a public constant or applying a k-ary function~$f$ to the state of $\Rind(\rid)$.
        We can simulate either transition in the abstract system~$\asys$ by performing the corresponding message deduction rule in $\MD_\Sigma$, which updates the abstract state in the same way after merging the states of the environment and of the components performing protocol-independent operations and applying the renaming function~\rnFn{}.
        Thus, $(s_2, s'_2) \in \R$.

        \item $e = \chi(skip, e')$ for some $e' \in \Renv^{e+}$:\\
        Then, $e' = e$ and, by definition of the composition operators, we obtain $\cejstate{1} \cestep{e} \cejstate{2}$, $\ccjstate{2}{i,\rid} = \ccjstate{1}{i,\rid}$, and $\cindjstate{2}{i,\rid} = \cindjstate{1}{i,\rid}$ for all $i, \rid$.
        By definition of $\csyncFn$, $e$ cannot have the shape~$F^e(\rid,x)$ for some $\rid$, $x$, and $F \in \{ \lambda_{\Outindfact}, \lambda_{\Inindfact}, \lambda_{\Frindfact} \}$, which rules out the transitions in $\Rbuf^{+}$.
        Thus, $e \in \Renv^{e}$.
        Since $(s_1, s'_1) \in \R$, we have $\cejstate{1} \subseteq^m \aejstate{1}$.
        Since $e$'s guard is stable under supermultiset, the rewrite rule~$e$ can be applied in state~$\aejstate{1}$, \ie $\aejstate{1} \aestep{e} \aejstate{2}$.
        As this abstract transition only modifies the submultiset~$\cejstate{1}$ by adding or removing facts for which the renaming function~\rnFn{} behaves as the identity function and leaves all other $\acjstate{1}{i,\rid}$ and $\aejstate{1} \setminus^m \cejstate{1}$ unchanged, we obtain $\aejstate{2} = \rnFn((\cupM_{i,\rid} \cindjstate{1}{i,\rid}) \cupM \cejstate{2})$.
        Thus, $s_1 \astep{e} s_2$ and $(s_2, s'_2) \in \R$.
    \end{itemize}    
\end{itemize}
\end{proof}

\begin{theorem}[Soundness]
    \label{thm:diodon-ioindependence-soundness}
    Suppose \asmref{diodon-verifier-assumption} holds and that we have verified, for each role~$i$, the Hoare triple $\vdash_{\pi'_\text{ext}} \hoaretriple{\psi_i(\rid)}{c_i(\rid)}{\true}$. Then
    \begin{equation*}
        (\interl_{i,\rid} \pi_\text{int}(\concretelts{i}{\rid})) \sync{\csyncrelabeledFn} \realEnv \tracePre_t \RR.
    \end{equation*}
\end{theorem}
\Thmref{diodon-ioindependence-soundness} states that composing unboundedly many instances of each role's LTS~\concretelts{i}{\rid} with the concrete environment~\realEnv{} refines the protocol model~\RR{}.
While this theorem is identical to \cite[Thm.~2]{DBLP:journals/corr/abs-2212-04171}, the proof differs since our \asmref{diodon-verifier-assumption} considers a larger set of traces per LTS~\concretelts{i}{\rid}.

\begin{proof}
We decompose the proof into a similar series of trace inclusions as \cite{DBLP:journals/corr/abs-2212-04171} but add an additional trace inclusion to abstract the protocol-independent I/O operations to the environment, which contains the \ac{DY} attacker (\cf \lemref{attacker-refinement}).

The first trace inclusion is
\begin{equation}
\begin{split}
    & \left( \interl_{i,\rid} \pi_\text{int}(\concretelts{i}{\rid}) \right) \sync{\csyncrelabeledFn} \realEnv \\
    \tracePre
    & \left( \interl_{i,\rid} \pi(\pi'_\text{ext}(\concretelts{i}{\rid})) \right) \sync{\csyncFn} \pi_\text{ext}(\pi'_\text{ext}(\realEnv)),
\end{split}
\end{equation}
where we obtain the first line from the second by pushing the relabeling $\pi_\text{ext} \circ \pi'_\text{ext}$ into the parallel composition, thus changing the set of synchronization labels from $\csyncFn$ to $\csyncrelabeledFn$.

\looseness=-1
By combining \asmref{diodon-verifier-assumption}, the assumption $\vdash_{\pi'_\text{ext}} \hoaretriple{\psi_i(\rid)}{c_i(\rid)}{\true}$, and \cite[Thm.~1]{DBLP:journals/corr/abs-2212-04171}, we obtain
\begin{equation}
\label{eq:implementation-inclusion}
    \pi(\pi'_\text{ext}(\concretelts{i}{\rid})) \tracePre \Rrole{i}(\rid) \interl \Rind(\rid),
\end{equation}
where $\Rind(\rid)$ is a \acf{MSR} system capturing the execution of protocol-independent I/O operations.
All facts in this \ac{MSR} system are instantiated with the thread id~$\rid$, which helps in distinguishing the facts belonging to each $\Rind$ instance.
Additionally, \eqref{eq:implementation-inclusion} implicitly specifies that the \ac{MSR} systems~$\Rrole{i}(\rid)$ and $\Rind(\rid)$ operate independently, \ie on different multisets of facts.
By performing the taint analysis, we ensure that $\Rrole{i}(\rid)$ does not influence $\Rind(\rid)$.
Checking the opposite, \ie that $\Rind(\rid)$ does not influence $\Rrole{i}(\rid)$ by performing a second taint analysis is not necessary.
We explicitly track throughout code-level verification the multiset of facts representing the state of $\Rrole{i}(\rid)$, which is only manipulated by internal and I/O library functions corresponding to state updates permitted by $\Rrole{i}(\rid)$.
Therefore, this state cannot be influenced by $\Rind(\rid)$.

We can leverage a general composition theorem~\cite[Thm.~2.3]{DBLP:journals/pacmpl/0001KEW0CB20} that implies that trace inclusion is compositional for a large class of composition operators including $\interl$ and $\sync{\Lambda}$.
Applying this theorem to \eqref{eq:implementation-inclusion} and \cite[Prop.~2]{DBLP:journals/corr/abs-2212-04171} establishes the trace inclusion
\begin{equation}
\begin{split}
    & \left( \interl_{i,\rid} \pi(\pi'_\text{ext}(\concretelts{i}{\rid})) \right) \sync{\csyncFn} \pi_\text{ext}(\pi'_\text{ext}(\realEnv)) \\
    \tracePre
    & \left( \interl_{i,\rid} \left( \Rrole{i}(\rid) \interl \Rind(\rid) \right) \right) \sync{\csyncFn} \Renv^{e+}.
\end{split}
\end{equation}

Applying \lemref{attacker-refinement} in connection with \cite[Lemma~1 \& Lemma~2]{DBLP:journals/corr/abs-2212-04171} completes the proof.
\end{proof}

\subsection{Combining Auto-Active Verification and Static Analyses}
\label{app:diodon-composition-soundness}
\setboolean{mainbody}{false} % set boolean such that figures, definitions, etc. get the correct label
In this subsection, we sketch soundness of our combination of auto-active program verification and fully automatic static analyses by constructing a proof in concurrent \gls{separation-logic}~\cite{DBLP:conf/lics/Reynolds02, DBLP:journals/entcs/Vafeiadis11} for the entire codebase.
More specifically, we give an invariant that is maintained by each statement in our programming language~(\appref{diodon-program-invariant}) and present proof rules that use, besides certain side conditions, only this invariant in their premises and conclusions~(\appref{diodon-proof-rules}).
We use \approach's static analyses to discharge these side conditions~(\appref{diodon-soundness-static-analyses}).
Therefore, we can compose the proof rules to prove $\shoare{\phi}{\prog}{\true}$ for an I/O specification~$\phi$ and an entire codebase~$\prog{}$~(\appref{diodon-soundness-composition-proof-construction}).

To focus on the main proof insights, we deliberately keep the considered programming language simple (\cf \defref{diodon-language}) and consider the case where an execution of the codebase~$\prog{}$ corresponds to at most one~execution of a protocol role, which is represented by the I/O specification~$\phi$.
We discuss limitations in \appref{diodon-soundness-composition-limitations} and extensions lifting these restrictions in \appref{diodon-soundness-composition-extensions}.

\mypar{Prerequisites}
We consider an imperative, concurrent, and heap-man\-ipulating programming language as shown in \defref{diodon-language}.
For simplicity, we omit function boundaries and statements creating complex control flow.
Furthermore, we assume that programs are in \ac{SSA} form such that we do not have to consider variable reassignments for the purpose of our proof.
Besides statements to allocate, read, and write a heap location, we make each auto-actively verified \ac{API} function of the \core a dedicated statement in the language even though these statements are themselves implemented as sequences of statements, which are considered by our static analyses.
\ccorealloc{c}{\bar{e}} corresponds to calling the \core's constructor and creating a new \core instance~$c$.
We use \ccorecall{k}{c}{\bar{e}}{\bar{r}} to represent invoking the $k$-th \ac{API} function\footnote{%
We assume the existence of some total order for \ac{API} functions, \eg based on their declarations' syntactical ordering.
} on a \core instance~$c$ using input arguments~$\bar{e}$ and return arguments~$\bar{r}$.
$s_1 \cseq s_2$ denotes standard sequential composition of two~statements and $\cfork{\bar{x}}{s}$ spawns a new thread executing statement~$s$ while passing variables~$\bar{x}$ to this thread.
We syntactically require that the newly spawned thread accesses only its own local variables and variables~$\bar{x}$.

\begin{definition}[Basic Programming Language]
    \label{def:diodon-language}
    We consider the following programming language, where $S$ ranges over labeled statements, \var{x} over variables, $\ell$ over statement labels, and $e$ over expressions.
    We have the usual side effect-free expressions.
    We use $\bar{y}$ as a shorthand notation to denote lists of kind~$y$.
    \begin{align*}
    S \triangleq&\; U^\ell
    \\
    U \triangleq
    &\;
    \cskip \mid
    \cheapalloc{x} \mid
    \cheapread{x}{e} \mid
    \cheapwrite{x}{e} \mid\\
    &\;
    \ccorealloc{c}{\bar{e}} \mid
    \ccorecall{k}{c}{\bar{e}}{\bar{r}} \mid\\
    &\;
    S \cseq S \mid
    \cfork{\bar{x}}{S}
    \end{align*}
    We call $S \cseq S$ and $\cfork{\bar{x}}{S}$ \emph{compound}~statements, while all other statements in our language are called \emph{simple}.
    When not relevant, we omit a statement's label~$\ell$, which uniquely identifies the statement in the program text. We use these labels to refer to the program points before and after each labeled statement.
    We abstractly treat \ccorealloc{c}{\bar{e}} and \ccorecall{k}{c}{\bar{e}}{\bar{r}} as first-class statements in our language despite being implemented as sequences of statement that are considered by our static analyses and the auto-active program verifier.
    This is possible because we can treat these statements as opaque boxes from a proof construction point of view as we prove a Hoare triple for each such statement using the auto-active program verifier.
\end{definition}

We assume that all memory accesses in the unverified \app neither cause crashes nor data races such that we can reason about their effects.
While we could have avoided assuming data race freedom by defining that all heap operations in our language are atomic, we try to stay faithful to most programming languages, which specify data races to cause undefined behavior, thus, making this assumption necessary.

\begin{assumption}[Crash freedom]
    \label{asm:crash-freedom}
    We assume that all heap accesses within the \app, \cheapread{x}{e} and \cheapwrite{x}{e} do not crash\footnote{%
        We require \asmref{crash-freedom} as we construct a Hoare triple for the entire codebase, whose definition includes that a program does not crash.
        We could avoid this assumption by altering the definition of a Hoare triple to guarantee the postcondition \emph{only} if the program does not crash.
        This alternative definition would be suitable for programming languages like Go in which dereferencing \nil is defined behavior and results in a crash, which does not invalidate our security guarantees.
    }, \ie the \app dereferences only pointers to allocated heap locations as opposed to \nil.
\end{assumption}

\begin{assumption}[Data race freedom]
    \label{asm:data-race-freedom}
    We assume that all heap accesses within the \app, \cheapread{x}{e} and \cheapwrite{x}{e}, are data race free.
    \Ie all accesses to the heap locations to which $e$ and $x$, respectively, point are linearizable and, thus, do not cause data races.
\end{assumption}

Note that \asmref{crash-freedom} and \asmref{data-race-freedom} apply to heap accesses within the \app only, as we auto-actively prove \gls{safety} for the \core.

By auto-actively verifying the \core, we prove a Hoare triple for each \ac{API} function.
This allows us to abstractly treat each \ac{API} function as a statement in our language as long as there are no callbacks; we discuss callbacks as an extension at the end of this subsection.
We syntactically restrict the specification of \core \ac{API} functions, \ie the assertions occurring in the auto-actively verified Hoare triples, such that we can discharge the side conditions using static analyses and, thus, construct a proof for the entire codebase.
We state these restrictions immediately after introducing some notational conventions.

\begin{definition}[Notation]
    \looseness=-1
    We introduce the following notation to simplify forthcoming definitions, explanations, and proofs.
    \accnilable{x} denotes full permission for the heap location to which $x$ points but only if $x$ is non-\nil{}.
    Analogously, we define \invnilable{x} for the \core{}~invariant.
    Lastly, we lift permissions for a heap location to lists thereof, internally using the iterated separating conjunction~$\forallstar_i$ ranging over $i$.
    \begin{align*}
        \accnilable{x} \triangleq \; & x \neq \nil \implies \acc{x}\\
        \invnilable{x} \triangleq \; & x \neq \nil \implies \inv{x}\\
        \acc{\bar{x}} \triangleq \; & \forallstar_{0 \leq i < \len{\bar{x}}} \acc{\at{\bar{x}}{i}}\\
        \accnilable{\bar{x}} \triangleq \; & \forallstar_{0 \leq i < \len{\bar{x}}} \accnilable{\at{\bar{x}}{i}}
    \end{align*}
    where $\len{\bar{x}}$ returns the length of list~$\bar{x}$ and $\at{\bar{x}}{i}$ the $i$-th element therein.
\end{definition}

\begin{assumption}[Syntactic restrictions for \core{} specification]
    \label{asm:syntactic-restrictions}
    \looseness=-1
    We make the following syntactical assumptions about the pre- and postconditions of \core{} \ac{API} functions, which ultimately enable us to apply static analyses.
    \begin{alignat*}{2}
        &\precorealloc{\bar{e}} &&\triangleq \; \phi \star R\\
        &\postcorealloc{c}{\bar{e}} &&\triangleq \; \inv{c} \star R'\\
        &\precorecall{k}{c}{\bar{e}} &&\triangleq \; \invnilable{c} \star S_k\\
        &\postcorecall{k}{c}{\bar{e}}{\bar{r}} &&\triangleq \; \invnilable{c} \star S'_k
    \end{alignat*}
    where $R$, $R'$, $S_k$, and $S'_k$ are \gls{separation-logic} assertions that specify permissions for the arguments~$\bar{e}$ and, if applicable, $\bar{r}$.
    Preconditions are free of \glspl{functional-property} and specify at most permissions for non-nil arguments, \ie $\accnilable{\bar{e}} \models R$ and $\accnilable{\bar{e}} \models S_k$.
    Each postcondition needs to specify the same or more permissions than the respective precondition, \ie $R' \models R$ and $S'_k \models S_k$.
    Additionally, postconditions need to specify full permission to every heap location that becomes accessible to the \app{} and that is created within the corresponding \core{} function or any function transitively called thereby.
    For simplicity, we disallow \ccoreallocshort{\bar{e}} to return such heap locations other than the \core{} instance itself and, thus, permissions to such heap locations can only occur in $S'_k$ for the return arguments~$\bar{r}$.
    Furthermore, we restrict the input arguments~$\bar{e}$ and output arguments~$\bar{r}$ to be shallow, \ie their transitive closure of reachable heap locations is the singleton set, \ie $\forall e \in \bar{e} \ldotp e \neq \nil \implies \reachable{e} = \setliteral{e}$ and analogously for $\bar{r}$.
    This restriction simplifies the reasoning about which heap locations are passed between the \core and \app.
    However, lifting this restriction is possible and would require that $S_k$ and $S'_k$ specify the permission for every reachable heap location.
\end{assumption}

\subsubsection{Program Invariant}
\label{app:diodon-program-invariant}

\looseness=-1
In order to define composable proof rules for our language, we define a program invariant that is maintained by each statement.
Our invariant conceptually partitions the heap among two~dimensions, namely whether a heap location is accessible by multiple threads and whether a heap location is owned by the \app as opposed to the \core.
As we will formalize later, we call a heap location~$h$ \emph{\app-managed} if $h$ is under the \app's control, which means that it is not covered by the \core~invariant.
Furthermore, we ensure that the \app only accesses memory that is \app-managed.

We make the heap partitioning explicit by introducing \glsdisp{ghost-code}{ghost variables} tracking the heap locations belonging to each partition.
We use a global \glsdisp{ghost-code}{ghost set} pointed to by~\setglobalheap{} tracking the set of heap locations that are accessible by multiple threads.
The thread-local \glsdisp{ghost-code}{ghost variable}~\setlocalheap{} tracks \app-managed heap locations that are only accessible by the current thread.
Lastly, the thread-local variable~\setinvariantheap{} tracks the \core instance if it is already allocated.

Relying on these \glsdisp{ghost-code}{ghost variables}, we can define the \emph{program invariants}~\localproginv{} and \globalproginv{} that specify the \gls{separation-logic} permissions held by a thread at each program point, as shown in \defref{proginv}, where \flagiospec{} is a pointer to a boolean specifying whether the I/O~permissions~$\phi$ have already been consumed to allocate a \core instance.

\begin{definition}[Program invariants]
    \label{def:proginv}
    \begin{align*}
        \localproginv \triangleq \;
            & \bigl(\forallstar_{l \in \setlocalheap} \acc{l}\bigr) \star \bigl(\forallstar_{i \in \setinvariantheap} \inv{i}\bigr)\\
        \globalproginv \triangleq \;
            & \acc{\setglobalheap} \star \bigl(\forallstar_{g \in \ederef{\setglobalheap}} \acc{g}\bigr) \star {}\\
            & \acc{\flagiospec} \star \bigl(\neg(\ederef{\flagiospec}) \implies \phi\bigr)
    \end{align*}
\end{definition}

\localproginv{} specifies permissions that are exclusively owned by each thread.
The first~conjunct specifies (full) permissions to every heap location in \setlocalheap{}, which, as we will see, holds every heap location that is accessible only by the current thread and is unrelated to \core instances.
Additionally, \localproginv{} specifies that the \core invariant~\inv{i} holds for each \core instance~$i$.
Note that \inv{i} is a \gls{separation-logic} predicate that specifies permissions for a subset of the transitively reachable heap locations starting from $i$ and possibly \glspl{functional-property} about these heap locations.
While the definition of \inv{i} matters for the \core's auto-active verification, we treat \inv{i} for the purpose of the program invariant as an opaque \gls{separation-logic} resource.

\globalproginv{} specifies permissions to heap locations that are potentially shared among multiple threads.
When accessing such a heap location, a thread can temporarily acquire the corresponding permission from \globalproginv{}, which is justified as long as all accesses to this location are linearizable.
Since we assume absence of data races (\cf \asmref{data-race-freedom}), there exists a linearization of heap accesses such that permission for manipulating $g$, \ie \acc{g}, can temporarily be obtained from \globalproginv{} for the manipulation's duration.
Furthermore, \globalproginv{} specifies the I/O permissions~$\phi$ if they have not been used yet to create a \core{} instance, in which case the pointer~\flagiospec{} points to a heap location storing the value \false{}.
As mentioned, we focus in this proof on the case of creating at most one~\core{} instance.
However, this conjunct can easily be adapted to provide a family of I/O permissions such that the creation of arbitrarily-many \core{} instances becomes possible, as we will detail in \secref{diodon-soundness-extensions}.

Since the presented program invariants rely on \glsdisp{ghost-code}{ghost variables} to specify permissions, we have to ensure that these \glsdisp{ghost-code}{ghost variables} stay in sync with a program's execution, \ie the effects of each statement.
Hence, we present algorithm~\ghostalgorithm{} in \figref{ghostalgorithm-appendix} that augments a program with \glsdisp{ghost-code}{ghost statements} updating the \glsdisp{ghost-code}{ghost variables} according to each statement's effects.
These \glsdisp{ghost-code}{ghost statements} manipulate only \glsdisp{ghost-code}{ghost variables} and aid verification without changing the input program's runtime behavior.
Thus, these \glsdisp{ghost-code}{ghost variables} and \glsdisp{ghost-code}{ghost statements} can be erased before compilation.

Common to all cases of algorithm~\ghostalgorithm{} is that for a statement~$s$, first, the current heap is partitioned into a heap~$H$ on which $s$ possibly operates and the remaining heap~$F$ that $s$ leaves untouched by removing the \gls{separation-logic} resources for $H$ via corresponding \glsdisp{ghost-code}{ghost set} subtractions.
The \gls{separation-logic} resources belonging to heap~$F$ remain in the \glsdisp{ghost-code}{ghost sets}.
Afterwards, statement~$s$ is executed that changes heap~$H$ to $H'$, followed by merging the heaps~$H'$ and $F$ via \glsdisp{ghost-code}{ghost set} union operations.

Allocating a heap location operates on an empty heap and produces a new heap location, which is added to the set of local heap locations as there is no way any other thread might already have gained access thereto.
Dereferencing a pointer~$e$ and reading the corresponding heap location requires a permission for the duration of this operation.
Therefore, we first subtract and afterwards add this location from the current heap by manipulating either the \glsdisp{ghost-code}{ghost set} of local or global heap locations depending on whether this heap location is contained in \setlocalheap{}.
In case the heap location is in the \glsdisp{ghost-code}{ghost set} of global heap locations, we insert an atomic block, which is justified by \asmref{data-race-freedom} stating that accesses to this heap location are linearizable.

Noteworthy are write operations to heap locations, especially in the case that a heap location is accessible by other threads as the written value becomes accessible by these threads.
Hence, we first remove all local heap locations that are transitively reachable from the written value and add them afterwards to the \glsdisp{ghost-code}{ghost set} of global heap location as these locations possibly escape the current thread via this write operation.
Similarly, when forking a thread, the heap locations that are reachable from the variables~$\bar{x}$ escape the current thread and, thus, the sets of local and global heap locations are updated accordingly.

For \ccoreallocshort{\bar{e}} and \ccorecall{k}{c}{\bar{e}}{\bar{r}}, the algorithm~\ghostalgorithm{} adds and subtracts only $\bar{e}$ and $\bar{r}$ as opposed to all transitively reachable heap locations.
This is sufficient because \asmref{syntactic-restrictions} restricts $\bar{e}$ and $\bar{r}$ to be shallow and, thus, no other heap locations are reachable.
However, extending algorithm~\ghostalgorithm{} to support non-shallow arguments would be straightforward by adding and removing $\reachable{\bar{e}}$ and $\reachable{\bar{r}}$ instead of $\bar{e}$ and $\bar{r}$ to and from \setlocalheap{}, respectively.

\subsubsection{Proof Rules}
\label{app:diodon-proof-rules}

Thanks to the program invariants and the \glsdisp{ghost-code}{ghost statements} that algorithm~\ghostalgorithm{} inserts into a program, we can define proof rules as shown in \figref{proof-rules-appendix}.
In particular, all proof rules share the same pre- and postcondition, namely the local and global program invariants~\localproginv{} and \globalproginv{}, resp., which allow us to compose the proof rules to obtain a whole program proof.
The proof rules' simplicity is enabled by their side conditions (\cf \figref{side-conditions-appendix}) that we discharge using our static analyses.

Besides containment of heap locations in particular \glsdisp{ghost-code}{ghost sets}, the side conditions rely on disjointness of input arguments, which we formally define next.
Informally, two~arguments are disjoint if they point to different heap locations or one of the arguments is \nil{}.

\begin{definition}[Variable value]
    \valueof{x}{\tau} denotes the value of variable~$x$ on trace~$\tau$.
    Since we assume that our programs are in \ac{SSA}-form, this definition is independent of a particular program point.
    However, $x$ must be declared such that \valueof{x}{\tau} is defined.
\end{definition}

\begin{definition}[Disjointness]
    Two~pointer variables~$x$ and $y$ are disjoint if their pointer value is different or \nil{} for all traces~$\tau$.
    \begin{align*}
        \disjoint{\setliteral{x, y}} \triangleq \; \forall \tau \ldotp &\valueof{x}{\tau} = \nil \lor \valueof{x}{\tau} \neq \valueof{y}{\tau}
    \end{align*}
    We straightforwardly lift this definition to lists of variables~$\bar{z}$, where \disjoint{\bar{z}} denotes pairwise disjointness between every element in $\bar{z}$.
\end{definition}

Next, we sketch the proof rules' soundness proof, which relies on the side conditions~$\omega$.
Afterwards, we define what properties our static analyses provide given that their execution succeeded and show that these properties imply the side conditions~$\omega$.
We conclude by proving a corollary stating that we construct a Hoare triple for the entire codebase.

\begin{theorem}[Soundness of proof rules]
    \label{thm:proof-rules-soundness}
    \begin{equation*}
        \text{If }\; \globalproginv \vdash \simpleHoare{\localproginv}{\ghostalgorithm(s)}{\localproginv} \text{, then }\; \globalproginv \models \simpleHoare{\localproginv}{\ghostalgorithm(s)}{\localproginv}
    \end{equation*}
\end{theorem}
\begin{proofsketchcomposition}
    We perform structural induction over the input statement~$s$ to algorithm~\ghostalgorithm{} and construct a proof tree in \gls{separation-logic} building up on the proof rules by Vafeiadis~\cite{DBLP:journals/entcs/Vafeiadis11}.
    We use a small caps font to denote proof rules, such as \rnskip{}. All rules in this theorem's proof are from Vafeiadis~\cite{DBLP:journals/entcs/Vafeiadis11}, except \rnfork{} and \rnseqmulti{} that are straightforward extensions from the parallel and sequential composition rules, respectively.
    Side conditions arising in the proof trees are marked in blue and form $\omega$ (\cf \figref{side-conditions-appendix}).
    \begin{itemize}
        \item $\ghostalgorithm(\cskip{})$:
            Since the algorithm~\ghostalgorithm{} does not insert any \glsdisp{ghost-code}{ghost commands} and \cskip{} does not alter the program state, \localproginv{} is trivially maintained.
            The \rnskip{}~rule is immediately applicable and completes the proof tree.

        \item $\ghostalgorithm(\cheapalloc{x})$:
            \Figref{cheapalloc-prooftree} shows the proof tree that uses \figref{setlocalheapadd-prooftree} as a sub-proof for inserting a heap location into the \glsdisp{ghost-code}{ghost set} of local heap locations.

        \item $\ghostalgorithm(\cheapread{x}{e})$:
            The side condition~$\omega$ ensures that $e \in \setlocalheap \cup \ederef{\setglobalheap}$ holds.
            If $e \in \setlocalheap$ then \figref{cheapread-prooftree-1} is a valid proof tree for this read operation.
            Otherwise, $e \in \ederef{\setglobalheap}$ holds and \figref{cheapread-prooftree-2} shows the corresponding proof tree.

        \item $\ghostalgorithm(\cheapwrite{x}{e})$:
            For write operations, we construct a proof tree similar to read operations, as explained in the case above, except that we extract permissions for \var{x} instead of $e$ from the program invariants and replace applications of the \rnread{}~rule by \rnwrite{}.
            We can apply these rules because we possess full permission (as opposed to only partial permission) to the heap location~(\ie \acc{x}).

        \item $\ghostalgorithm(\ccorealloc{c}{\bar{e}})$:
            \Figref{ccorealloc-prooftree} shows the proof tree extending the subproof that the auto-active program verifier implicitly constructs (in \figref{ccorealloc-subprooftree}) while verifying the Hoare triple for \ccoreallocshort{\bar{e}}.

        \item $\ghostalgorithm(\ccorecall{k}{c}{\bar{e}}{\bar{r}})$:
            We construct a proof tree in \figref{ccorecall-prooftree} using \figref{ccorecall-subprooftree} as a subtree that is similar to the one for the \core{} allocation command with the main difference that the precondition requires \invnilable{c} instead of the I/O~permissions~$\phi$.
            The side condition $c \in \setinvariantheap \lor c = \nil$ ensures that we can obtain \invnilable{c} from \localproginv{} within the proof.

        \item $\ghostalgorithm(s_1 \cseq s_2)$:
            We apply the standard \rnseq{}~rule from \gls{separation-logic} to combine the proof subtrees for $\ghostalgorithm(s_1)$ and $\ghostalgorithm(s_2)$ that we obtain by applying our induction hypothesis.

        \item $\ghostalgorithm(\cfork{\bar{x}}{s})$:
            \Figref{cfork-prooftree} shows the proof tree that applies the induction hypothesis to $\ghostalgorithm(s)$.
            Since the algorithm~\ghostalgorithm{} removes the permissions for heap locations only in $\reachable{\bar{x}} \cap \setlocalheap$, the resulting side condition ($(\reachable{\bar{x}} \cap \setlocalheap) \subseteq \setlocalheap$) is trivial since these heap locations are by definition contained in \setlocalheap{}.

            The main proof insight is that we ensure that the global invariant covers the permissions for all heap locations that become accessible by the spawned thread and establish the local invariant for the spawned thread by initializing the set of local heap locations and (local) \core{}~instances to the empty set.
            \reachable{\bar{x}} forms an upper bound on the heap locations that command~$s$ might access because we syntactically require that $s$ accesses only $\bar{x}$ and its own local variables.
    \end{itemize}
\end{proofsketchcomposition}

\ifthenelse{\boolean{show_composition_soundness_proofs}}{%
    \begin{prooftreefigure*}
\begin{prooftree}
\[
    \[
        \justifies
        \ctxhoare
            {\globalproginv}
            {\forall l \in \setlocalheap \cupnil \setliteral{\var{x}} \ldotp \acc{l}}
            {\csetadd{\setlocalheap}{\var{x}}}
            {\forall l \in \setlocalheap \ldotp \acc{l}}
        \using\rnassign
    \]
    \justifies
    \ctxhoare
        {\globalproginv}
        {(\forall l \in \setlocalheap \ldotp \acc{l}) \star \accnilable{x}}
        {\csetadd{\setlocalheap}{\var{x}}}
        {\forall l \in \setlocalheap \ldotp \acc{l}}
    \using\rnconseq
\]
\justifies
\ctxhoare
    {\globalproginv}
    {\localproginv \star \accnilable{x}}
    {\csetadd{\setlocalheap}{\var{x}}}
    {\localproginv}
\using\rnframe
\end{prooftree}
\caption{%
    Proof tree for $\csetadd{\setlocalheap}{\var{x}}$, where $\accnilable{e} \triangleq \; e \neq \nil \implies \acc{e}$.
}
\label{fig:setlocalheapadd-prooftree}
\end{prooftreefigure*}

\begin{prooftreefigure*}
\begin{prooftree}
\[
    \[
        \[
            \justifies
            \ctxhoare
                {\emp}
                {\acc{\setglobalheap} \star \ederef{\setglobalheap} = v}
                {\csetadd{\ederef{\setglobalheap}}{x}}
                {\acc{\setglobalheap} \star \ederef{\setglobalheap} = v \cupnil \setliteral{x}}
            \using\rnwrite
        \]
        \justifies
        \ctxhoare
            {\emp}
            {\acc{\setglobalheap} \star \ederef{\setglobalheap} = v \star R}
            {\csetadd{\ederef{\setglobalheap}}{x}}
            {\acc{\setglobalheap} \star \ederef{\setglobalheap} = v \cupnil \setliteral{x} \star R}
        \using\rnframe
    \]
    \justifies
    \ctxhoare
        {\emp}
        {\acc{\setglobalheap} \star (\forall g \in \ederef{\setglobalheap} \ldotp \acc{g}) \star \accnilable{x}}
        {\csetadd{\ederef{\setglobalheap}}{x}}
        {\acc{\setglobalheap} \star \forall g \in \ederef{\setglobalheap} \ldotp \acc{g}}
    \using\rnconseq
\]
\justifies
\ctxhoare
    {\emp}
    {\globalproginv \star \accnilable{x}}
    {\csetadd{\ederef{\setglobalheap}}{x}}
    {\globalproginv}
\using\rnframe
\end{prooftree}
\vspace{12pt}
\begin{align*}
    \text{with} \quad
    R &\triangleq \; \forall g \in (v \cupnil \setliteral{x}) \ldotp \acc{g}
\end{align*}
\caption{%
    Proof tree for $\csetadd{\ederef{\setglobalheap}}{x}$ given that \globalproginv{} is already local, where $v$ is a \emph{fresh} variable and the \rnwrite~rule has been naturally extended to internally perform a heap read operation returning the value~$v$ for \ederef{\setglobalheap} as specified in the precondition.
}
\label{fig:csetglobaladd-subprooftree}
\end{prooftreefigure*}

\begin{prooftreefigure*}
\begin{prooftree}
\[
    \leadsto
    \ctxhoare
        {\emp}
        {\globalproginv \star \accnilable{x}}
        {\csetadd{\ederef{\setglobalheap}}{x}}
        {\globalproginv}
    \using\figref{csetglobaladd-subprooftree}
\]
\justifies
\ctxhoare
    {\globalproginv}
    {\accnilable{x}}
    {\csetadd{\ederef{\setglobalheap}}{x}}
    {\emp}
\using\rnatomic
\end{prooftree}
\caption{%
    Proof tree for $\csetadd{\ederef{\setglobalheap}}{x}$.
}
\label{fig:csetglobaladd-prooftree}
\end{prooftreefigure*}

\begin{prooftreefigure*}
\begin{prooftree}
\[
    \[
        \justifies
        \ctxhoare
            {\globalproginv}
            {\forall i \in \setinvariantheap \cupnil \setliteral{\var{c}} \ldotp \inv{i}}
            {\csetadd{\setinvariantheap}{\var{c}}}
            {\forall i \in \setinvariantheap \ldotp \inv{i}}
        \using\rnassign
    \]
    \justifies
    \ctxhoare
        {\globalproginv}
        {(\forall i \in \setinvariantheap \ldotp \inv{i}) \star \invnilable{c}}
        {\csetadd{\setinvariantheap}{\var{c}}}
        {\forall i \in \setinvariantheap \ldotp \inv{i}}
    \using\rnconseq
\]
\justifies
\ctxhoare
    {\globalproginv}
    {\localproginv \star \invnilable{c}}
    {\csetadd{\setinvariantheap}{\var{c}}}
    {\localproginv}
\using\rnframe
\end{prooftree}
\caption{%
    Proof tree for $\csetadd{\setinvariantheap}{\var{c}}$, where $\invnilable{c} \triangleq \; c \neq \nil \implies \inv{c}$.
}
\label{fig:setinvariantheapadd-prooftree}
\end{prooftreefigure*}

\begin{prooftreefigure*}
\begin{prooftree}
\[
    \sidecondition{e \in \setlocalheap}
    \[
        \[
            \justifies
            \ctxhoare
                {\globalproginv}
                {\forall l \in \setlocalheap \setminus \setliteral{e} \ldotp    \acc{l}}
                {\csetrem{\setlocalheap}{e}}
                {\forall l \in \setlocalheap \ldotp \acc{l}}
            \using\rnassign
        \]
        \justifies
        \ctxhoare
            {\globalproginv}
            {(\forall l \in \setlocalheap \setminus \setliteral{e} \ldotp \acc{l}) \star \acc{e}}
            {\csetrem{\setlocalheap}{e}}
            {(\forall l \in \setlocalheap \ldotp \acc{l}) \star \acc{e}}
        \using\rnframe
    \]
    \justifies
    \ctxhoare
        {\globalproginv}
        {\forall l \in \setlocalheap \ldotp \acc{l}}
        {\csetrem{\setlocalheap}{e}}
        {(\forall l \in \setlocalheap \ldotp \acc{l}) \star \acc{e}}
    \using\rnconseq
\]
\justifies
\ctxhoare
    {\globalproginv}
    {\localproginv}
    {\csetrem{\setlocalheap}{e}}
    {\localproginv \star \acc{e}}
\using\rnframe
\end{prooftree}
\caption{%
    Proof tree for $\csetrem{\setlocalheap}{e}$ if $e \in \setlocalheap$.
}
\label{fig:setlocalheaprem-prooftree}
\end{prooftreefigure*}

\begin{prooftreefigure*}
\begin{prooftree}
\[
    \sidecondition{e \in \setlocalheap \lor e = \nil}
    \[
        \[
            \justifies
            \ctxhoare
                {\globalproginv}
                {\forall l \in \setlocalheap \setminus \setliteral{e} \ldotp    \acc{l}}
                {\csetrem{\setlocalheap}{e}}
                {\forall l \in \setlocalheap \ldotp \acc{l}}
            \using\rnassign
        \]
        \justifies
        \ctxhoare
            {\globalproginv}
            {(\forall l \in \setlocalheap \setminus \setliteral{e} \ldotp \acc{l}) \star \accnilable{e}}
            {\csetrem{\setlocalheap}{e}}
            {(\forall l \in \setlocalheap \ldotp \acc{l}) \star \accnilable{e}}
        \using\rnframe
    \]
    \justifies
    \ctxhoare
        {\globalproginv}
        {\forall l \in \setlocalheap \ldotp \acc{l}}
        {\csetrem{\setlocalheap}{e}}
        {(\forall l \in \setlocalheap \ldotp \acc{l}) \star \accnilable{e}}
    \using\rnconseq
\]
\justifies
\ctxhoare
    {\globalproginv}
    {\localproginv}
    {\csetrem{\setlocalheap}{e}}
    {\localproginv \star \accnilable{e}}
\using\rnframe
\end{prooftree}
\caption{%
    Alternative proof tree for $\csetrem{\setlocalheap}{e}$ that permits $e$ being \nil{}.
}
\label{fig:setlocalheaprem-nilable-prooftree}
\end{prooftreefigure*}

\begin{prooftreefigure*}
\begin{prooftree}
\[
    \sidecondition{e \in \ederef{\setglobalheap}}
    \[
        \[
            \justifies
            \ctxhoare
                {\emp}
                {\acc{\setglobalheap} \star \ederef{\setglobalheap} = v}
                {\csetrem{\ederef{\setglobalheap}}{e}}
                {\acc{\setglobalheap} \star \ederef{\setglobalheap} = v \setminus \setliteral{e}}
            \using\rnwrite
        \]
        \justifies
        \ctxhoare
            {\emp}
            {\acc{\setglobalheap} \star \ederef{\setglobalheap} = v \star R}
            {\csetrem{\ederef{\setglobalheap}}{e}}
            {\acc{\setglobalheap} \star \ederef{\setglobalheap} = v \setminus \setliteral{e} \star R}
        \using\rnframe
    \]
    \justifies
    \ctxhoare
        {\emp}
        {\acc{\setglobalheap} \star \forall g \in \ederef{\setglobalheap} \ldotp \acc{g}}
        {\csetrem{\ederef{\setglobalheap}}{e}}
        {\acc{\setglobalheap} \star (\forall g \in \ederef{\setglobalheap} \ldotp \acc{g}) \star \acc{e}}
    \using\rnconseq
\]
\justifies
\ctxhoare
    {\emp}
    {\globalproginv}
    {\csetrem{\ederef{\setglobalheap}}{e}}
    {\globalproginv \star \acc{e}}
\using\rnframe
\end{prooftree}
\vspace{12pt}
\begin{align*}
    \text{with} \quad
    R &\triangleq \; (\forall g \in (v \setminus \setliteral{e}) \ldotp \acc{g}) \star \acc{e}
\end{align*}
\caption{%
    Proof tree for $\csetrem{\ederef{\setglobalheap}}{e}$ that requires \globalproginv{} to be local.
}
\label{fig:setglobalheaprem-prooftree}
\end{prooftreefigure*}

\begin{prooftreefigure*}
\begin{prooftree}
\[
    \sidecondition{c \in \setinvariantheap \lor c = \nil}
    \[
        \[
            \justifies
            \ctxhoare
                {\globalproginv}
                {\forall i \in \setinvariantheap \setminus \setliteral{c} \ldotp \inv{l}}
                {\csetrem{\setinvariantheap}{c}}
                {\forall i \in \setinvariantheap \ldotp \inv{i}}
            \using\rnassign
        \]
        \justifies
        \ctxhoare
            {\globalproginv}
            {(\forall i \in \setinvariantheap \setminus \setliteral{c} \ldotp \inv{l}) \star \invnilable{c}}
            {\csetrem{\setinvariantheap}{c}}
            {(\forall i \in \setinvariantheap \ldotp \inv{i}) \star \invnilable{c}}
        \using\rnframe
    \]
    \justifies
    \ctxhoare
        {\globalproginv}
        {\forall i \in \setinvariantheap \ldotp \inv{i}}
        {\csetrem{\setinvariantheap}{c}}
        {(\forall i \in \setinvariantheap \ldotp \inv{i}) \star \invnilable{c}}
    \using\rnconseq
\]
\justifies
\ctxhoare
    {\globalproginv}
    {\localproginv}
    {\csetrem{\setinvariantheap}{c}}
    {\localproginv \star \invnilable{c}}
\using\rnframe
\end{prooftree}
\caption{%
    Proof tree for $\csetrem{\setinvariantheap}{c}$.
}
\label{fig:setinvariantheaprem-prooftree}
\end{prooftreefigure*}

\begin{prooftreefigure*}
\begin{prooftree}
\[
    \[
        \[
            \[
                \sidecondition{\neg(\ederef{\flagiospec})}
                \[
                    \[
                        \justifies
                        \ctxhoare
                            {\emp}
                            {\acc{\flagiospec}}
                            {\cassign{\ederef{\flagiospec}}{\true}}
                            {\acc{\flagiospec} \star \ederef{\flagiospec} = true}
                        \using\rnwrite
                    \]
                    \justifies
                    \ctxhoare
                        {\emp}
                        {\acc{\flagiospec} \star \phi}
                        {\cassign{\ederef{\flagiospec}}{\true}}
                        {\acc{\flagiospec} \star \ederef{\flagiospec} = true \star \phi}
                    \using\rnframe
                \]
                \justifies
                \ctxhoare
                    {\emp}
                    {\acc{\flagiospec} \star \neg(\ederef{\flagiospec}) \implies \phi}
                    {\cassign{\ederef{\flagiospec}}{\true}}
                    {\acc{\flagiospec} \star \ederef{\flagiospec} = true \star \phi}
                \using\rnconseq
            \]
            \justifies
            \ctxhoare
                {\emp}
                {\acc{\flagiospec} \star \neg(\ederef{\flagiospec}) \implies \phi}
                {\cassign{\ederef{\flagiospec}}{\true}}
                {\acc{\flagiospec} \star (\neg(\ederef{\flagiospec}) \implies \phi) \star \phi}
            \using\rnconseq
        \]
        \justifies
        \ctxhoare
            {\emp}
            {\globalproginv}
            {\cassign{\ederef{\flagiospec}}{\true}}
            {\globalproginv \star \phi}
        \using\rnframe
    \]
    \justifies
    \ctxhoare
        {\globalproginv}
        {\emp}
        {\catomic{\cassign{\ederef{\flagiospec}}{\true}}}
        {\phi}
    \using\rnatomic
\]
\justifies
\ctxhoare
    {\globalproginv}
    {\localproginv}
    {\catomic{\cassign{\ederef{\flagiospec}}{\true}}}
    {\localproginv \star \phi}
\using\rnframe
\end{prooftree}
\caption{%
    Proof tree for $\catomic{\cassign{\ederef{\flagiospec}}{\true}}$.
}
\label{fig:setflagiospec-prooftree}
\end{prooftreefigure*}

\begin{prooftreefigure*}
\begin{prooftree}
\[
    \[
        \justifies
        \ctxhoare
            {\globalproginv}
            {\emp}
            {\cheapalloc{x}}
            {\acc{x}}
    \using\rnalloc
    \]
    \justifies
    \ctxhoare
        {\globalproginv}
        {\localproginv}
        {\cheapalloc{x}}
        {\localproginv \star \acc{x}}
    \using\rnframe
\]
\[
    \[
        \leadsto
        \ctxhoare
            {\globalproginv}
            {\localproginv \star \accnilable{x}}
            {\csetadd{\setlocalheap}{\var{x}}}
            {\localproginv}
        \using\figref{setlocalheapadd-prooftree}
    \]
    \justifies
    \ctxhoare
        {\globalproginv}
        {\localproginv \star \acc{x}}
        {\csetadd{\setlocalheap}{\var{x}}}
        {\localproginv}
    \using\rnconseq
\]
\justifies
\ctxhoare
    {\globalproginv}
    {\localproginv}
    {\gacheapallocout}
    {\localproginv}
\using\rnseq
\end{prooftree}
\caption{%
    Proof tree for $\ghostalgorithm(\gacheapallocin)$.
}
\label{fig:cheapalloc-prooftree}
\end{prooftreefigure*}

\begin{prooftreefigure*}
\begin{prooftree}
\[
    \leadsto
    \ctxhoare
        {\globalproginv}
        {\localproginv}
        {s_1}
        {\localproginv \star \acc{e}}
    \using\figref{setlocalheaprem-prooftree}
\]
\[
    \[
        \justifies
        \ctxhoare
            {\globalproginv}
            {\acc{e}}
            {s_2}
            {\acc{e}}
        \using\rnread
    \]
    \justifies
    \ctxhoare
        {\globalproginv}
        {\localproginv \star \acc{e}}
        {s_2}
        {\localproginv \star \acc{e}}
    \using\rnframe
\]
\[
    \[
        \leadsto
        \ctxhoare
            {\globalproginv}
            {\localproginv \star \accnilable{e}}
            {s_3}
            {\localproginv}
        \using\figref{setlocalheapadd-prooftree}
    \]
    \justifies
    \ctxhoare
        {\globalproginv}
        {\localproginv \star \acc{e}}
        {s_3}
        {\localproginv}
    \using\rnconseq
\]
\justifies
\ctxhoare
    {\globalproginv}
    {\localproginv}
    {s_1 \cseq s_2 \cseq s_3 \showghostalgorithmoutput{\gacheapreadoutlocal}}
    {\localproginv}
\using\rnseqmulti
\end{prooftree}
\vspace{12pt}
\begin{align*}
    \text{with} \quad
    s_1 &\triangleq \; \csetrem{\setlocalheap}{e} &
    s_2 &\triangleq \; \cheapread{x}{e} &
    s_3 &\triangleq \; \csetadd{\setlocalheap}{e}
\end{align*}
\caption{%
    Proof tree for $\ghostalgorithm(\gacheapreadin)$ if $e \in \setlocalheap$, where \rnseqmulti{} represents repeated application of the \rnseq~rule.
    We discharge the side condition from \figref{setlocalheaprem-prooftree} as $e \in \setlocalheap$ holds by definition.
}
\label{fig:cheapread-prooftree-1}
\end{prooftreefigure*}

\begin{prooftreefigure*}
\begin{prooftree}
\[
    \[
        \[
            \sidecondition{e \in \ederef{\setglobalheap}}
            \leadsto
            \justifies
            \ctxhoare
                {\emp}
                {\globalproginv}
                {s_1}
                {\globalproginv \star \acc{e}}
            \using\figref{setglobalheaprem-prooftree}
        \]
        \[
            \[
                \justifies
                \ctxhoare
                    {\emp}
                    {\acc{e}}
                    {s_2}
                    {\acc{e}}
                \using\rnread
            \]
            \justifies
            \ctxhoare
                {\emp}
                {\globalproginv \star \acc{e}}
                {s_2}
                {\globalproginv \star \acc{e}}
            \using\rnframe
        \]
        \[
            \[
                \leadsto
                \ctxhoare
                    {\emp}
                    {\globalproginv \star \accnilable{e}}
                    {s_3}
                    {\globalproginv}
                \using\figref{csetglobaladd-subprooftree}
            \]
            \justifies
            \ctxhoare
                {\emp}
                {\globalproginv \star \acc{e}}
                {s_3}
                {\globalproginv}
            \using\figref{csetglobaladd-subprooftree}
        \]
        \justifies
        \ctxhoare
            {\emp}
            {\globalproginv}
            {s_1 \cseq s_2 \cseq s_3}
            {\globalproginv}
        \using\rnseqmulti
    \]
    \justifies
    \ctxhoare
        {\globalproginv}
        {\emp}
        {\catomic{s_1 \cseq s_2 \cseq s_3}}
        {\emp}
    \using\rnatomic
\]
\justifies
\ctxhoare
    {\globalproginv}
    {\localproginv}
    {\catomic{s_1 \cseq s_2 \cseq s_3} \showghostalgorithmoutput{\gacheapreadoutglobal}}
    {\localproginv}
\using\rnframe
\end{prooftree}
\vspace{12pt}
\begin{align*}
    \text{with} \quad
    s_1 &\triangleq \; \csetrem{\ederef{\setglobalheap}}{e} &
    s_2 &\triangleq \; \cheapread{x}{e} &
    s_3 &\triangleq \; \csetadd{\ederef{\setglobalheap}}{e}
\end{align*}
\caption{%
    Proof tree for $\ghostalgorithm(\gacheapreadin)$ if $e \not\in \setlocalheap$.
}
\label{fig:cheapread-prooftree-2}
\end{prooftreefigure*}

\begin{prooftreefigure*}
\begin{prooftree}
\[
    \sidecondition{\asmref{syntactic-restrictions}}
    \[
        \[
            \leadsto
            \ctxhoare
                {\globalproginv}
                {\precorealloc{\bar{e}}}
                {\ccorealloc{c}{\bar{e}}}
                {\postcorealloc{c}{\bar{e}}}
            \using\text{auto-active verification}
        \]
        \justifies
        \ctxhoare
            {\globalproginv}
            {\precorealloc{\bar{e}} \star F}
            {\ccorealloc{c}{\bar{e}}}
            {\postcorealloc{c}{\bar{e}} \star F}
        \using\rnframe
    \]
    \justifies
    \ctxhoare
        {\globalproginv}
        {\accnilable{\bar{e}} \star \phi}
        {\ccorealloc{c}{\bar{e}}}
        {\accnilable{\bar{e}} \star \inv{c}}
    \using\rnconseq
\]
\justifies
\ctxhoare
    {\globalproginv}
    {\localproginv \star \accnilable{\bar{e}} \star \phi}
    {\ccorealloc{c}{\bar{e}}}
    {\localproginv \star \accnilable{\bar{e}} \star \inv{c}}
\using\rnframe
\end{prooftree}
\caption{%
    Proof tree for $\ccorealloc{c}{\bar{e}}$ using the subproof that we extract from the auto-active program verifier.
    The side condition (\asmref{syntactic-restrictions}) states that $\precorealloc{\bar{e}} = \phi \star R$ and $\accnilable{\bar{e}} \models R$.
    We call $F$ the permissions that are framed around, \ie $\accnilable{\bar{e}} = R \star F$.
    The side condition further specifies that $\postcorealloc{c}{\bar{e}} = \inv{c} \star R'$ and $R' \models R$ hold, allowing us to derive $R' \star F \models \accnilable{\bar{e}}$.
    Thus, we can apply the \rnconseq{}~rule.
    We abuse the notation~\accnilable{\bar{e}} to denote the iterated separating conjunction expressing \accnilable{e} for each element~$e$ in $\bar{e}$, \ie $\forall i \ldotp 0 \leq i < \len{\bar{e}} \implies \accnilable{\at{\bar{e}}{i}}$, where $\len{\bar{e}}$ and $\at{\bar{e}}{i}$ return the length and the $i$-th element of the list~$\bar{e}$, respectively.
}
\label{fig:ccorealloc-subprooftree}
\end{prooftreefigure*}

\begin{prooftreefigure*}
\begin{prooftree}
\[
    \sidecondition{\neg(\ederef{\flagiospec})}
    \leadsto
    \ctxhoare
        {\globalproginv}
        {\localproginv}
        {s_1}
        {\localproginv \star \phi}
    \using\figref{setflagiospec-prooftree}
\]
\[
    \[
        \sidecondition{
            (\set{\bar{e}} \setminus \nil) \subseteq \setlocalheap \land{}\\
            \disjoint{\bar{e}}
        }
        \leadsto
        \ctxhoare
            {\globalproginv}
            {\localproginv}
            {s_2}
            {R'_2}
        \using\figref{setlocalheaprem-nilable-prooftree}
    \]
    \justifies
    \ctxhoare
        {\globalproginv}
        {\localproginv \star \phi}
        {s_2}
        {R_2}
    \using\rnframe
\]
\[
    \sidecondition{\asmref{syntactic-restrictions}}
    \leadsto
    \ctxhoare
        {\globalproginv}
        {R_2}
        {s_3}
        {R_3}
    \using\figref{ccorealloc-subprooftree}
\]
\[
    \[
        \leadsto
        \ctxhoare
            {\globalproginv}
            {R'_2}
            {s_4}
            {\localproginv}
        \using\figref{setlocalheapadd-prooftree}
    \]
    \justifies
    \ctxhoare
        {\globalproginv}
        {R_3}
        {s_4}
        {R_4}
    \using\rnframe
\]
\[
    \[
        \leadsto
        \ctxhoare
            {\globalproginv}
            {R'_4}
            {s_5}
            {\localproginv}
        \using\figref{setinvariantheapadd-prooftree}
    \]
    \justifies
    \ctxhoare
        {\globalproginv}
        {R_4}
        {s_5}
        {\localproginv}
    \using\rnseq
\]
\justifies
\ctxhoare
    {\globalproginv}
    {\localproginv}
    {s_1 \cseq s_2 \cseq s_3 \cseq s_4 \cseq s_5 \showghostalgorithmoutput{\gaccoreallocout}}
    {\localproginv}
\using\rnseqmulti
\end{prooftree}
\begin{flushleft}
    with
\end{flushleft}
\begin{align*}
    s_1 &\triangleq \; \catomic{\cassign{\ederef{\flagiospec}}{\true}} &
    s_2 &\triangleq \; \csetremmult{\setlocalheap}{\bar{e}} &
    s_3 &\triangleq \; \ccorealloc{c}{\bar{e}} &
    s_4 &\triangleq \; \csetaddmult{\setlocalheap}{\bar{e}} &
    s_5 &\triangleq \; \csetadd{\setinvariantheap}{\var{c}}\\
    R'_2 &\triangleq \; \localproginv \star \accnilable{\bar{e}} &
    R_2 &\triangleq \; R'_2 \star \phi &
    R_3 &\triangleq \; R'_2 \star \inv{c} &
    R_4 &\triangleq \; \localproginv \star \inv{c} &
    R'_4 &\triangleq \; \localproginv \star \invnilable{c}
\end{align*}
\caption{%
    Proof tree for $\ghostalgorithm(\gaccoreallocin)$. We naturally extend \figref{setlocalheapadd-prooftree} and \figref{setlocalheaprem-prooftree} to adding and removing \emph{lists} of heap locations to and from the \glsdisp{ghost-code}{ghost set} \setlocalheap{}, respectively.
    The latter requires their disjointness.
}
\label{fig:ccorealloc-prooftree}
\end{prooftreefigure*}

\begin{prooftreefigure*}
\begin{prooftree}
\[
    \sidecondition{\asmref{syntactic-restrictions}}
    \[
        \[
            \leadsto
            \ctxhoare
                {\globalproginv}
                {\precorecall{k}{c}{\bar{e}}}
                {\ccorecall{k}{c}{\bar{e}}{\bar{r}}}
                {\postcorecall{k}{c}{\bar{e}}{\bar{r}}}
            \using\text{auto-active verification}
        \]
        \justifies
        \ctxhoare
            {\globalproginv}
            {\precorecall{k}{c}{\bar{e}} \star F}
            {\ccorecall{k}{c}{\bar{e}}{\bar{r}}}
            {\postcorecall{k}{c}{\bar{e}}{\bar{r}} \star F}
        \using\rnframe
    \]
    \justifies
    \ctxhoare
        {\globalproginv}
        {\invnilable{c} \star \accnilable{\bar{e}}}
        {\ccorecall{k}{c}{\bar{e}}{\bar{r}}}
        {\invnilable{c} \star \accnilable{\bar{e}} \star \accnilable{\bar{r}}}
    \using\rnconseq
\]
\justifies
\ctxhoare
    {\globalproginv}
    {\localproginv \star \invnilable{c} \star \accnilable{\bar{e}}}
    {\ccorecall{k}{c}{\bar{e}}{\bar{r}}}
    {\localproginv \star \invnilable{c} \star \accnilable{\bar{e}} \star \accnilable{\bar{r}}}
\using\rnframe
\end{prooftree}
\caption{%
    Proof tree for \ccorecall{k}{c}{\bar{e}}{\bar{r}}.
}
\label{fig:ccorecall-subprooftree}
\end{prooftreefigure*}

\begin{prooftreefigure*}
\begin{prooftree}
\[
    \sidecondition{c \in \setinvariantheap \lor c = \nil}
    \leadsto
    \ctxhoare
        {\globalproginv}
        {\localproginv}
        {s_1}
        {R_1}
    \using\figref{setinvariantheaprem-prooftree}
\]
\[
    \[
        \sidecondition{(\set{\bar{e}} \setminus \nil) \subseteq \setlocalheap \land{}\\ \disjoint{\bar{e}}}
        \leadsto
        \ctxhoare
            {\globalproginv}
            {\localproginv}
            {s_2}
            {R'_2}
        \using\figref{setlocalheaprem-nilable-prooftree}
    \]
    \justifies
    \ctxhoare
        {\globalproginv}
        {R_1}
        {s_2}
        {R_2}
    \using\rnframe
\]
\[
    \sidecondition{\asmref{syntactic-restrictions}}
    \leadsto
    \ctxhoare
        {\globalproginv}
        {R_2}
        {s_3}
        {R_3}
    \using\figref{ccorecall-subprooftree}
\]
\[
    \[
        \leadsto
        \ctxhoare
            {\globalproginv}
            {R'_3}
            {s_4}
            {\localproginv}
        \using\figref{setlocalheapadd-prooftree}
    \]
    \justifies
    \ctxhoare
        {\globalproginv}
        {R_3}
        {s_4}
        {R_1}
        \using\figref{setlocalheapadd-prooftree}
    \using\rnframe
\]
\[
    \leadsto
    \ctxhoare
        {\globalproginv}
        {R_1}
        {s_5}
        {\localproginv}
    \using\figref{setinvariantheapadd-prooftree}
\]
\justifies
\ctxhoare
    {\globalproginv}
    {\localproginv}
    {s_1 \cseq s_2 \cseq s_3 \cseq s_4 \cseq s_5 \showghostalgorithmoutput{\gaccorecallout}}
    {\localproginv}
\using\rnseqmulti
\end{prooftree}
\begin{flushleft}
    with
\end{flushleft}
\begin{align*}
    s_1 &\triangleq \; \csetrem{\setinvariantheap}{\var{c}} &
    s_2 &\triangleq \; \csetremmult{\setlocalheap}{\bar{e}} &
    s_3 &\triangleq \; \ccorecall{k}{c}{\bar{e}}{\bar{r}} &
    s_4 &\triangleq \; \csetaddmult{\setlocalheap}{\bar{e} \cupnil \bar{r}} &
    s_5 &\triangleq \; \csetadd{\setinvariantheap}{\var{c}}\\
    R_1 &\triangleq \; \localproginv \star \invnilable{c} &
    R'_2 &\triangleq \; \localproginv \star \accnilable{\bar{e}} &
    R_2 &\triangleq \; R'_2 \star \invnilable{c} &
    R'_3 &\triangleq \; R'_2 \star \accnilable{\bar{r}} &
    R_3 &\triangleq \; R'_3 \star \invnilable{c}
\end{align*}
\caption{%
    Proof tree for $\ghostalgorithm(\gaccorecallin)$.
}
\label{fig:ccorecall-prooftree}
\end{prooftreefigure*}

\begin{prooftreefigure*}
\begin{prooftree}
\[
    \justifies
    \ctxhoare
        {\proginv_g}
        {\emp}
        {\cassign{\setlocalheap}{\emptyset}}
        {\setlocalheap = \emptyset}
    \using\rnassign
\]
\[
    \[
        \[
            \justifies
            \ctxhoare
                {\proginv_g}
                {\emp}
                {\cassign{\setinvariantheap}{\emptyset}}
                {\setinvariantheap = \emptyset}
            \using\rnassign
        \]
        \justifies
        \ctxhoare
            {\proginv_g}
            {\setlocalheap = \emptyset}
            {\cassign{\setinvariantheap}{\emptyset}}
            {\setlocalheap = \emptyset \star \setinvariantheap = \emptyset}
        \using\rnframe
    \]
    \justifies
    \ctxhoare
        {\proginv_g}
        {\setlocalheap = \emptyset}
        {\cassign{\setinvariantheap}{\emptyset}}
        {\proginv_l}
    \using\rnconseq
\]
\[
    \leadsto
    \ctxhoare
        {\proginv_g}
        {\proginv_l}
        {\ghostalgorithm(s)}
        {\proginv_l}
    \using\text{\acs{IH}}
\]
\justifies
\ctxhoare
    {\proginv_g}
    {\emp}
    {\cassign{\setlocalheap}{\emptyset} \cseq \cassign{\setinvariantheap}{\emptyset} \cseq \ghostalgorithm(s)}
    {\proginv_l}
\using\rnseqmulti
\end{prooftree}
\caption{%
    Proof tree for the sequence of statements that is executed as the newly spawned thread, where $s$ represents an arbitrary input statement and \acs{IH} denotes an application of the induction hypothesis.
    We omit trivial applications of the \rnconseq{}~rule.
}
\label{fig:cfork-body-prooftree}
\end{prooftreefigure*}

\begin{prooftreefigure*}
\begin{prooftree}
\[
    \leadsto
    \ctxhoare
        {\proginv_g}
        {\proginv_l}
        {s_1}
        {\proginv_l \star R}
    \using\figref{setlocalheaprem-prooftree}
\]
\[
    \[
        \[
            \[
                \leadsto
                \ctxhoare
                    {\proginv_g}
                    {R'}
                    {s_2}
                    {\emp}
                \using\figref{csetglobaladd-prooftree}
            \]
            \justifies
            \ctxhoare
                {\proginv_g}
                {R}
                {s_2}
                {\emp}
            \using\rnconseq
        \]
        \[
            \[
                \leadsto
                \ctxhoare
                    {\proginv_g}
                    {\emp}
                    {\cassign{\setlocalheap}{\emptyset} \cseq \cassign{\setinvariantheap}{\emptyset} \cseq \ghostalgorithm(s)}
                    {\proginv_l}
                \using\figref{cfork-body-prooftree}
            \]
            \justifies
            \ctxhoare
                {\proginv_g}
                {\emp}
                {\cfork{\bar{x}}{\cassign{\setlocalheap}{\emptyset} \cseq \cassign{\setinvariantheap}{\emptyset} \cseq \ghostalgorithm(s)}}
                {\emp}
            \using\rnfork
        \]
        \justifies
        \ctxhoare
            {\proginv_g}
            {R}
            {s_2 \cseq \cfork{\bar{x}}{\cassign{\setlocalheap}{\emptyset} \cseq \cassign{\setinvariantheap}{\emptyset} \cseq \ghostalgorithm(s)}}
            {\emp}
        \using\rnseq
    \]
    \justifies
    \ctxhoare
        {\proginv_g}
        {\proginv_l \star R}
        {s_2 \cseq \cfork{\bar{x}}{\cassign{\setlocalheap}{\emptyset} \cseq \cassign{\setinvariantheap}{\emptyset} \cseq \ghostalgorithm(s)}}
        {\proginv_l}
    \using\rnframe
\]
\justifies
\ctxhoare
    {\proginv_g}
    {\proginv_l}
    {s_1 \cseq s_2 \cseq \cfork{\bar{x}}{\cassign{\setlocalheap}{\emptyset} \cseq \cassign{\setinvariantheap}{\emptyset} \cseq \ghostalgorithm(s)} \showghostalgorithmoutput{\gacforkout}}
    {\proginv_l}
\using\rnseq
\end{prooftree}
\vspace{12pt}
\begin{align*}
    \text{with} \quad
    s_1 &\triangleq \; \csetremmult{\setlocalheap}{(\reachable{\bar{x}} \cap \setlocalheap)} &
    s_2 &\triangleq \; \csetaddmult{\ederef{\setglobalheap}}{(\reachable{\bar{x}} \cap \setlocalheap)}\\
    R &\triangleq \; \forall l \in (\reachable{\bar{x}} \cap \setlocalheap) \ldotp \acc{l} &
    R' &\triangleq \; \forall l \in (\reachable{\bar{x}} \cap \setlocalheap) \ldotp \accnilable{l}
\end{align*}
\caption{%
    Proof tree for $\ghostalgorithm(\gacforkin)$ that assumes the existence of a \rnfork{}~rule.
    The side conditions stemming from \figref{csetglobaladd-prooftree} hold trivially as $\reachable{\bar{x}} \cap \setlocalheap \subseteq \setlocalheap$ and since $\reachable{\bar{x}}$ returns a set of heap locations, which is by definition free of duplicates and, thus, its elements are pairwise disjoint.
}
\label{fig:cfork-prooftree}
\end{prooftreefigure*}

}{%
}

\subsubsection{Static Analyses}
\label{app:diodon-soundness-static-analyses}
Since our proof rules rely on the side conditions~$\omega$ (\cf \figref{side-conditions-appendix}), we introduce next our static analyses, cover the properties we assume they provide, and show that these properties imply $\omega$.
We end by proving a corollary that we can construct a whole program proof for a codebase given that we have auto-actively verified the \core and successfully executed the static analyses.

\mypar{Pointer analysis}
A pointer analysis computes for each pointer~$x$ a set of heap locations~$L$ to where $x$ \emph{may} point, which we formalize as a judgement~$\pointstoset{x} = L$.
Each heap location in $L$ is identified by its allocation site, which corresponds to the label of a particular statement in the program's text.
Note that this analysis over-approximates the set of heap locations that actually change when writing to $x$.
The pointer analysis we are using is context insensitive, \ie ignores control flow and ordering of statements.
Thus, we omit the program location at which such a judgement holds as it holds for all program locations within a given codebase.
If necessary, we could employ a context-sensitive pointer analysis to increase precision.

To formalize what the pointer analysis computes, let us first state several definitions before stating the pointer analysis' soundness, which we assume.

\begin{definition}[Reachability]
    \label{def:reachability}
    \reachabletrace{x}{\tau}{p} returns the set of addresses for all heap locations that are transitively reachable from variable~$x$ at program point~$p$ on trace~$\tau$.
    Hence, $\forall x, \tau, p \ldotp \valueof{x}{\tau} \in \reachabletrace{x}{\tau}{p}$ holds for all program points~$p$ after $x$ is defined.
\end{definition}

\begin{definition}[Allocation site]
    \allocationsite{h}{\tau} returns the allocation site for a heap location~$h$ on trace~$\tau$, which is the label of the statement that allocated this heap location.
\end{definition}

\begin{assumption}[Soundness of pointer analysis]
    \label{asm:pointer-analysis-soundness}
    The pointer analysis computes for a variable~$x$ the heap locations~\pointstoset{x} to which $x$ \emph{may} point on all possible traces.
    These heap locations are identified by their allocation site.
    We assume that the pointer analysis is sound, \ie computes an over-approximation of the heap locations to which $x$ actually points when looking at concrete traces.
    \begin{align*}
        \forall x, \tau \ldotp \valueof{x}{\tau} \neq \nil \implies \allocationsite{\valueof{x}{\tau}}{\tau} \in \pointstoset{x}
    \end{align*}
\end{assumption}

\begin{lemma}[Disjointness from pointer analysis]
    \label{lem:pts-empty-intersection-implies-disjoint}
    We can use the pointer analysis' may-point-to judgments to derive disjointness.
    \begin{align*}
        \forall x, y \ldotp &\pointstoset{x} \cap \pointstoset{y} = \emptyset \implies \disjoint{\setliteral{x, y}}
    \end{align*}
\end{lemma}
\begin{proofsketchcomposition}
    If $x$ or $y$ store the value~\nil{}, $\disjoint{\setliteral{x, y}}$ holds.
    Otherwise, $x$ and $y$ are non-\nil{}.
    We apply \asmref{pointer-analysis-soundness} to our premise and obtain $\forall \tau \ldotp \allocationsite{\valueof{x}{\tau}}{\tau} \neq \allocationsite{\valueof{y}{\tau}}{\tau}$.
    Since $x$ and $y$ point on all possible traces to heap locations that were allocated at different allocation sites, the heap locations themselves must be different, \ie $\valueof{x}{\tau} \neq \valueof{y}{\tau}$.
\end{proofsketchcomposition}

\mypar{Pass-through analysis}
\looseness=-1
As hinted at by our \glsdisp{ghost-code}{ghost sets}, we distinguish two~types of heap locations, namely heap locations that make up \core instances and heap locations that the \app might access.
Heap locations of the former type are tracked by collecting the respective \core instances in \setinvariantheap.
The latter type encompasses heap locations that are either allocated within the \app by \cheapallocshort{} or allocated within the \core and returned from a \core \ac{API} call.

To distinguish these types of heap locations, we run a pass-through analysis that provides the judgments
\passthroughcoretrace{a}{\tau}{p} and \passthroughreturntrace{a}{\tau}{p} denoting that a heap location allocated at allocation site~$a$ passed through ($\mathit{pt}$) the return argument~$c$ of a \ccorealloc{c}{\bar{e}} statement and through one of the return arguments~$\bar{r}$ of a \ccorecall{k}{c}{\bar{e}}{\bar{r}} statement, respectively, between label~$a$ and program point~$p$ on trace~$\tau$.
\Ie we have that $\passthroughcoretrace{\allocationsite{\valueof{c}{\tau}}{\tau}}{\tau}{p}$ and $\forall r \in \set{\bar{r}} \ldotp \passthroughreturntrace{\allocationsite{\valueof{r}{\tau}}{\tau}}{\tau}{p}$ hold at the program point~$p$ on trace~$\tau$ after executing the statement \ccorealloc{c}{\bar{e}} and \ccorecall{k}{c}{\bar{e}}{\bar{r}}, respectively.

\begin{definition}[\app-managed heap locations]
    We call a heap location~$h$ \emph{\app-managed} at program point~$p$ on trace~$\tau$ if $h$ is either allocated within the \app or has been returned from a \ccorecall{k}{c}{\bar{e}}{\bar{r}} statement.
    \begin{align*}
        \appmanaged{h}{\tau}{p} \triangleq \; \isapplabel{\allocationsite{h}{\tau}} \lor \passthroughreturntrace{\allocationsite{h}{\tau}}{\tau}{p}
    \end{align*}
\end{definition}

\mypar{Escape analysis}
The goal of the escape analysis is to correctly place heap locations into \setlocalheap{}, \ederef{\setglobalheap}, and \setinvariantheap{}.
In particular, we want to establish globally that an \app-managed heap location and a \core instance are in \setlocalheap{} and \setinvariantheap{}, respectively, if they are \emph{local}.

\looseness=-1
We first define what it means for a heap location to be local (\cf \defref{heap-locality}).
\Ie this definition takes all threads into account and states that a heap location~$h$ is local to a thread~$t$ if and only if $t$ is the only thread that can potentially access $h$.

Locality of a heap location is approximated by our escape analysis.
The result of the escape analysis is formalized in a judgement $\local{x}{p}$ for some variable $x$ and program point~$p$.
The intuition is that a variable that is local points to heap locations (\ie $*x$) that are accessible \emph{only} by the current thread and, thus, can be modified or even referred to only by the current thread.
The escape analysis is sound in that no heap location that is accessible by another thread will ever be reported as local (\cf \asmref{escape-analysis-soundness}), but potentially imprecise in that some locations that are not accessible by other threads will fail to be local.

\begin{definition}[Accessibility]
    We write \accessible{h}{t}{p} to denote that heap location~$h$ is \emph{accessible} by thread~$t$ at program point~$p$.
    A thread may access such a heap location either directly via variables or indirectly by dereferencing other heap locations.
    We define accessibility independently of variables and, thus, accessibility of $h$ does not change when variables go out of scope.
    Instead, accessibility is monotonic for a thread's execution.
\end{definition}

\begin{definition}[Locality]
    \label{def:heap-locality}
    A heap location~$h$ is \emph{local} at program point~$p$ if it is accessible by a single thread~$t$.
    \begin{align*}
        \localheaplocationthread{h}{t}{p} \triangleq \; &\accessible{h}{t}{p} \land {}\\
        & (\forall t' \ldotp t' \neq t \implies \neg\accessible{h}{t'}{p})
    \end{align*}
\end{definition}

\begin{lemma}[Uniqueness of locality]
    \label{lem:local-unique}
    The thread~$t$ having access to a \emph{local} heap location~$h$ is unique, \ie
    \begin{align*}
        \forall h, t, t', p \ldotp &\localheaplocationthread{h}{t}{p} \land \localheaplocationthread{h}{t'}{p} \implies t = t'.
    \end{align*}
\end{lemma}
\begin{proofsketchcomposition}
    The lemma follows directly from \defref{heap-locality}.
\end{proofsketchcomposition}

\begin{lemma}[Locality is reverse monotonic]
    \label{lem:local-monotonic}
    A \emph{local} heap location~$h$ at program point~$p'$ must be local at every \emph{earlier} program point~$p$ if $h$ is accessible at $p$, \ie
    \begin{align*}
        \forall h, t, p, p' \ldotp & p \preceq p' \land \accessible{h}{t}{p} \land \localheaplocationthread{h}{t}{p'} \implies {}\\
        & \localheaplocationthread{h}{t}{p}.
    \end{align*}
\end{lemma}
\begin{proofsketchcomposition}
    We prove this lemma by contradiction for arbitrary $h$, $t$, $p$, and $p'$.
    $\neg\localheaplocationthread{h}{t}{p}$ implies that $h$ is accessible by another thread~$t'$, \ie $t' \neq t \land \accessible{h}{t'}{p}$.
    Since accessibility is monotonic, $h$ remains accessible by $t'$ at $p'$ contradicting \localheaplocationthread{h}{t}{p'}.
\end{proofsketchcomposition}

\begin{lemma}[Locality is reverse transitive]
    \label{lem:local-transitive}
    If a \emph{local} heap location~$h'$ is transitively reachable from another heap location~$h$ then $h$ must also be local.
    \begin{align*}
        \forall h, h', t, \tau, p \ldotp &\localheaplocationthread{h'}{t}{p} \land h' \in \reachabletrace{h}{\tau}{p} \implies {}\\
        &\localheaplocationthread{h}{t}{p}
    \end{align*}
\end{lemma}
\begin{proofsketchcomposition}
    We prove this lemma by contradiction for arbitrary $h$, $h'$, $t$, $\tau$, and $p$.
    \Ie a thread~$t'$ exists such that $h$ is accessible by $t'$.
    $h'$ is accessible by $t'$ via reachability from $h$, thus, contradicting \localheaplocationthread{h'}{t}{p}.
\end{proofsketchcomposition}

\begin{assumption}[Soundness of escape analysis]
    \label{asm:escape-analysis-soundness}
    We assume that the escape analysis is sound, \ie reports a heap location to which variable $x$ points as being local \emph{only} if the corresponding heap location is indeed local (or $x$ is \nil{}) for every possible trace~$\tau$, \ie
    \begin{align*}
        \forall x, \tau, p \ldotp &\local{x}{p} \implies {}\\
        &\valueof{x}{\tau} = \nil \lor \exists t \ldotp \localheaplocationthread{\valueof{x}{\tau}}{t}{p}.
    \end{align*}
\end{assumption}

Based on these definitions and the soundness of our analyses, we prove several lemmata that relate accessible heap locations to our \glsdisp{ghost-code}{ghost sets} and corollaries that lift these properties to variables and the judgments we obtain from our static analyses.
We will later use these corollaries to show that these judgments discharge our proof rules' side conditions~$\omega$.

\begin{lemma}[Inaccessability implies set absence]
    \label{lem:ghost-set-absence}
    All heap locations stored in the \glsdisp{ghost-code}{ghost sets} are accessible by at least one~thread.
    \begin{align*}
        \forall h, \tau, p \ldotp &h \neq \nil \land p \in \tau \land (\forall t \ldotp \neg\accessible{h}{t}{p}) \implies{}\\
        &\forall t \ldotp h \not\in \programpointparen{\threadset{\setlocalheap}{t} \cup \ederef{\setglobalheap} \cup \threadset{\setinvariantheap}{t}}{p}
    \end{align*}
    where \programpointwoparen{e}{p} denotes evaluating expression~$e$ at program point~$p$.
\end{lemma}
\begin{proofsketchcomposition}
    We prove this lemma by induction over program traces.
    The base case for the empty trace holds trivially as \setlocalheap{} and \setinvariantheap{} for every thread~$t$ and \ederef{\setglobalheap} are initialized to the empty set.
    In the inductive step, we prove this lemma for an arbitrary heap location~$h'$, program point~$p'$, and trace~$\tau$.
    We assume the premise and apply the induction hypothesis for the immediately preceding program point~$p$ as $\forall t \ldotp \neg\accessible{h'}{t}{p'}$ implies $\forall t \ldotp \neg\accessible{h'}{t}{p}$ due to monotonicity.
    We show that $\forall t \ldotp h' \not\in \programpointparen{\threadset{\setlocalheap}{t} \cup \ederef{\setglobalheap} \cup \threadset{\setinvariantheap}{t}}{p'}$ holds by analyzing the \glsdisp{ghost-code}{ghost operations} that \ghostalgorithm{} inserts for a statement~$s$.
    We assume without loss of generality that thread~$t_s$ executes $\ghostalgorithm(s)$, which transitions from $p$ to $p'$.
    We observe that every element that is added to \threadset{\setlocalheap}{t_s}, \ederef{\setglobalheap} or \threadset{\setinvariantheap}{t} is either the heap location to which a variable accessible by $t_s$ points or a set of heap locations that are reachable from such a variable.
    Since $h'$ by assumption is not accessible from any thread at $p'$, \ghostalgorithm{} does not add $h'$ to any \glsdisp{ghost-code}{ghost set}.
\end{proofsketchcomposition}

The next lemmata depend on certain requirements for a codebase, which we define next.
As we will see, successfully executing the static analyses implies that a codebase meets these requirements.

\begin{figure*}[t]
    \begin{align*}
        \requirements{p}{\tau} \triangleq \;
        % begin write constraints
        &(\forall s, x, e, \ell \ldotp s^{\ell} = \cheapwrite{x}{e} \land \ell \prec p \implies \appmanaged{\valueof{x}{\tau}}{\tau}{\prestatesuperscript{\ell}}) \land{}\\
        % end write constraints
        % begin core alloc constraints
        &(\forall s, c, e, \bar{e}, \ell \ldotp s^{\ell} = \ccorealloc{c}{\bar{e}} \land \ell \prec p \land e \in \set{\bar{e}} \implies{}\\
        &\phantom{(\forall s, c,} \valueof{e}{\tau} = \nil \lor \appmanaged{\valueof{e}{\tau}}{\tau}{\prestatesuperscript{\ell}} \land \local{e}{\poststatesuperscript{\ell}}) \land{}\\
        % end core alloc constraints
        % begin core api constraints
        &(\forall s, k, c, e, \bar{e}, r, \bar{r}, \ell \ldotp s^{\ell} = \ccorecall{k}{c}{\bar{e}}{\bar{r}} \land \ell \prec p \land e \in \set{\bar{e}} \land r \in \set{\bar{r}} \implies{}\\
        &\phantom{\forall s, k,} (\valueof{e}{\tau} = \nil \lor \appmanaged{\valueof{e}{\tau}}{\tau}{\prestatesuperscript{\ell}} \land \local{e}{\poststatesuperscript{\ell}}) \land{}\\
        &\phantom{\forall s, k,} (\valueof{c}{\tau} = \nil \lor \neg\appmanaged{\valueof{c}{\tau}}{\tau}{\prestatesuperscript{\ell}}) \land{}\\
        &\phantom{\forall s, k,} (\valueof{r}{\tau} = \nil \lor \local{r}{\poststatesuperscript{\ell}}))
        % end core api constraints
    \end{align*}
    \caption{%
        \requirements{p}{\tau} expresses requirements that all statements in a codebase before program point~$p$ on trace~$\tau$ must satisfy.
        These requirements allow us to relate properties of heap locations to containment in the \glsdisp{ghost-code}{ghost sets}.
        In particular, heap write statements must write to \app-managed heap locations only, arguments that are passed to the \core (\ie $\bar{e}$ in \ccoreallocshort{\bar{e}} and \ccorecall{k}{c}{\bar{e}}{\bar{r}} statements) must be \app-managed and local \emph{after} executing the statement unless they are \nil{}, the \core instance~$c$ must \emph{not} be \app-managed, and return arguments from the \core, \ie $\bar{r}$ in \ccorecall{k}{c}{\bar{e}}{\bar{r}}, must be local or \nil{}.
    }
    \label{fig:requirements}
\end{figure*}

\begin{lemma}[Locality implies set containment for \app-managed locations]
    \label{lem:ghost-set-containment-for-local-app-locations}
    An \app-managed heap location~$h$ is in thread~$t$'s \setlocalheap{} at program point~$p$ if $h$ is local, and in \ederef{\setglobalheap} if $h$ is accessible by multiple threads.
    Both cases hold if a codebase meets the requirements~$\requirements{p}{\tau}$ (\cf \figref{requirements}).
    \begin{align*}
        \forall h, t, &\tau, p \ldotp (h \neq \nil \land p \in \tau \land \accessible{h}{t}{p} \land {}\\
        &\phantom{\tau, p \ldotp }\appmanaged{h}{\tau}{p} \land \requirements{p}{\tau}) \implies {}\\
        &\left((\forall t' \ldotp t' = t \lor \neg\accessible{h}{t'}{p}) \iff h \in \programpointwoparen{\threadset{\setlocalheap}{t}}{p}\right) \land{}\\
        &\left((\exists t' \ldotp t' \neq t \land \accessible{h}{t'}{p}) \iff h \in \programpointwoparen{\ederef{\setglobalheap}}{p}\right)
    \end{align*}
\end{lemma}
\begin{proofsketchcomposition}
    We prove this lemma by induction over program traces.
    The base case for the empty trace holds trivially as there are no allocated and, thus, accessible heap locations yet.
    In the inductive step, we prove this lemma for an arbitrary heap location~$h'$, thread~$t$, program point~$p'$, and trace~$\tau$ by applying the induction hypothesis to the immediately preceding program point~$p$ and showing that we obtain the specified set containment for $p'$.
    \Ie we assume the premise and show that
    \begin{equation}
        \label{eq:ghost-set-containment-for-local-app-locations-goal}
        \begin{split}
            &((\forall t' \ldotp t' = t \lor \neg\accessible{h'}{t'}{p'}) \iff h' \in \programpointwoparen{\threadset{\setlocalheap}{t}}{p'}) \land{}\\
            &((\exists t' \ldotp t' \neq t \land \accessible{h'}{t'}{p'}) \iff h' \in \programpointwoparen{\ederef{\setglobalheap}}{p'})
        \end{split}
    \end{equation}
    holds.
    We case split on statement~$s$ (before applying \ghostalgorithm{}) such that executing $\ghostalgorithm(s)$ on thread~$t_s$ transitions from $p$ to $p'$.
    We first note that the restrictions~$r$ are monotonic when going backwards on a trace, \ie $\requirements{p}{\tau}$ follows from $\requirements{p'}{\tau}$.
    \begin{itemize}
        \item $s = \cskip$:
        Since \cskip{} does not allocate any heap locations and leaves accessibility unchanged, we get $\accessible{h'}{t}{p}$ and apply the induction hypothesis.
        Because algorithm~\ghostalgorithm{} leaves all \glsdisp{ghost-code}{ghost sets} unmodified, \eqref{eq:ghost-set-containment-for-local-app-locations-goal} holds.

        \item $s = \cheapalloc{x}$:
        If $h' = \valueof{x}{\tau}$, then $t = t_s$ as \accessible{h'}{t}{p'} holds and no other thread can access $h'$ yet.
        \ghostalgorithm{} adds $h'$ to \threadset{\setlocalheap{}}{t} and \eqref{eq:ghost-set-containment-for-local-app-locations-goal} holds as $h'$ is in no other \glsdisp{ghost-code}{ghost set} (by \lemref{ghost-set-absence}).
        Otherwise, $h'$ is already allocated at~$p$, and we apply the induction hypothesis to obtain \eqref{eq:ghost-set-containment-for-local-app-locations-goal} as \ghostalgorithm{} neither adds nor removes $h'$ to and from any \glsdisp{ghost-code}{ghost set}.

        \item $s = \cheapread{x}{e}$:
        Since $s$ neither allocates new heap locations nor changes accessibility of $h'$, \accessible{h'}{t}{p} holds, and we apply the induction hypothesis.
        If $h' = \valueof{e}{\tau}$, then \accessible{h'}{t_s}{p} and, thus, $h' \in \programpointparen{\threadset{\setlocalheap}{t_s} \cup \ederef{\setglobalheap}}{p}$ hold.
        Hence, \ghostalgorithm{} ensures $\forall t' \ldotp \programpointwoparen{\threadset{\setlocalheap}{t'}}{p'} = \programpointwoparen{\threadset{\setlocalheap}{t'}}{p}$ and $\programpointwoparen{\setlocalheap}{p'} = \programpointwoparen{\setlocalheap}{p}$.
        Otherwise, \ghostalgorithm{} neither adds nor removes $h'$ to and from any \glsdisp{ghost-code}{ghost set}.

        \item $s = \cheapwrite{x}{e}$:
        Since $s$ does not allocate new heap locations, \appmanaged{h'}{\tau}{p} holds.
        If $\valueof{x}{\tau} = h'$, then $h'$ is accessible by $t_s$, and we apply the induction hypothesis. Since \ghostalgorithm{} leaves $h'$ in the same \glsdisp{ghost-code}{ghost set}, \eqref{eq:ghost-set-containment-for-local-app-locations-goal} holds.
        Otherwise, we focus on the case $\valueof{x}{\tau} \in \programpointwoparen{\ederef{\setglobalheap}}{p} \land h' \in \reachabletrace{e}{\tau}{p} \cap \programpointwoparen{\threadset{\setlocalheap}{t_s}}{p}$ as \ghostalgorithm{} removes in this case $h'$ from \threadset{\setlocalheap}{t_s} and for all other cases guarantees that $h'$ remains in the same \glsdisp{ghost-code}{ghost set}.
        From $h' \in \programpointwoparen{\threadset{\setlocalheap}{t_s}}{p}$ and our induction hypothesis, we get $t = t_s$ as $h'$ is accessible only by a single thread.
        Since \accessible{\valueof{x}{\tau}}{t_s}{p} and \appmanaged{\valueof{x}{\tau}}{\tau}{p}  (from \requirements{p'}{\tau}) hold, we apply the induction hypothesis and obtain that another thread~$t'$ with $t' \neq t_s$ exists that can access \valueof{x}{\tau}.
        However, by writing $e$ to \valueof{x}{\tau}, all from $e$ reachable heap locations including $h'$ become accessible from $t'$ at $p'$.
        Since $h'$ is accessible at $p'$ from at least two~different threads, namely $t_s$ and $t'$, we have to show that $h' \in \programpointwoparen{\ederef{\setglobalheap}}{p'}$ and that $h'$ is removed from \threadset{\setlocalheap}{t_s}, which is guaranteed by \ghostalgorithm{}.

        \item $s = \ccorealloc{c}{\bar{e}}$:
        If $\exists e \ldotp e \in \bar{e} \land \valueof{e}{\tau} = h'$, we get \local{e}{p'} from $\requirements{p'}{\tau}$.
        Thus, $t_s = t$ as only a single thread can access $h'$.
        From \asmref{escape-analysis-soundness} and \lemref{local-monotonic}, \localheaplocationthread{h'}{t}{p} holds, and we apply the induction hypothesis to obtain $h' \in \programpointwoparen{\threadset{\setlocalheap}{t}}{p}$.
        \ghostalgorithm{} guarantees that $h'$ remains in \threadset{\setlocalheap}{t} and that $h'$ is not inserted into any other \glsdisp{ghost-code}{ghost set} since $h' \neq \valueof{c}{\tau}$.
        Otherwise, \accessible{h'}{t}{p} holds because $s$ cannot change $h'$'s accessibility as the arguments~$\bar{e}$ are shallow (\cf \asmref{syntactic-restrictions}) and, thus, $s$ internally does not have access to $h'$.
        We apply the induction hypothesis and observe that \ghostalgorithm{} does not change set containment of $h'$.

        \item $s = \ccorecall{k}{c}{\bar{e}}{\bar{r}}$:
        We reason similarly as in the case of \ccoreallocshort{\bar{e}} except that we consider a third case, namely $\exists r \ldotp r \in \bar{r} \land \valueof{r}{\tau} = h'$.
        In this case, $\requirements{p'}{\tau}$ guarantees that $h'$ is local and from \accessible{h'}{t}{p'} follows that $t = t_s$.
        $h'$ is a heap location newly allocated by $s$ and \ghostalgorithm{} guarantees that $h'$ is inserted into \threadset{\setlocalheap}{t}.
        From \lemref{ghost-set-absence}, we get that $h'$ is in no other \glsdisp{ghost-code}{ghost set}.

        \item $s = \cfork{\bar{x}}{s'}$:
        Let us call the newly spawned thread~$t'_s$ with $t'_s \neq t_s$.
        Since $t'_s$ can access the variables~$\bar{x}$, we have $\forall h \ldotp h \in \reachabletrace{\bar{x}}{\tau}{p} \implies \accessible{h}{t_s}{p'} \land \accessible{h}{t'_s}{p'}.$
        If $h' \not\in \reachabletrace{\bar{x}}{\tau}{p}$, then accessibility of $h'$ does not change by executing $s$, and we apply the induction hypothesis and note that \ghostalgorithm{} does not modify set containment of $h'$.
        In particular, $h'$ is not accessible by $t'_s$ and, thus, $h' \not\in \programpointwoparen{\threadset{\setlocalheap}{t'_s}}{p'}$ holds as required by \eqref{eq:ghost-set-containment-for-local-app-locations-goal}.
        Otherwise ($h' \in \reachabletrace{\bar{x}}{\tau}{p}$), we have to prove that $\forall t' \ldotp h' \not\in \programpointwoparen{\threadset{\setlocalheap}{t'}}{p'}$ and $h' \in \programpointwoparen{\ederef{\setglobalheap}}{p'}$ hold.
        Since \accessible{h'}{t_s}{p} holds, we apply the induction hypothesis and case split on whether $h' \in \programpointwoparen{\threadset{\setlocalheap}{t_s}}{p}$ holds.
        If so, \ghostalgorithm{} moves $h'$ from \threadset{\setlocalheap}{t_s} to \ederef{\setglobalheap}, which is sufficient as $\forall t' \ldotp t' \neq t_s \implies h' \not\in \programpointwoparen{\threadset{\setlocalheap}{t'}}{p}$ holds.
        Otherwise, $h' \in \programpointwoparen{\ederef{\setglobalheap}}{p}$ holds and \ghostalgorithm{} ensures $h' \in \programpointwoparen{\ederef{\setglobalheap}}{p'}$.
    \end{itemize}
\end{proofsketchcomposition}

\begin{corollary}[Set containment in $\setlocalheap \cup \ederef{\setglobalheap}$]
    \label{cor:ghost-set-containment-for-heap-locations}
    A variable~$x$ is in a thread~$t$'s \threadset{\setlocalheap}{t} or \ederef{\setglobalheap} at program point~$p$ if $x$ is a defined variable, all heap locations $x$ may point to are \app{}-managed, and the requirements~\requirements{p}{\tau} hold.
    \begin{align*}
        \forall x, t, &p, \tau \ldotp p \in \tau \land \defined{x}{t}{p} \land \requirements{p}{\tau} \land{}\\
        &\phantom{p, \tau \ldotp } (\forall h \ldotp \allocationsite{h}{\tau} \in \pointstoset{x} \implies \appmanaged{h}{\tau}{p}) \implies {}\\
        &\valueof{x}{\tau} = \nil \lor \valueof{x}{\tau} \in \programpointparen{\threadset{\setlocalheap}{t} \cup \setglobalheap}{p}
    \end{align*}
    where \defined{x}{t}{p} expresses that $x$ is defined at $p$ for thread~$t$.
\end{corollary}
\begin{proofsketchcomposition}
    Let $x$, $t$, $p$, and $\tau$ be arbitrary and assume the corollary's premise.
    If $\valueof{x}{\tau} = \nil$ holds, then the corollary holds trivially.
    Otherwise, $x$ points at $p$ to an allocated heap location, which we call $h'$, that is, thus, accessible from thread~$t$ \ie $h' = \valueof{x}{\tau} \land \accessible{h'}{t}{p}$.
    From \asmref{pointer-analysis-soundness} we obtain $\allocationsite{h'}{\tau} \in \pointstoset{x}$ and, thus, $\appmanaged{h'}{\tau}{p}$ holds.
    We apply \lemref{ghost-set-containment-for-local-app-locations} and observe that one of the equivalences' left-hand sides must be satisfied.
    Therefore, $h'$ is either in \programpointwoparen{\threadset{\setlocalheap}{t}}{p} or \programpointwoparen{\ederef{\setglobalheap}}{p}.
\end{proofsketchcomposition}

\begin{lemma}[Locality implies set containment for \core~instances]
    \label{lem:ghost-set-containment-for-core-instances}
    A heap location~$h$ at program point~$p$ that corresponds to a \core instance returned from an earlier \ccoreallocshort{\bar{e}} statement is in a thread $t$'s \setinvariantheap{} if $h$ is local and the restrictions~$\requirements{p}{\tau}$~(\figref{requirements}) hold.
    \begin{align*}
        \forall h, t, \tau, p \ldotp &h \neq \nil \land p \in \tau \land \localheaplocationthread{h}{t}{p} \land{}\\
        &\passthroughcoretrace{h}{\tau}{p} \land \requirements{p}{\tau} \implies h \in \programpointwoparen{\threadset{\setinvariantheap}{t}}{p}
    \end{align*}
\end{lemma}
\begin{proofsketchcomposition}
    We prove this lemma by induction over program traces.
    The base case for the empty trace holds trivially as there are no allocated heap locations yet.
    In the inductive step, we prove this lemma for an arbitrary heap location~$h'$, thread~$t$, program point~$p'$, and trace~$\tau$ by applying the induction hypothesis to the immediately preceding program point~$p$ and showing that we obtain the specified set containment for $p'$.
    \Ie we assume the premise and show that $h' \in \programpointwoparen{\threadset{\setinvariantheap}{t}}{p'}$ holds.
    We case split on statement~$s$ (before applying \ghostalgorithm{}) such that executing $\ghostalgorithm(s)$ on thread~$t_s$ transitions from $p$ to $p'$.
    We first note that the restrictions~$r$ are monotonic when going backwards on a trace, \ie $\requirements{p}{\tau}$ follows from $\requirements{p'}{\tau}$.
    \begin{itemize}
        \item $s = \cskip$:
        Since \cskip{} does not allocate \core instances and leaves accessibility unchanged, we get $\localheaplocationthread{h'}{t}{p}$ and apply the induction hypothesis.
        We get $h' \in \programpointwoparen{\threadset{\setinvariantheap}{t}}{p'}$ as \ghostalgorithm{} leaves all \glsdisp{ghost-code}{ghost sets} unmodified.

        \item $s = \cheapalloc{x}$:
        $h' \neq \valueof{x}{\tau}$ holds because $x$ points to a newly allocated heap location that has not been passed through the return argument of \ccoreallocshort{\bar{e}}.
        Thus, \localheaplocationthread{h'}{t}{p} holds, and we apply the induction hypothesis.
        We observe that \ghostalgorithm{} leaves \threadset{\setinvariantheap}{t} unchanged.

        \item $s = \cheapread{x}{e}$:
        Since $s$ does not allocate \core instances, \passthroughcoretrace{h'}{t}{p} holds, and we apply the induction hypothesis.
        The lemma holds as \ghostalgorithm{} does not modify \threadset{\setinvariantheap}{t}.

        \item $s = \cheapwrite{x}{e}$:
        Identical reasoning as for reading a heap location.

        \item $s = \ccorealloc{c}{\bar{e}}$:
        If $\valueof{c}{\tau} = h'$, then \localheaplocationthread{h'}{t}{p'} implies $t = t_s$. \ghostalgorithm{} guarantees that $h' \in \programpointwoparen{\threadset{\setinvariantheap}{t}}{p'}$.
        Otherwise, \localheaplocationthread{h'}{t}{p} and \passthroughcoretrace{h'}{\tau}{p} hold, and we apply the induction hypothesis.
        $h' \in \programpointwoparen{\threadset{\setinvariantheap}{t}}{p'}$ holds as \ghostalgorithm{} does not remove elements from \setinvariantheap{}.

        \item $s = \ccorecall{k}{c}{\bar{e}}{\bar{r}}$:
        Since $s$ does not allocate \core instances, \localheaplocationthread{h'}{t}{p} and \passthroughcoretrace{h'}{\tau}{p} hold, and we apply the induction hypothesis.
        Furthermore, \ghostalgorithm{} does not remove elements from \threadset{\setinvariantheap}{t'} for any thread~$t'$.

        \item $s = \cfork{\bar{x}}{s'}$:
        Let us call the newly spawned thread~$t'_s$ with $t'_s \neq t_s$.
        If $\accessible{h'}{p}{t'_s}$, then $t = t'_s$ as $h'$ is local.
        However, $h'$ can only be accessible to $t'_s$ if $h'$ is reachable from $\bar{x}$, which is accessible from thread~$t_s$ too.
        \Ie \accessible{h'}{p}{t_s} holds contradicting \localheaplocationthread{h'}{t}{p'}.
        Otherwise, \ghostalgorithm{} initializing \threadset{\setinvariantheap}{t'_s} to the empty set does not violate the lemma as $t'_s$ cannot access $h'$.
        Furthermore, we apply the induction hypothesis as \passthroughcoretrace{h'}{\tau}{p} holds, and we note that \ghostalgorithm{} does not remove any element from \threadset{\setinvariantheap}{t}.
    \end{itemize}
\end{proofsketchcomposition}

\begin{corollary}[Escape analysis implies set containment in \setinvariantheap{}]
    \label{cor:ghost-set-containment-for-core-instances}
    A variable~$x$ is in a thread~$t$'s \threadset{\setinvariantheap}{t} at program point~$p$ if $x$ is a defined variable, local, all heap locations $x$ may point to passed through the return parameter of some \ccoreallocshort{\bar{e}}, and the requirements~\requirements{p}{\tau} hold.
    \begin{align*}
        \forall x, t, &p, \tau \ldotp p \in \tau \land \local{x}{p} \land \defined{x}{t}{p} \land \requirements{p}{\tau} \land{}\\
        &\phantom{p, \tau \ldotp }(\forall h \ldotp \allocationsite{h}{\tau} \in \pointstoset{x} \implies \passthroughcoretrace{h}{\tau}{p}) \implies {}\\
        &\valueof{x}{\tau} = \nil \lor \valueof{x}{\tau} \in \programpointwoparen{\threadset{\setinvariantheap}{t}}{p}
    \end{align*}
\end{corollary}
\begin{proofsketchcomposition}
    Let $x$, $t$, $p$, and $\tau$ be arbitrary and assume the corollary's premise.
    If $\valueof{x}{\tau} = \nil$ holds, then the corollary holds trivially.
    Otherwise, $x$ points at $p$ to an allocated heap location, which we call $h'$, which is, thus, accessible from thread~$t$, \ie $h' = \valueof{x}{\tau} \land \accessible{h'}{t}{p}$.
    \localheaplocationthread{h'}{t}{p} follows from \asmref{escape-analysis-soundness}.
    From \asmref{pointer-analysis-soundness} we obtain $\allocationsite{h'}{\tau} \in \pointstoset{x}$ and, thus, $\passthroughcoretrace{h'}{\tau}{p}$.
    Applying \lemref{ghost-set-containment-for-core-instances} completes the proof.
\end{proofsketchcomposition}

Having defined the properties that successfully executing our static analyses provides, we present next how we apply the static analyses in \approach (\defref{diodon-judgements-appendix}) and prove in \lemref{side-conditions-hold} that this application discharges the side conditions~$\omega$ (\cf \figref{side-conditions-appendix}).

\looseness=-1
As shown in \defref{diodon-judgements-appendix}, we check for every heap read operation~\cheapread{x}{e} that $e$ points to \app-managed heap locations, which are identified by their allocation site~$a$.
Analogously, we check for heap writes~\cheapwrite{x}{e} that $x$ satisfies the same property.
For every \ccoreallocshort{\bar{e}} and \ccorecall{k}{c}{\bar{e}}{\bar{r}}, we check that the arguments~$\bar{e}$ point to disjoint heap locations and that these heap locations are local and \app-managed.
Additionally, we check for \ccorecall{k}{c}{\bar{e}}{\bar{r}} that $c$ points to a local \core instance, \ie a local heap location that has been returned by an earlier \core allocation call, and that the outputs~$\bar{r}$ are local.

\begin{definition}[Static analyses for \approach]
    \label{def:diodon-judgements-appendix}
    In \approach, we execute the static analyses on a codebase to obtain the following judgments for every statement~$s$ at label~$\ell$ therein, denoted as $\judgements{s^\ell}$.
    \begin{align*}
        \judgements{\cheapread{x}{e}} \triangleq \;
            &\forall a, \tau \ldotp a \in \pointstoset{e} \implies {}\\
            &\phantom{\forall a, \tau \ldotp } \appmanaged{a}{\tau}{\prestatesuperscript{\ell}}\\
        \judgements{\cheapwrite{x}{e}} \triangleq \;
            &\forall a, \tau \ldotp a \in \pointstoset{x} \implies {}\\
            &\phantom{\forall a, \tau \ldotp} \appmanaged{a}{\tau}{\prestatesuperscript{\ell}}\\
        \judgements{\ccorealloc{c}{\bar{e}}} \triangleq \;
            &\disjointallocationsites{\bar{e}} \land \localappmanaged{\bar{e}}{\ell}\\
        \judgements{\ccorecall{k}{c}{\bar{e}}{\bar{r}}} \triangleq \;
            &\disjointallocationsites{\bar{e}} \land \localappmanaged{\bar{e}}{\ell}\\
            &\land \localcore{c}{\ell} \land \localreturn{\bar{r}}{\ell}
    \end{align*}
    where
    \begin{align*}
        \disjointallocationsites{\bar{e}} \triangleq \;
            &\forall i, j \ldotp 0 \leq i < j < \len{\bar{e}} \implies {}\\
            &\phantom{\forall i, j \ldotp} \pointstoset{\at{\bar{e}}{i}} \cap \pointstoset{\at{\bar{e}}{j}} = \emptyset\\
        \localappmanaged{\bar{e}}{\ell} \triangleq \;
            &\forall e, h, \tau \ldotp e \in \set{\bar{e}} \land \allocationsite{h}{\tau} \in \pointstoset{e} \implies {}\\
            &\phantom{\forall e, h, \tau \ldotp} \local{e}{\poststatesuperscript{\ell}} \land \appmanaged{h}{\tau}{\prestatesuperscript{\ell}}\\
        \localcore{c}{\ell} \triangleq \;
            &\forall h, \tau \ldotp \allocationsite{h}{\tau} \in \pointstoset{c} \implies {}\\
            &\phantom{\forall h, \tau \ldotp} \local{c}{\prestatesuperscript{\ell}} \land \passthroughcoretrace{h}{\tau}{\prestatesuperscript{\ell}}\\
        \localreturn{\bar{r}}{\ell} \triangleq \;
            &\forall r, \tau \ldotp r \in \set{\bar{r}} \implies \local{r}{\poststatesuperscript{\ell}}
    \end{align*}
\end{definition}

\begin{lemma}[Discharging the requirements]
    \label{lem:requirements-hold}
    We show that the judgments provided by our static analyses~$\judgements{s^\ell}$ (\cf \defref{diodon-judgements-appendix}) for every statement~$s$ at label~$\ell$ before program point~$p$ and our assumptions are sufficient to discharge the requirements~\requirements{p}{\tau} (\cf \figref{requirements}).
    \begin{align*}
        \forall p, \tau \ldotp p \in \tau \land (\forall s, \ell \ldotp \ell \prec p \land \judgements{s^\ell}) \implies \requirements{p}{\tau}
    \end{align*}
\end{lemma}
\begin{proofsketchcomposition}
    We prove this lemma by induction over program traces.
    The base case for the empty trace holds trivially as there are no preceding statements~$s^\ell$.
    In the inductive step, we prove this lemma for an arbitrary program point~$p'$ and trace~$\tau$ by applying the induction hypothesis to the immediately preceding program point~$p$.
    \Ie we show that $\forall s', \ell' \ldotp \ell' \prec p' \land \judgements{s'^{\ell'}}$ and \requirements{p}{\tau} imply \requirements{p'}{\tau} by case splitting on statement~$s^\ell$, whose execution transitions from $p$ to $p'$.
    \begin{itemize}
        \item $s^\ell = \cheapwrite{x}{e}$:
        We have to prove that \appmanaged{\valueof{x}{\tau}}{\tau}{p} holds.
        From \judgements{s^{\ell}} and \asmref{pointer-analysis-soundness}, we get $\valueof{x}{\tau} = \nil \lor \appmanaged{\valueof{x}{\tau}}{\tau}{p}$.
        $x \neq \nil$ holds as the statement would otherwise crash (\cf \asmref{crash-freedom}).

        \item $s^\ell = \ccorealloc{c}{\bar{e}}$:
        We have to show for an arbitrary argument~$e \in \set{\bar{e}}$ that $\valueof{e}{\tau} = \nil \lor \appmanaged{\valueof{e}{\tau}}{\tau}{p} \land \local{e}{p'}$ holds.
        If $\valueof{e}{\tau} \neq \nil$, then we apply \asmref{pointer-analysis-soundness} to obtain \appmanaged{\valueof{e}{\tau}}{\tau}{p} from \localappmanaged{\bar{e}}{\ell}.

        \item $s^\ell = \ccorecall{k}{c}{\bar{e}}{\bar{r}}$:
        We proceed identically as in the case of \ccoreallocshort{\bar{e}}.
        Additionally, we have to show $\valueof{c}{\tau} = \nil \lor \neg\appmanaged{\valueof{c}{\tau}}{\tau}{p}$ and $\valueof{r}{\tau} = \nil \lor \local{r}{p'}$ for an arbitrary return argument~$r \in \bar{r}$, which we get from \localcore{c}{\ell} by applying \asmref{pointer-analysis-soundness} and \localreturn{\bar{r}}{\ell}.

        \item Otherwise:
        \requirements{p'}{\tau} holds because no requirements for $s^\ell$ must be met.
    \end{itemize}
\end{proofsketchcomposition}

\begin{lemma}[Discharging the side conditions~$\omega$]
    \label{lem:side-conditions-hold}
    We show that the judgments provided by our static analyses~$\judgements{s}$ (\cf \defref{diodon-judgements-appendix}) for every statement~$s$ in a codebase~$c$ together with our assumptions are sufficient to discharge the side conditions~$\omega(s)$ (\cf \figref{side-conditions-appendix}).
    \begin{align*}
        \forall s \in c \ldotp (\forall s' \in c \ldotp \judgements{s'}) \implies \omega(s)
    \end{align*}
\end{lemma}
\begin{proofsketchcomposition}
    We prove this lemma for an arbitrary statement~$s$ at label~$\ell$ such that $s\in c$, assume $\forall s' \in c \ldotp \judgements{s'}$ and show that $\omega(s)$ holds by case splitting on statement~$s$.
    Throughout the proof, we use program point~$p$ to refer to $s$'s pre-state, \ie $p \triangleq \; \prestatesuperscript{\ell}$.
    We obtain $\forall \tau \ldotp \requirements{p}{\tau}$ from \lemref{requirements-hold}.
    \begin{itemize}
        \item $s = \cheapread{x}{e}$:
        From \corref{ghost-set-containment-for-heap-locations}, we get $\valueof{e}{\tau} = \nil \lor \valueof{e}{\tau} \in \programpointparen{\setlocalheap \cup \setglobalheap}{p}$.
        $e$ points to an allocated heap location and cannot be \nil as the statement would otherwise crash (\cf \asmref{crash-freedom}).

        \item $s = \cheapwrite{x}{e}$:
        Analogous to heap reads but for $x$ instead of $e$.

        \item $s = \ccorealloc{c}{\bar{e}}$:
        From \disjointallocationsites{\bar{e}}, we obtain $\disjoint{\bar{e}}$ by applying \lemref{pts-empty-intersection-implies-disjoint}.
        \localappmanaged{\bar{e}}{\ell} allows us to apply \lemref{ghost-set-containment-for-local-app-locations} providing $\forall e \in \set{\bar{e}} \ldotp \valueof{e}{\tau} = \nil \lor \valueof{e}{\tau} \in \programpointwoparen{\setlocalheap}{p}$.
        Lastly, $\ederef{\flagiospec} = \false$ holds by our assumption that we have a single \core{} allocation statement in the codebase~$c$. We lift this assumption in \secref{diodon-soundness-extensions}. 

        \item $s = \ccorecall{k}{c}{\bar{e}}{\bar{r}}$:
        Likewise to the previous case, we obtain $\disjoint{\bar{e}}$ and $\forall e \in \set{\bar{e}} \ldotp \valueof{e}{\tau} = \nil \lor \valueof{e}{\tau} \in \programpointwoparen{\setlocalheap}{p}$.
        Left to show is $\valueof{c}{\tau} = \nil \lor \valueof{c}{\tau} \in \programpointwoparen{\setinvariantheap}{p}$, which we obtain from \localcore{c}{\ell} by applying \corref{ghost-set-containment-for-core-instances}.

        \item Otherwise: $\omega(s) = \true$ and, thus, the lemma holds trivially.
    \end{itemize}
\end{proofsketchcomposition}

\ifthenelse{\boolean{show_composition_soundness_proofs}}{%
    \begin{prooftreefigure*}
\begin{prooftree}
\[
    \[
        \[
            \leadsto
            \ctxhoare
                {\emp}
                {\emp}
                {s_\text{init}}
                {R_g \star R_l}
        \]
        \justifies
        \ctxhoare
            {\emp}
            {\phi}
            {s_\text{init}}
            {R_g \star R_l \star \phi}
        \using\rnframe
    \]
    \justifies
    \ctxhoare
        {\emp}
        {\phi}
        {s_\text{init}}
        {\globalproginv \star \localproginv}
    \using\rnconseq
\]
\[
    \[
        \[
            \[
                \leadsto
                \ctxhoare
                    {\globalproginv}
                    {\localproginv}
                    {\ghostalgorithm(p)}
                    {\localproginv}
                \using\figref{proof-rules-appendix}
            \]
            \justifies
            \ctxhoare
                {\globalproginv}
                {\localproginv}
                {\ghostalgorithm(p)}
                {\true}
            \using\rnconseq
        \]
        \justifies
        \ctxhoare
            {\emp}
            {\globalproginv \star \localproginv}
            {\ghostalgorithm(p)}
            {\globalproginv}
        \using\rnshare
    \]
    \justifies
    \ctxhoare
        {\emp}
        {\globalproginv \star \localproginv}
        {\ghostalgorithm(p)}
        {\true}
    \using\rnconseq
\]
\justifies
\ctxhoare
    {\emp}
    {\phi}
    {s_\text{init} \cseq \ghostalgorithm(p)}
    {\true}
\using\rnseq
\end{prooftree}
\vspace{12pt}
\begin{align*}
    \text{with} \quad
    R_l &\triangleq \; \setlocalheap = \emptyset \star \setinvariantheap = \emptyset &
    R_g &\triangleq \; \acc{\setglobalheap} \star \ederef{\setglobalheap} = \emptyset \star \acc{\flagiospec} \star \ederef{\flagiospec} = \false
\end{align*}
\caption{%
    Proof tree showing the initial establishment of \localproginv{} and \globalproginv{} for a codebase~$p$.
    We assume that the \glsdisp{ghost-code}{ghost statement}~$s_\text{init}$ initializes \setlocalheap{} and \setinvariantheap{} to the empty set, as stated in $R_l$, and allocates two~heap locations on the \glsdisp{ghost-code}{ghost heap} storing $\emptyset$ and \false{} to which \setglobalheap{} and \flagiospec{} point, respectively (\cf $R_g$).
}
\label{fig:whole-program-prooftree}
\end{prooftreefigure*}

}{}%

\subsubsection{Proof Construction}
\label{app:diodon-soundness-composition-proof-construction}
While we showed that we can compose the proof rules in \figref{proof-rules-appendix} and discharge their side conditions~$\omega$, it remains to show that we initially establish the global context~\globalproginv{} and the local program invariant~\localproginv{}, such that we obtain a proof for the entire codebase~\prog{}.
We close this gap in \corref{proof-construction-appendix}.

We show that we obtain the desired proof for the entire codebase, namely that the codebase satisfies the I/O specification~$\phi$ expressed as the Hoare triple $\emp \vdash \simpleHoare{\phi}{s_\text{init} \cseq \ghostalgorithm(\prog)}{\true}$.
This Hoare triple relies on $s_\text{init}$ that initializes \setlocalheap{}, \setinvariantheap{}, and \ederef{\setglobalheap} to empty sets, as well as sets the \glsdisp{ghost-code}{ghost flag} \ederef{\flagiospec} to \false.
$s_\text{init}$ is similar in spirit to the \glsdisp{ghost-code}{ghost statements} that algorithm~\ghostalgorithm{} inserts as these statements are necessary to construct a proof for the codebase~\prog{}.
\Corref{proof-construction-mainbody}'s premise states that our static analyses succeed on the codebase~\prog{}, such that we obtain \judgements{s} for each statement~$s$ therein, and that we prove a Hoare triple for each \core function satisfying the syntactic restrictions.

We combine the proof for the entire codebase that we obtain from \corref{proof-construction-appendix} with the result of \appref{diodon-soundness-io-independence} to obtain \approach's overall soundness result.
This result states that successfully executing our static analyses on codebase~$\prog{}$ and auto-actively verifying its \core suffices to prove that the traces of executing $\prog{}$ together with other verified implementations and the environment are contained in the traces described by the abstract protocol model.

\begin{theorem}[Overall soundness]
    \label{thm:diodon-soundness-proof-overall-soundness}
    Suppose \asmref{diodon-verifier-assumption} holds and that we have established, for each role~$i$, \ioindependence and \corref{proof-construction-appendix}'s antecedent for a codebase~$c_i(\rid)$ and I/O specification $\psi_i(\rid)$.
    Then
    \begin{equation*}
        (\interl_{i,\rid} \pi_\text{int}(\concretelts{i}{\rid})) \sync{\csyncrelabeledFn} \realEnv \tracePre_t \RR.
    \end{equation*}
\end{theorem}
\begin{proofsketch}
    We apply \corref{proof-construction-appendix} to obtain for each role~$i$ $\emp \vdash \simpleHoare{\psi_i(\rid)}{s_\text{init} \cseq \ghostalgorithm(c_i(\rid))}{\true}$.
    Since we omitted the turnstile subscript~$\alpha$ (\cf \asmref{diodon-verifier-assumption}) throughout \appref{diodon-composition-soundness} for brevity and $\emp$ on the turnstile's left-hand side is a notational difference only (\asmref{diodon-verifier-assumption} could be adapted accordingly), we apply \thmref{diodon-ioindependence-soundness} to obtain the desired result.
\end{proofsketch}

\subsubsection{Limitations}
\label{app:diodon-soundness-composition-limitations}
Our formalization defines a simple programming language to focus on the main ideas of our soundness proof and to show that successfully executing our static analyses discharges all side conditions.
We believe this language covers the most critical features like heap manipulations and concurrency as these features are relevant for the results of our static analyses.
In addition, we abstract each function making up the \core's \ac{API} to a dedicated statement in our language, and assume that the specification of each such function satisfies our syntactic restrictions~\refsyntacticrestrictionsassumption.
However, there is a slight risk that this language misses Go features that would be a threat to soundness such as function boundaries, complex control flow, and callbacks;
the former two features would be straightforward to add, and we discuss in \secref{diodon-soundness-extensions} how to add the latter.

To prove a Hoare triple for the entire codebase, we assume that the \app is free of crashes~\refcrashfreedomassumption and data races~\refdataracefreedomassumption.
While our soundness proof does not make any statement in the case that the program crashes, our compositional proof informally guarantees that the trace inclusion holds for the program's prefix up to the program point at which a crash occurs, such that the crash freedom assumption could be dropped, which we leave to future work.
However, data race freedom remains an assumption;
more generally, we assume the absence of undefined behavior for programming languages other than Go and our formalized one.
This assumption can be mitigated by performing additional static analyses.

\subsubsection{Extensions}
\label{app:diodon-soundness-composition-extensions}

}{}

\ifthenelse{\boolean{show_full_ssm_agent_protocol}}{%
    \newpage
\section{Secure Shell Session Protocol}
\label{app:diodon-full-protocol}

\begin{figure}[t]
\footnotesize % a bit smaller than "\small"
% reduce spacing above and below:
\setlength{\abovedisplayskip}{0pt}
\setlength{\belowdisplayskip}{0pt}
\setboolean{complete_protocol}{true}
\input{sections/06_protocol}
\begin{alignat*}{2}
    \text{where} \qquad & \sigX && \triangleq \; \sigXfull \\
    & \sigY && \triangleq \; \sigYfull \\
    & \sessciphertext && \triangleq \; \sessciphertextfull \\
    & \sigSS && \triangleq \; \sigSSfull
\end{alignat*}
\caption{%
    Signed \acs{DH} key exchange for deriving the symmetric keys~$\kdfssone$ and $\kdfsstwo$ that are used during the transport phase, \ie in  messages~\symb{M15} and \symb{M16}. We use $\rightarrow$ and $\Rightarrow$ to denote communication via the untrusted network and a secure channel, respectively.
}
\label{fig:secure-sessions-full}
\end{figure}

\Figref{secure-sessions-full} shows the protocol for establishing interactive shell sessions between an \ac{SSM Agent}~(A) and a customer~(B).
The protocol includes two~additional roles namely \ac{KMS}~(S) and an optional, trusted monitor~(M) that is allowed to inspect the established shell sessions, \eg for compliance reasons.

Since A and B do not personally possess their secret keys for creating signatures, we explicitly model the presence of and the interactions with \ac{KMS} that remotely creates and checks signatures.
We model these interactions as happening on a secure channel, indicated by $\Rightarrow$, because each role instance of A and B establishes a TLS connection to \ac{KMS}.

On a high-level, this protocol performs a signed elliptic-curve \ac{DH} key exchange establishing two~symmetric keys~$\kdfssone$ and $\kdfsstwo$.
These keys are used in the transport phase, \ie \symb{M15} and \symb{M16}, to symmetrically encrypt (\symb{senc}) payloads for sending in a particular direction.
In \tamarin, we model the transport phase as a non-deterministic loop that allows each role~A and~B to send and receive an unbounded number of transport messages and interleave them arbitrarily.

\looseness=-1
More specifically, the protocol proceeds as follows.
Role~A first generates an elliptic-curve public-private key pair, which we model in \tamarin as generating a fresh term~$x$ and computing the corresponding public key via modular exponentiation denoted by $g^x$.
Then, A sends message~\symb{M1} to instruct \ac{KMS} to use a particular signing key belonging to A, identified by $\agentltkeyid$, to sign the triple~$\langle g^x, \readerid, \clientid \rangle$.
This triple includes the monitor's and B's identity to prevent \acf{MITM} attacks.
\ac{KMS} checks whether the requested signing key actually belongs to A before creating and sending the signature in \symb{M2} back to A.
This allows A to send a session request~(\symb{M3}) to B, which includes $g^x$, the signature, and the signing key's and monitor's identities.

After receiving a session request, B first checks the received signature via \ac{KMS}.
For this purpose, B sends the signature itself and the components over which the signature is computed in a signature check request~(\symb{M4}) to \ac{KMS}.
If the signature is valid, \ac{KMS} replies with a signature check response~(\symb{M5}).
Otherwise, \ac{KMS} aborts the protocol, which we model as not sending any response.
Afterwards, B generates its elliptic-curve public-private key pair~($g^y$, $y$) and uses \ac{KMS} to sign $g^y$ and A's identity.
B then sends a session response~(\symb{M8}) to A that contains B's public curve point, the signature, the identity of B's signing key, and a hash of the shared secret~$\sesshash$.
The latter allows A to detect early on if A and B computed different shared secrets, \eg due to an attempted replay attack.

After receiving a session response, A computes the shared secret and checks that it derives the same shared secret's hash value.
Additionally, A checks the received signature using \ac{KMS} and derives the two~symmetric session keys from the shared secret by applying two~different \acfp{KDF}~\symb{kdf1} and~\symb{kdf2}.
To enable a trusted monitor~M to audit the shell session, A computes $\sessciphertext$ by asymmetrically encrypting the two~session keys using the monitor's public key~\readerpk.
Next, role~A requests a signature from \ac{KMS} for $\sessciphertext$ and B's identity to bind these identities to the session keys.
The handshake ends by sending the encrypted session keys to the monitor~(\symb{M13}) and confirming the session keys to B~(\symb{M14}).
The latter message includes some version information, which we model as an attacker-chosen payload~\symb{z}.

Message~\symb{M13} enables M, a trusted third party, to monitor the transmitted shell commands should this be necessary for regulatory reasons (otherwise sending message~\symb{M13} can simply be skipped).
For this purpose, role~A sends the asymmetrically encrypted session keys to the monitor~M such that M can obtain the session keys and, thus, decrypt and audit the transport messages.
Note that the monitor does not need to be online during the handshake or transport phase; it is sufficient for the monitor to come online at a later time as an untrusted log server could store message~\symb{M13} and all messages sent during the transport phase until M becomes online and fetches these messages from the log server.

}{}

%%%%%%%%%%%%%%%%%%%%%%%%%%%%%%%%%%%%%%%%%%%%%%%%%%%%%%%%%%%%%%%%%%%%%%%%%%%%%%%%
\end{document}
%%%%%%%%%%%%%%%%%%%%%%%%%%%%%%%%%%%%%%%%%%%%%%%%%%%%%%%%%%%%%%%%%%%%%%%%%%%%%%%%